\documentclass[preprint,review,3p,times]{elsarticle}

\usepackage{amssymb}
\usepackage[fleqn]{amsmath}
\biboptions{sort&compress}
\usepackage{lipsum}
\usepackage{graphicx}
\usepackage{epstopdf}
\usepackage{hyperref}       
\usepackage{url}            
\usepackage{booktabs}       
\usepackage{amsfonts}       
\usepackage{nicefrac}       
\usepackage[dvipsnames]{xcolor}         
\usepackage{graphicx}
\usepackage{tikz}
\usetikzlibrary{fit,calc}
\usepackage{tikz-network}
\usepackage[noend]{algpseudocode}
\usepackage{algorithm}
\usepackage{multirow}
\usepackage{placeins}
\usepackage{titlesec}
\usepackage{amsthm}
\usepackage[mathscr]{eucal}
\usepackage[capitalize]{cleveref}
\usepackage[caption=false]{subfig}
\usepackage[export]{adjustbox}
\usepackage{xspace}
\ifpdf
  \DeclareGraphicsExtensions{.eps,.pdf,.png,.jpg}
\else
  \DeclareGraphicsExtensions{.eps}
\fi

\usepackage[inline,shortlabels]{enumitem}
\setlist[enumerate]{leftmargin=.5in}
\setlist[itemize]{leftmargin=.5in}

\newcommand{\ie}{\emph{i.e.}\xspace}

\newcommand{\eg}{\emph{e.g.}\xspace}
\newcommand{\etal}{\emph{et al.}\xspace}

\newcommand{\nat}{\mathbb{N}}
\newcommand{\real}{\mathbb{R}}

\newcommand{\ten}[1]{\mathscr{#1}}
\newcommand{\matric}[2]{#1^{(#2)}}

\newcommand{\size}[1]{\mathtt{size}\left(#1\right)}
\newcommand{\nodes}{\mathcal{V}}
\newcommand{\edges}{\mathcal{E}}

\newcommand{\data}[2]{\mathscr{R}_{#1}(#2)}
\newcommand{\kargmin}{\operatorname{K-argmin}}

\newcommand{\norm}[1]{\lVert#1\rVert_{F}}
\newcommand{\bigo}[1]{\smash{\mathcal{O}\left(#1\right)}}
\newcommand{\sv}[2]{\sigma_{#1}\left(#2\right)}
\newcommand{\effrank}[1]{\mathcal{H}\left(#1\right)}

\newcommand{\net}{\mathcal{N}}
\newcommand{\subnet}{\mathcal{S}}
\newcommand{\netwithnodes}[2]{#1\left({#2}\right)}
\newcommand{\error}{\varepsilon}
\newcommand{\indsi}{\mathcal{I}}
\newcommand{\indsj}{\mathcal{J}}

\newcommand{\searchres}[1]{#1^{\star}}

\newcommand{\algoname}{\textsc{HiST}\xspace}
\newcommand{\cadet}{\textsc{Cadet}\xspace}
\newcommand{\dimthreshold}{d_{\theta}}
\newcommand{\recsearch}{RecSearch\xspace}
\newcommand{\mergeindices}{ClusterIndices\xspace}
\newcommand{\mergeindicescall}{\textsc{\mergeindices}\xspace}
\newcommand{\structsearch}{StructureSearch}
\newcommand{\structsearchcall}{\textsc{\structsearch}\xspace}

\newcommand{\topreshape}{TopReshape\xspace}
\newcommand{\topreshapecall}{\textsc{\topreshape}\xspace}

\newcommand{\svd}[1]{\texttt{SVD}\left(#1\right)}
\newcommand{\reshape}[2]{\texttt{reshape}\left(#1, \left[#2\right]\right)}
\newcommand{\bipar}[1]{#1 \mid \overline{#1}}
\newcommand{\factthreshold}{k_{\theta}}

\newcommand{\sample}{StochasticSample\xspace}
\newcommand{\tnround}{TNSR\xspace}
\newcommand{\subnetsize}{N_\theta}
\newcommand{\clustercost}{\mathscr{E}}
\newcommand{\clustercands}{C_\theta}
\newcommand{\psample}{PrioritySample\xspace}
\newcommand{\maxiters}{T_\theta}

\algrenewcommand\algorithmicrequire{\textbf{Input}}
\algrenewcommand\algorithmicensure{\textbf{Output}}
\algnewcommand{\Yield}[1]{\textbf{yield} #1}
\algnewcommand{\IfReturn}[2]{\State \algorithmicif\ {#1} \algorithmicthen\ \algorithmicreturn\ {#2}}
\algnewcommand{\IfThen}[2]{\State \algorithmicif\ {#1} \algorithmicthen\ {#2}}
\algnewcommand{\ElsIfThen}[2]{\State \algorithmicelsif\ {#1} \algorithmicthen\ {#2}}

\newcommand{\mypara}[1]{\textbf{\textit{#1.}}}

\definecolor{splitcolor}{RGB}{190,174,212}
\definecolor{nodecolor}{RGB}{86,180,233}

\newcommand{\dixon}{\textsc{Dixon}}
\newcommand{\pathological}{\textsc{Pathological}}
\newcommand{\pinter}{\textsc{Pinter}}
\newcommand{\qing}{\textsc{Qing}}
\newcommand{\schaffer}{\textsc{Schaffer}}
\newcommand{\trigonometric}{\textsc{Trigonometric}}
\newcommand{\hilbert}{\textsc{Hilbert}}
\newcommand{\sqsum}{\textsc{ReciprocalL2}}

\newcommand{\mincr}{2.5}
\newcommand{\maxcr}{100}
\newcommand{\peakcr}{1000}

\crefname{subsection}{section}{sections}
\Crefname{subsection}{Section}{Sections}
\Crefname{ALC@unique}{Line}{Lines}

\theoremstyle{plain}
\newtheorem{problem}{Problem}

\definecolor{zgcolor}{RGB}{0,114,178}

\definecolor{agcolor}{RGB}{213,94,0}

\definecolor{revonecolor}{RGB}{240,135,0}
\newcommand{\revone}[1]{#1}

\definecolor{revtwocolor}{RGB}{4,77,200}
\newcommand{\revtwo}[1]{#1}

\journal{Journal Of Computational Physics}

\begin{document}

\begin{frontmatter}



\title{\revone{Hierarchical Search of Tree Tensor Networks for High-Dimensional Data}}


\author[midas,eecs]{Zheng Guo} 
\author[aero]{Aditya Deshpande}
\author[eecs]{Xinyu Wang}
\author[ners]{Brian C. Kiedrowski}
\author[aero]{Alex A. Gorodetsky}
\affiliation[midas]{
    organization={Michigan Institute for Data, AI, \& Society, University of Michigan},
    city={Ann Arbor},
    postcode={48105},
    state={Michigan},
    country={USA}}
\affiliation[eecs]{
    organization={Electrical Engineering and Computer Science Department, University of Michigan},
    city={Ann Arbor},
    postcode={48107},
    state={Michigan},
    country={USA}}
\affiliation[ners]{
    organization={Nuclear Engineering \& Radiological Sciences Department, University of Michigan},
    city={Ann Arbor},
    postcode={48107},
    state={Michigan},
    country={USA}}
\affiliation[aero]{
    organization={Aerospace Engineering Department, University of Michigan},
    city={Ann Arbor},
    postcode={48107},
    state={Michigan},
    country={USA}}

\begin{abstract}
Tensor network methods provide a scalable solution to represent high-dimensional data.
However, their efficacy is often limited by static, expert-defined structures that fail to adapt to evolving data correlations.
\revone{We address this limitation by formalizing the structural rounding problem for tree tensor networks and introducing a hierarchical search algorithm
\algoname, which automatically identifies optimized structures with index reshaping for input tree tensor networks.}
To navigate the combinatorial explosion of the structural search space, \algoname integrates stochastic sub-network sampling with hierarchical refinement.
This approach utilizes entropy-guided index clustering to reduce dimensionality and targeted reshaping to expose latent data correlations.
Numerical experiments on analytical functions and real-world physics applications, including thermal radiation transport, neutron diffusion, and computational fluid dynamics, demonstrate that \algoname exhibits empirical polynomial scaling with dimensionality relative to the sampling budget, bypassing the scalability barriers in prior work.
\algoname achieves compression ratios $\mincr\times$ to $\maxcr\times$ higher than standard fixed formats such as Tensor Trains and Hierarchical Tuckers~(peaking at $\peakcr\times$).
Furthermore, \algoname discovers structures that generalize effectively: applying a structure optimized for one data instance to a related target data typically maintains compression performance within $10\%$ of the result obtained by performing structure search on that target data.
These results highlight \algoname as a robust, automated tool for adaptive data representation and high-dimensional simulation compression with tensor network methods.
\end{abstract}


\begin{highlights}
\item Defines structural rounding to jointly optimize \revone{tree network structures} and bond dimensions
\item New method finds well-performed network structures to compress high-dimensional data
\item Entropy-guided heuristic navigates complex search spaces without checking every option
\item Achieves up to 100x higher compression than fixed formats on real-world engineering data
\item Discovered structures maintain high compression efficiency across similar data
\end{highlights}

\begin{keyword}
tensor network structure search \sep tensor networks \sep low-rank tensor approximation \sep adaptive data compression \sep simulation compression


\end{keyword}

\end{frontmatter}


\section{Introduction}\label{sec:intro}
High-dimensional data representations are crucial to modern computational science and engineering.
Whether representing spatial-temporal fields in computational fluid dynamics~\cite{takamoto2022pdebench,zawawi2018review}, parametric neutron transport~\cite{bassett2019comparison,cox2022monte}, or high-fidelity radiation transport solutions~\cite{gorodetsky2025thermal}, such data are naturally expressed as multi-dimensional tensors~\cite{kolda2009tensor}.
However, the utility of these representations is fundamentally challenged by the curse of dimensionality~\cite{oseledets2009breaking,altman2018curse}, where storage requirements and computational complexity scale exponentially with the number of dimensions.

Low-rank tensor approximations have emerged as a transformative solution to the scalability problem~\cite{bachmayr2023low,dolgov2012tt,rodgers2024tensor,truong2024tensor}.
By decomposing a high-order tensor into a network of smaller, interconnected core tensors, formats such as Tucker~\cite{Tucker_1966}, Tensor Train (TT)~\cite{oseledets2011tensor}, and Hierarchical Tucker (HT)~\cite{ht,falco2021tree} achieve drastic compression while preserving essential structural information.
Despite their success in fine-tuning of foundation models~\cite{veeramacheneni2025canonical,yang2024loretta,ghiasvand2025decentralized}, signal processing~\cite{zheng2022subttd,xie2025coarray,wang2023tensor,ravishankar2022hierarchical}, and solving linear and nonlinear systems~\cite{sands2025high,dolgov2012tt, ghahremani2024deim,ghahremani2024cross}, a persistent challenge remains: no single tensor network format is universally optimal.
The efficacy of a specific topology depends sensitively on the underlying data correlations, which are rarely known \emph{a priori}.

This search for optimality is particularly evident in the physics community, which often moves beyond rigid formats to utilize more heterogeneous network structures.
In applications such as quantum circuit simulation~\cite{pan2022simulation,kalachev2021multi,huang2020classical}, quantum chemical problems~\cite{nakatani2013efficient,murg2015tree,gunst2019three}, and many-body systems~\cite{ferrari2022adaptive,seki2020tensor,hikihara2023automatic,ke2023tree}, researchers must navigate complex, non-standard graphs to select suitable structures.
In this context, the contraction that converts a tensor network into the corresponding data tensor becomes a bottleneck; consequently, substantial effort has been concentrated on improving contraction efficiency for arbitrary topologies~\cite{yang2017loop,gray2024hyperoptimized,gao2025fermionic}.
While these hand-crafted structures are designed to mirror specific physical interactions, they rely heavily on expert intuition and are not guaranteed to accurately represent the desired quantum states.
\revtwo{Recent efforts on automatic topology optimization represent a step forward, but they are largely constrained to bond-by-bond updates under fixed rank bounds for binary trees~\citep{hikihara2023automatic,watanabe2025ttnopt}. While these methods demonstrated practical success in generative modeling by using mutual information to guide structural changes~\citep{akamatsu2026plastic,harada2025tensor}, their search procedures are primarily tailored to binary trees.}

Such structural rigidity is particularly problematic in dynamic systems governed by partial differential equations~(PDEs).
While a fixed format like TT may effectively capture data entanglement at initial timesteps, it often loses compression power as the solution evolves and the underlying correlation structure shifts~\cite{gorodetsky2025thermal, dektor2025coordinate}.
Current mitigation strategies, such as adaptive coordinate transformations~\cite{dektor2023tensor,dektor2025coordinate,rodgers2022adaptive}, attempt to realign the basis to the network cores.
However, these transformations often come at a high cost: they can destroy the multi-linear structure of the problem, making analytical computations complicated.
Ultimately, if the network topology itself is ill-suited to the evolving data, even an optimal basis transformation yields suboptimal results.

A similar limitation exists in traditional tensor network rounding~\cite{oseledets2011tensor,watanabe2025ttnopt}.
Standard rounding techniques reduce internal bond dimensions, or ranks, of an input network within a prescribed error tolerance.
This operation is a crucial component in tensor network arithmetic.
For instance, tensor train addition and multiplication typically yield results with inflated ranks that necessitate subsequent truncation~\cite{oseledets2011tensor}.
Another example is adaptive rank search in cross-approximation methods~\cite{oseledets2010tt,ghahremani2024cross}, which can lead to overestimated internal ranks if the adaptive step size is overly conservative.
However, conventional rounding is limited to its static structure, focusing exclusively on rank reduction while lacking the structural flexibility to transition between different topologies.
We argue that a more robust approach is required: an automated process that not only seeks lower ranks to fit an error bound but actively searches for new, more efficient network topologies.
The aforementioned necessity has motivated the emerging field of Tensor Network Structure Search (TN-SS), which aims at automatically discovering the optimal network structure for a given data tensor and error tolerance~\cite{handschuh2015numerical}.

\revone{In this work, we address a generalized and challenging instance of TN-SS that incorporates index factorization into the structural search for tree tensor networks:}

\begin{problem}[TN-SS with Index Reshaping]\label{problem:tnss}
    Given a data tensor $\ten{T}$ with indices $\indsi$ and a relative error $\error$, find an optimal network $\searchres{\net}$ with indices factorized from $\indsi$ that minimizes the total parameter count.
    The resulting tensor network $\searchres{\net}$ satisfies $\norm{\data{\indsi}{\searchres{\net}} - \ten{T}} \leq \error \norm{\ten{T}}$, where $\data{\indsi}{\searchres{\net}}$ is the reconstruction of the data tensor from $\searchres{\net}$ by mapping its reshaped indices back to $\indsi$.
\end{problem}

\begin{figure}[t]
    \centering
    \includegraphics[width=.9\linewidth]{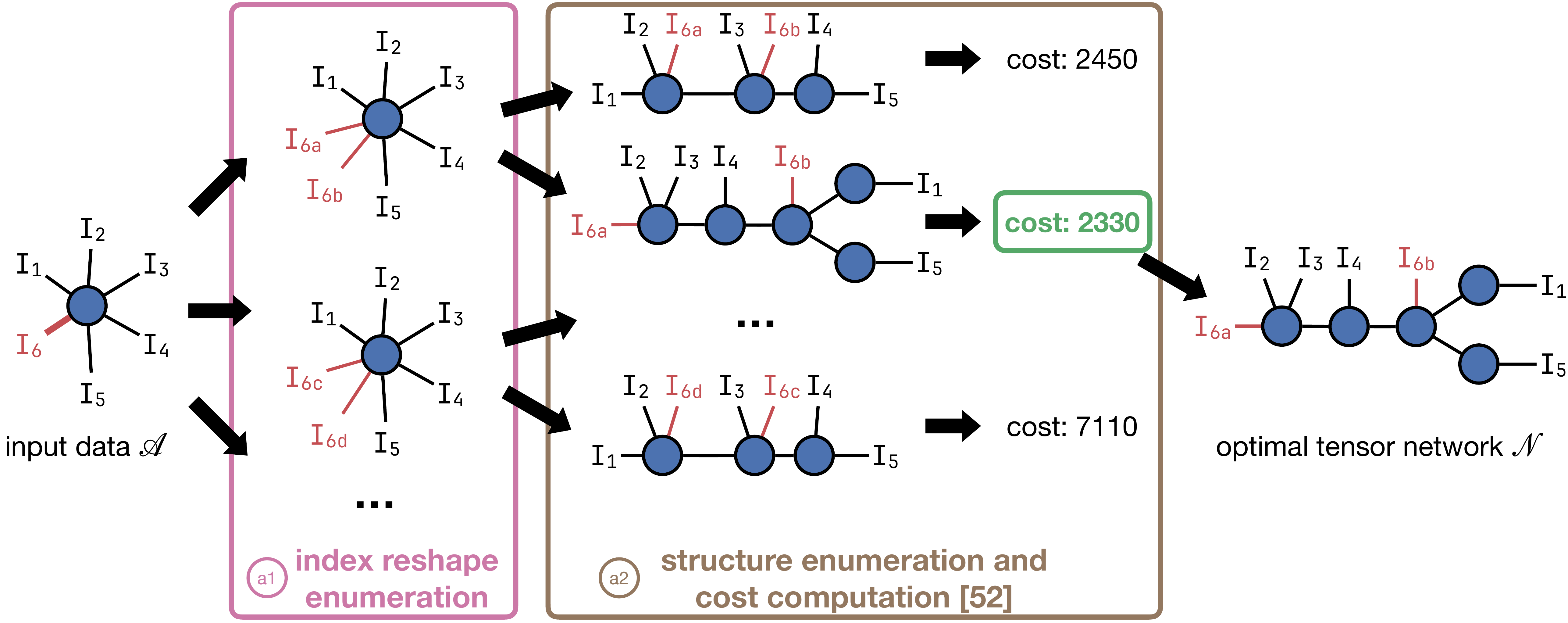}
    \caption{
        Traditional tensor network structure search~\cite{guo2025tensor} with the extension to support index reshaping.
        It enumerates different reshapings of indices and searches for optimal structures for every reshaping configuration.
        This na\"ive extension often fails as dimensionality increases.
        }\label{fig:overview:traditional}
\end{figure}

Unlike prior studies~\cite{li2020evolutionary,li2022permutation,li2023alternating,hashemizadeh2020adaptive,guo2025tensor} that focus on fixed indices, we explicitly incorporate index reshaping (step \textcircled{\tiny a1} in \cref{fig:overview:traditional}) into the search space.
Strategically factorizing and regrouping indices can expose latent low-rank structures invisible in native indices~\cite{borm2003introduction,chen2024learning,anandkumar2014tensor,linderman2014discovering}, and the quantics tensor train format~\cite{khoromskij2011d} is a typical example.

Beyond structural search for raw data tensors, we consider a more general problem where the input itself is a tree tensor network~\cite{falco2021tree}.
This leads to the task of \emph{structural rounding}, where we seek to compress a given \revone{tree tensor network} by both adjusting bond dimensions and reconfiguring its topology:

\begin{problem}[Tensor Network Structural Rounding (TNSR)]\label{problem:tnsr}
    \revtwo{Given a tensor network $\net$ with a set of indices $\indsi$ and a prescribed relative error tolerance $\error$, find an optimal network $\searchres{\net}$ that minimizes the total parameter count subject to $\norm{\data{\indsi}{\searchres{\net}} - \data{\indsi}{\net}} \leq \error \norm{\data{\indsi}{\net}}$, where $\data{\indsi}{\searchres{\net}}$ is the reconstruction of the data tensor from $\searchres{\net}$ by mapping its reshaped indices back to $\indsi$.}
\end{problem}

Notably, \cref{problem:tnss} is a special case of \cref{problem:tnsr} where the input is a single-core tensor network.
By focusing on \cref{problem:tnsr}, we develop an algorithmic framework that applies to both.

However, TNSR inherently suffers from the scalability challenge. \revone{Prior work proves that choosing optimal dimension trees~\citep{kaya2019computing} and determining optimal rank assignments in Tuckers~\citep{hillar2013most,ghadiri2023approximately} are independently NP-hard problems. Simultaneously searching the combinatorial space of general tree topologies, executing index reshaping, and resolving these underlying NP-hard subproblems compounds this complexity. Consequently, an exhaustive search over all valid tree tensor networks is computationally intractable, even for moderately high-dimensional data.}
Most existing techniques~\cite{li2020evolutionary,li2022permutation,li2023alternating,hashemizadeh2020adaptive,guo2025tensor} only consider TN-SS without index reshaping and adopt a sampling-evaluation paradigm.
As illustrated in \cref{fig:overview:traditional}, this traditional approach can be extended to support index reshaping, but only through a nested enumeration:
for every possible index reshaping, a vast space of candidate structures must be sampled and individually validated via certain assessment methods.
Consequently, the computational cost scales with the product of the number of sampled structures and the cost of evaluating a single candidate.
Accordingly, prior work has pursued two complementary directions to reduce computational costs.
One line of work attempts to reduce the number of sampled structures through heuristics~\cite{li2020evolutionary,hashemizadeh2020adaptive,li2023alternating} or constrained search strategies~\cite{zheng2024svdinstn,wang2025renormalization,li2022permutation}, such as greedy selection~\cite{hashemizadeh2020adaptive} or local neighborhood exploration~\cite{li2023alternating}.
The other direction focuses on accelerating candidate evaluation, for example by replacing expensive tensor decompositions with cheaper surrogate estimators~\cite{guo2025tensor}.
Nevertheless, as tensor dimensionality increases, the required sample size grows combinatorially~\cite{hillar2013most}, making existing TN-SS methods that rely on exhaustive or flat search intractable.

\begin{figure}[t]
    \centering
    \includegraphics[width=.95\linewidth]{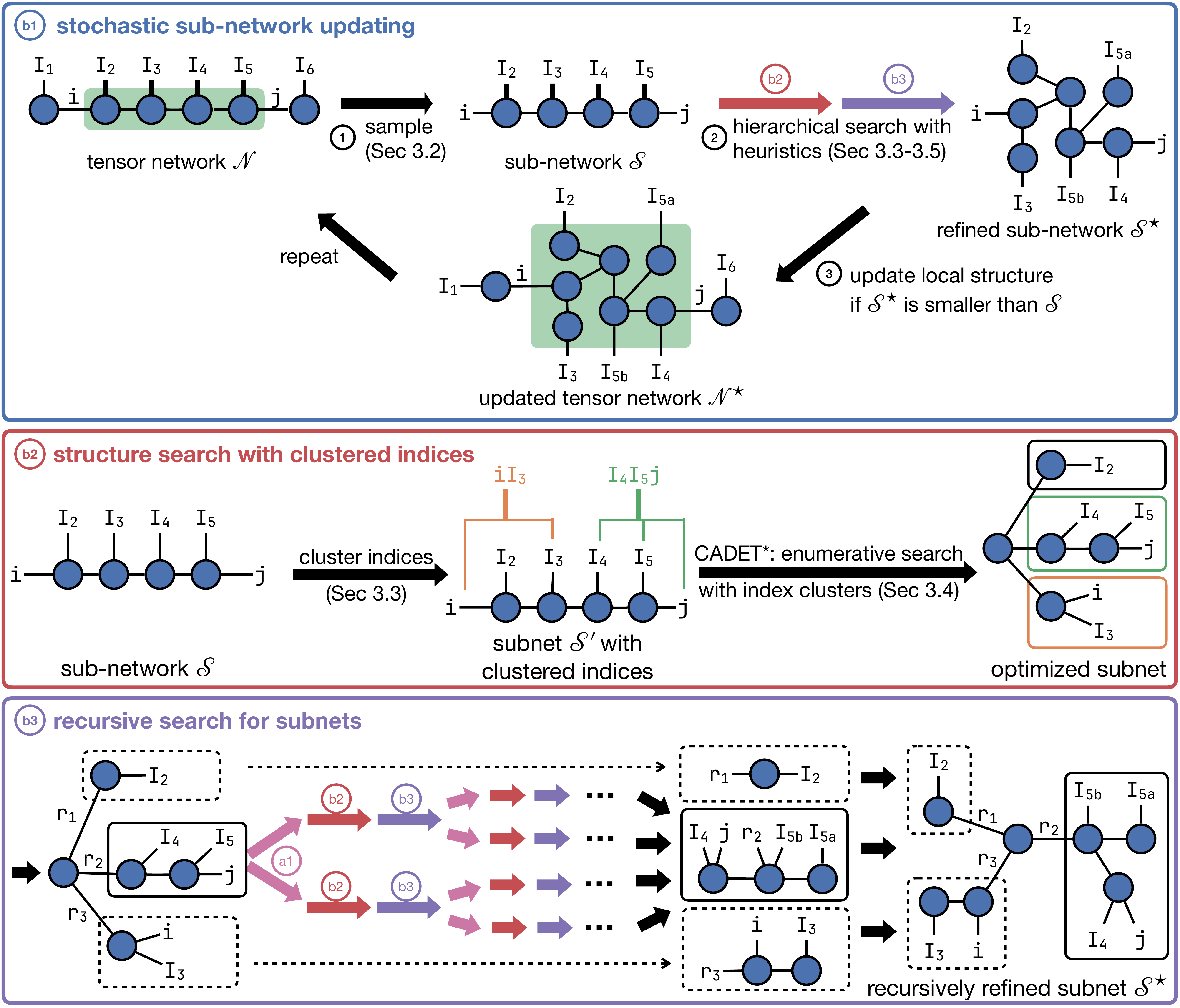}
    \caption{
    Hierarchical tensor network structure search (\algoname): \textcircled{\tiny b1} stochastically samples a sub-network from the input network, refines the local structure, and substitute the resulting network into the input;  \textcircled{\tiny b2} groups indices into sets and performs structure search on clustered indices to get an optimized network; \textcircled{\tiny b3} partitions the network into subnets for recursive optimization, eventually reintegrating them into a refined tree.
    }
    \label{fig:overview:ours}
\end{figure}

\subsection{Overview of \algoname}
The scalability problem in TN-SS also occurs in \tnround.
\revone{To address this issue, we propose \algoname, a scalable hierarchical search algorithm for tree structures.} The central design principle is to alleviate the combinatorial explosion through a combined optimization strategy: global stochastic sampling and local hierarchical structure refinement.
This framework is specifically engineered to balance exploration and exploitation.
At the global scale, the algorithm employs stochastic sampling to obtain sub-networks, which effectively avoids prohibitive costs of exhaustive search and prevents premature convergence to local optima.
At the local scale, the refinement of sub-networks 
shifts from blind enumeration to a targeted approach guided by the decay of the singular values.
Leveraging this property, \algoname rapidly identifies network structures that offer high compression performance, even within a complex search space.
\revtwo{Though localized refinement is also adopted by several existing frameworks~\cite{hikihara2023automatic,harada2025tensor,akamatsu2026plastic}, our method extends their bond-by-bond updates to a flexible sub-network optimization with index reshaping. This extension enables broader structural transformations, such as introducing new internal nodes and splitting original data dimensions.}

An overview of the proposed workflow is illustrated in \cref{fig:overview:ours}.
At the topmost level, the algorithm preserves diversity by repeatedly sampling sub-networks $\subnet$
from the global tensor network (step \textcircled{\tiny b1}).
These sub-networks are optimized and subsequently reintegrated into the global network to enhance overall compression performance.
Within the refinement phase of each sub-network, the algorithm transitions from stochastic exploration to heuristic-guided search.
Indices are first grouped into clusters, and each cluster is treated as an aggregated dimension (step \textcircled{\tiny b2}).
This clustering effectively reduces the tensor order and enables a more efficient structure search to produce an optimized sub-network structure.
Following this, each constituent network fragment defined by these index clusters undergoes recursive refinement (step \textcircled{\tiny b3}), where the indices are reshaped (step \textcircled{\tiny a1}) to uncover fine-grained compression opportunities.
This hierarchical search strategy allows \algoname to maintain a balance between computational scalability and compression quality.

\revone{While this workflow builds upon our prior tensor network structure search framework \cadet~\citep{guo2025tensor} (denoted as \cadet* in \textcircled{\tiny b2} because we slightly modify the original algorithm), \algoname introduces fundamental algorithmic advancements to address its core limitations. Specifically, \cadet is restricted to single input data tensors of relatively low dimensionality (up to 6-D). In contrast, \algoname extends the input scope to encompass tree tensor networks and scales effectively to higher-dimensional data. To further optimize compression performance, \algoname expands the search space by integrating index reshaping. These extensions enable \algoname to discover high-quality tensor network topologies that are inherently unreachable by \cadet.}

\subsection{Summary of Contributions}
In summary, the \algoname framework introduces the following contributions to the field of tensor network structure search for high-dimensional data:
\begin{itemize}
    \item \algoname extends traditional structure search, which is typically restricted to raw data tensors, to the refinement of tree tensor networks.
    This structural rounding capability allows for the automated optimization of pre-defined or sub-optimal network topologies under a prescribed error tolerance.
    \item Unlike standard TN-SS methods, \algoname simultaneously optimizes both network structures and index reshaping.
    This allows the framework to uncover latent low-rank correlations that are otherwise obscured by the original data indices.
    \item \algoname overcomes combinatorial explosion of search space by integrating stochastic sub-network sampling with hierarchical structure refinement.
    This integration ensures broad exploration of the search space while maintaining scalable computational cost.
    \item \algoname utilizes an efficient heuristic based on singular value entropy to identify promising index clusters and reshapes.
    This eliminates the need for exhaustive enumeration of all index clustering and reshaping configurations during the search process.
\end{itemize}

\subsection{Overview of Empirical Results}
We demonstrate the efficacy of \algoname across a diverse suite of benchmarks, ranging from analytical functions to high-fidelity engineering applications, including thermal radiation transport, parametric multi-group neutron diffusion, and computational fluid dynamics (CFD)
\footnote{\revone{The implementation is available at \url{https://github.com/gorodetsky-umich/tensor_networks}}}
\footnote{\revone{The evaluation scripts are available at \url{https://github.com/aaronguo1996/TNSS_scripts}}}.

Experiment results indicate that the search time of \algoname scales mostly linearly---and at most polynomially---with the input dimensionality at a fixed number of sampling steps, providing a sustainable path for high-dimensional problems where prior methods scale combinatorially.
The discovered tensor network structures consistently achieve compression ratios around $\mincr\times$ to $\maxcr\times$ higher than conventional fixed structures such as TT and HT.
In the special case of thermal radiation transport, the compression performance is improved by up to three orders of magnitude.
Additionally, index reshaping yields visible compression gains as dimensionality increases.

Furthermore, we observe that these optimized structures exhibit strong generalization capabilities.
Even in dynamic systems where specific network structures across timesteps may vary, they frequently share common sub-structures that facilitate consistently high compression ratios.
This structural consistency confirms that the discovered networks capture intrinsic data correlations, effectively reducing the need for repeated searches.

The remainder of this paper is organized as follows. In \cref{sec:background}, we provide the mathematical preliminaries on tensors and tensor networks. \Cref{sec:search} details the proposed \algoname algorithm, including stochastic sub-network sampling (\cref{sec:method:refine}), sub-network structure refinement (\cref{sec:method:search}), and entropy-based clustering (\cref{sec:method:cluster}) and reshaping (\cref{sec:method:reshape}) heuristics. Extensive numerical experiments and performance benchmarks are presented in \cref{sec:eval}. Finally, we offer concluding remarks and discuss future work in \cref{sec:conclusion}.
\section{Background}\label{sec:background}
In this section, we provide the background knowledge on tensor networks.
\subsection{Tensors and Tensor Operations}
A tensor $\ten{A} \in \real^{n_1 \times n_2 \times \cdots \times n_d}$ is a $d$-dimensional array, each dimension (or mode) $\mu$ of size $n_\mu$.
The unfolding (or matricization) of $\ten{A}$ at modes $\indsi \subseteq \{1,2,\ldots, d\}$, denoted by $\matric{\ten{A}}{\indsi}$, is a matrix such that the entries of $\matric{\ten{A}}{\indsi}$ are mapped from $\ten{A}$ as follows:
\begin{equation}
    \matric{\ten{A}}{\indsi}(i_{\indsi_1}, i_{\indsi_2}, \ldots, i_{\indsi_k}; i_{\indsj_1}, i_{\indsj_2}, \ldots, i_{\indsj_{d-k}}) = \ten{A}(i_1, i_2, \ldots, i_d),
\end{equation}
where $\indsj = \{1,2,\ldots, d\} - \indsi$, $|\indsi| = k$, and $|\indsj| = d-k$.
For instance, a tensor $\ten{A}$ of shape $2 \times 3 \times 4$ can be unfolded along the first and last modes, resulting in a matrix $\matric{\ten{A}}{\{1,3\}}$ with shape $8 \times 3$.
The size of a tensor $\ten{A}$ is defined as $\size{\ten{A}} = \prod_{\mu=1}^d n_\mu$.
The Frobenius norm of a tensor $\ten{A}$ is represented as $\norm{\ten{A}}$, defined as 
\begin{equation}
   \norm{\ten{A}} = \sqrt{\sum_{i_1=1}^{n_1}\sum_{i_2=1}^{n_2}\cdots\sum_{i_d=1}^{n_d} \ten{A}^2(i_1, i_2, \ldots, i_d)}
\end{equation}
We use $U, \Sigma, V = \svd{M}$ to represent the singular value decomposition (SVD) of the matrix $M$, where $U$ and $V$ are the left and right singular vectors, and $\Sigma$ are the singular values.
The $i^{th}$ largest singular value of a matrix $M$ is denoted as $\sigma_i(M)$.

\subsection{Tensor Networks}\label{sec:preliminary:tn}
A tensor network $\net$ is a graph $(\nodes, \edges)$ where nodes are tensors and edges are how to contract the connected tensors.
The contraction of its node tensors conforming to the edge connections is its represented data tensor.
We call the dangling edges in a tensor network \emph{free indices}, and the edges that connect at least two nodes \emph{internal indices}.
A tensor network without cycles are also named \emph{tree tensor networks}~\cite{falco2021tree}.
In this work, we only consider optimization of tree tensor networks.

For a given tensor network $\net$ with a set of free indices $\indsi$, we denote the network as $\net_{\indsi}$.
The size of the network is defined as the total number of entries across all its constituent nodes: $\size{\net} = \sum_{\ten{A} \in \nodes} \size{\ten{A}}$.
The data tensor reconstructed from $\net_{\indsi}$ is denoted by $\data{\indsj}{\net_{\indsi}}$.
This represents the result of contracting all internal edges of $\net_{\indsi}$ and subsequently mapping the free indices $\indsi$ of $\net_{\indsi}$ to the target indices $\indsj$ via index reshaping.
The unfolding of a tree tensor network $\net_{\indsi}$ at indices $\indsj \subset \indsi$, written as $\matric{\net_{\indsi}}{\indsj}$, is well-defined if and only if the indices in $\indsj$ corresponds to a valid cut in the network topology.
In other words, $\indsj$ must either reside entirely on a single node in $\net_{\indsi}$, or be separable from the remaining indices $\indsi \backslash \indsj$ by the removal of a single internal edge.

As an example, consider the input tensor network in \cref{fig:overview:ours} \textcircled{\scriptsize b1}.
This structure is a tree tensor network with 6 nodes and 11 edges.
Among the edges, the named edges $I_1$, $I_2$, $I_3$, $I_4$, $I_5$, $I_6$ are free edges, and the other ones such as $i$ and $j$ are internal edges.
The unfoldings $\matric{\net}{I_1}$ and $\matric{\net}{I_1, I_2}$ are valid for $\net$, but $\matric{\net}{I_1, I_3}$ is an invalid unfolding under the current network topology.
Representing $\matric{\net}{I_1, I_3}$ would require either merging the three leftmost nodes into a single core, or moving the index $I_3$ to the left of $I_2$ such that $I_1$ and $I_3$ are in a contiguous range that corresponds to a single cut in the network.
\section{\revone{Hierarchical Search of Tree Structures With Index Reshaping}}\label{sec:search}

This section introduces the algorithm, \algoname, to solve the \tnround problem.
The algorithm demonstrates enhanced computational efficiency for high-dimensional data, and integrates index reshaping into the search space.

\subsection{High-level Algorithm}\label{sec:method:high-level}

\begin{algorithm}[t]
\caption{\revone{The high-level hierarchical search algorithm}}\label{alg:hiss}
\begin{algorithmic}[1]
    \Require Tensor network $\net$ with indices $\indsi$, error tolerance $\error$, subnet node number $\subnetsize$, sample iterations $\maxiters$, dimensionality threshold $\dimthreshold$, number of cluster candidates $\clustercands$, exploration probability $\epsilon$, and factorization threshold $\factthreshold$
    \Ensure An optimized network $\searchres{\net}$ such that $\norm{\data{\indsi}{\searchres{\net}} - \data{\indsi}{\net}} / \norm{\data{\indsi}{\net}} \leq \error$
    \Function{\algoname}{$\net, \error, \subnetsize, \maxiters, \dimthreshold, \clustercands, \epsilon, \factthreshold$}
    \State $\searchres{\net} \gets \net$ \label{alg:hiss:global-init}
    \State $f \gets \{u: 0 \mid u \in \nodes\}$ where $\net = (\nodes, \edges)$
    \Comment{Initialize visit frequencies}
    \label{alg:hiss:freq}
    \For{$t = 1, 2, \ldots, \maxiters$}\label{alg:hiss:sample-start}
        \State $\subnet \gets \Call{\sample}{\searchres{\net}, \subnetsize, f}$\Comment{Sample a subnet}
        \label{alg:hiss:line-sample}
        \State \revtwo{Orthonormalize $\searchres{\net}$ to any node inside $\subnet$}\Comment{Orthogonalization for error propagation}
        \label{alg:hiss:line-orthonormal}
        \State $\searchres{\subnet} \gets \Call{\recsearch}{\subnet, \error / \sqrt{\maxiters}, \clustercands, \epsilon, \factthreshold}$\Comment{Optimize the subnet}
        \label{alg:hiss:line-optimize}
        \State $\searchres{\net} \gets \left[\subnet \mapsto \searchres{\subnet}\right]\searchres{\net}$\Comment{Replace the subnet if improved}
        \label{alg:hiss:line-replace}
    \EndFor\label{alg:hiss:sample-end}
    \State \Return{$\searchres{\net}$}
    \EndFunction
\end{algorithmic}
\end{algorithm}

\begin{algorithm}[t]
\caption{The recursive sub-network optimization algorithm}\label{alg:recsearch}
\begin{algorithmic}[1]
    \Require Tensor network $\net_{\indsi}$, error tolerance $\error$, dimensionality threshold $\dimthreshold$, number of cluster candidates $\clustercands$, exploration probability $\epsilon$, and factorization threshold $\factthreshold$
    \Ensure An optimized network $\searchres{\net}$ such that $\norm{\data{\indsi}{\searchres{\net}} - \data{\indsi}{\net}} / \norm{\data{\indsi}{\net}} \leq \error$
    \Function{\recsearch}{$\net_\indsi, \error, \clustercands, \epsilon, \factthreshold$}\label{alg:hiss:rec-begin}
        \State $\net_{\{\indsi_i\}_{i=1}^{\dimthreshold}}\gets$ \Call{\mergeindices}{$\net_\indsi, \dimthreshold, \clustercands, \epsilon$}
        \label{alg:hiss:line-cluster}
        \State Split $\error$ into $\error_1,\error_2$ such that $\error_1^2 + \error_2^2 = \error^2$
        \label{alg:hiss:line-error}
        \State $\netwithnodes{\searchres{\net}}{G_1, G_2, \ldots, G_k} \gets$ \Call{\structsearch}{$\net_{\{\indsi_i\}_{i=1}^{\dimthreshold}}, \error_1$}
        \label{alg:hiss:line-search}
        \For{$i = 1, 2, \ldots, k$}\label{alg:hiss:line-rec-start}
            \State \revtwo{Orthonormalize $\searchres{\net}$ to $G_i$}\label{alg:hiss:line-rec-orthonormal}
            \State $\searchres{G_{i}} \gets \min_{G_{i,j} \in \Call{\topreshape}{G_{i}, \factthreshold}} \Call{\recsearch}{G_{i,j}, \error_2 / \sqrt{k}}$
            \label{alg:hiss:line-reshape}
        \EndFor\label{alg:hiss:line-rec-end}
        \State \Return{$\netwithnodes{\net^{\star}}{G^{\star}_1, G^{\star}_2, \ldots, \searchres{G_{k}}}$}
    \EndFunction\label{alg:hiss:rec-end}
\end{algorithmic}
\end{algorithm}

As detailed in \cref{alg:hiss}, \algoname reduces the search complexity by operating through two principal layers:

\mypara{Stochastic Sub-Network Sampling} (Lines \ref{alg:hiss:global-init}--\ref{alg:hiss:sample-end})
To maintain scalability and compression performance across high-dimensional networks, the algorithm adopts the stochastic hill climbing strategy~\cite{juels1995stochastic}.
This approach maintains a global network $\searchres{\net}$ (Line \ref{alg:hiss:global-init}) and a node visit frequency map $f$ (Line \ref{alg:hiss:freq}).
It iteratively samples a random sub-network $\subnet$ according to node visit frequencies $f$ (Line \ref{alg:hiss:line-sample}).
\revone{Then, the network $\searchres{\net}$ is orthonormalized to a node selected within $\subnet$~(Line \ref{alg:hiss:line-orthonormal}), transforming the network into a canonical form where the environment of $\subnet$ is norm-preserving~\citep{shi2006classical}.} \revone{Subsequently, the algorithm executes the enumerative structure search oracle \textsc{\recsearch} (\cref{alg:recsearch}) on each sub-network $\subnet$ using a local relative error $\error / \sqrt{\maxiters}$ (Line \ref{alg:hiss:line-optimize}). Because of the norm-preserving environment, local truncation errors map exactly to independent contributions in the total global error squared, ensuring that the cumulative effect of these localized refinements remains strictly within the global threshold $\error$.}
If $\searchres{\subnet}$ reduces the size of $\subnet$, $\subnet$ is replaced by $\searchres{\subnet}$ within the global network $\searchres{\net}$ (Line \ref{alg:hiss:line-replace}).
We denote this sub-network update as $[\subnet \mapsto \searchres{\subnet}]\searchres{\net}$.
The sampling loop repeats $\maxiters$ times and the final global network $\searchres{\net}$ is returned.
This sampling mechanism allows for global exploration of the network topology while restricting the computational intensity to manageable sub-networks.

\mypara{Recursive Sub-Network Optimization} (Line \ref{alg:hiss:line-optimize})
For each sampled sub-network $\subnet$, the algorithm optimizes it using a recursive procedure \textsc{\recsearch}.
This process, presented in \cref{alg:recsearch}, consists of three major steps:
\begin{enumerate}[(1)]
    \item \textbf{Index clustering} (Line \ref{alg:hiss:line-cluster}) To efficiently support high-dimensional inputs, we first reduce the input sub-network $\net_{\indsi}$ of $|\indsi|$ dimensions into a representation $\net_{\{\indsi_i\}_{i=1}^{\dimthreshold}}$ of $d_\theta$ dimensions, where each dimension corresponds to a cluster of indices $\indsi_i$.
    This clustering process is guided by an entropy-based heuristic to avoid exhaustive enumeration.
    
    \item \textbf{Enumerative structure search} (Lines \ref{alg:hiss:line-error}--\ref{alg:hiss:line-search}).
    Next, the error tolerance $\error$ is divided into $\error_1$ and $\error_2$ (Line \ref{alg:hiss:line-error}), where $\error_1$ is used in the structure search for the network with clustered indices $\net_{\{\indsi_i\}_{i=1}^{\dimthreshold}}$, and $\error_2$ is reserved for recursive optimization.
    An enumerative structure search~\cite{guo2025tensor} is then performed on this reduced $d_\theta$-dimensional representation using error budget $\error_1$.
    By optimizing the arrangement of these clustered indices, the algorithm identifies a sub-network replacement $\searchres{\net}(G_1, G_2, \ldots, G_k)$.
    \item \textbf{Recursive search with index reshaping} (Lines \ref{alg:hiss:line-rec-start}--\ref{alg:hiss:line-rec-end}). Each resulting network fragment $G_i$ inside $\searchres{\net}$ is further refined. \revone{To ensure that local truncation errors propagate precisely to the global relative error tolerance} \revone{as that in \Cref{alg:hiss} Line \ref{alg:hiss:line-orthonormal}, the nodes in $\searchres{\net}$ external to $G_i$ are orthonormalized to establish canonical forms for index splitting and recursive fragment optimization.}
    Then, the indices in each $G_i$ are split into smaller dimensions via \topreshapecall (Line \ref{alg:hiss:line-reshape}), and the enumerative search \textsc{\recsearch} is applied recursively to exploit compression opportunities within the index clusters and obtain a refined network fragment $\searchres{G}_i$.
\end{enumerate}

Note that the separation into two distinct layers is a deliberate response to the different characteristics of the global and local search spaces.
For the global network, the number of possible structures is so vast that any enumerative approach would succumb to the combinatorial explosion.
While deterministic heuristics offer a path toward tractability, such greedy approaches are frequently trapped in local minima, overlooking globally optimal network structures.
Our stochastic sampling layer augments purely greedy search with global exploration, providing a diverse set of candidates for heuristic-based refinement.

Conversely, when the search is localized to a sampled sub-network, the search space is small enough for deterministic optimization.
We utilize a heuristic-guided recursive algorithm \textsc{\recsearch} to achieve a well-performing sub-network structure.
At this scale, stochasticity introduces unnecessary variance and potentially miss the exact index arrangement required for maximal compression; instead, a systematic search provides a reliable guarantee of near-optimality.
By alternating between stochastic sampling and recursive sub-network refinement, \algoname effectively balances computational efficiency with the ability to discover complex topologies.

In the following sections, we detail the core components of this framework: the stochastic sampling process \textsc{\sample} (\cref{sec:method:refine}), the index clustering mechanics \mergeindicescall (\cref{sec:method:cluster}), the modified structure search oracle \structsearchcall (\cref{sec:method:search}), and the index reshaping heuristics \topreshapecall (\cref{sec:method:reshape}).

\begin{algorithm}[t]
\caption{Stochastic Sampling of Sub-Networks}\label{alg:sample}
\begin{algorithmic}[1]
    \Require The global tensor network $\net = (\nodes, \edges)$, number of nodes in a subnet $\subnetsize$, and visit frequencies $f$.
    \Ensure A sub-network $\subnet = (\nodes_s, \edges_s)$ where $|\nodes_s| = \subnetsize$, $\nodes_s \subseteq \nodes$ and $\edges_s \subseteq \edges.$
    \Function{\sample}{$\net, \subnetsize, f$}
        \State $u \gets \Call{\psample}{\nodes, f}$ \Comment{Initialize with a random seed node}
        \label{alg:sample:init-node}
        \State $\nodes_s \gets \{u\} \quad f[u] \gets f[u] + 1$
        \label{alg:sample:init-subnet}
        \While{$|\nodes_s| \leq \subnetsize$}
            \State $u \gets \Call{\psample}{\nodes_s, f}$ \Comment{Select a node to expand}
            \label{alg:sample:expansion-node}
            \State $v \gets \Call{\psample}{\textsc{Neighbors}(u), f}$\Comment{Add a neighbor}
            \label{alg:sample:expansion-nbr}
            \State $\nodes_s \gets \nodes_s \cup \{v\} \quad \edges_s \gets \edges_s \cup \{(u, v)\}$
            \label{alg:sample:update-subnet}
            \State $f[v] \gets f[v] + 1$ \Comment{Update the visit frequency}
            \label{alg:sample:update-freq}
        \EndWhile
        \State \Return{$\subnet=(\nodes_s, \edges_s)$}
    \EndFunction
\end{algorithmic}
\end{algorithm}

\subsection{\textsc{\sample}: Stochastic Sampling of Sub-Networks}\label{sec:method:refine}

In this section, we explain how a sub-network $\subnet$ is sampled from a global network $\net$, which is summarized as a procedure \textsc{\sample} in \cref{alg:sample}.
The procedure takes three inputs: the global tensor network $\net$, the desired number of nodes $\subnetsize$, and the visit frequency map $f$ that tracks how many times each node has been selected. 
The output is a sub-network $\subnet$, which is constructed through a priority-biased expansion.

First, a seed node $u$ is sampled from all nodes $\nodes$ in the global network via \textsc{\psample}, and added to $\subnet$ (Lines \ref{alg:sample:init-node}--\ref{alg:sample:init-subnet}).
The method \textsc{\psample} selects a random node from the input set based on their visit frequencies $f$.
To prevent the search from stagnating in recently explored regions, nodes with lower visit counts are prioritized.
Then, the algorithm enters a sequence of expansion steps to grow the sub-network $\subnet$.
In each step, a node $u$ is sampled from the existing subnet nodes $\nodes_s$ and a random neighboring node $v$ is incorporated into $\subnet$ (Lines \ref{alg:sample:expansion-node}--\ref{alg:sample:update-subnet}).
The visit frequency of $v$ is incremented at the end of each expansion step (Line \ref{alg:sample:update-freq}).
This iterative addition continues until $\subnet$ reaches the target node count $\subnetsize$, isolating a connected component in the global network for optimization.

Each sampled $\subnet$ is optimized via the recursive procedure \textsc{\recsearch} and substituted into the global network $\net$, as outlined in \cref{alg:recsearch} (Lines \ref{alg:hiss:line-optimize}--\ref{alg:hiss:line-replace}).
Although the number of nodes in $\subnet$ is small, it can still be high-order and complex.
To optimize this sub-network efficiently, we cluster the indices, refine the sub-network with clustered indices, and recursively optimize the nodes in the resulting network while considering index reshaping.
\subsection{\mergeindicescall: Reduction to Lower-Dimensional Data}
\label{sec:method:cluster}

Given a sub-network $\net_{\indsi}$
with free indices $\indsi$, the search space for optimal structures
grows combinatorially with its dimensionality. To maintain tractability, we perform \emph{index
clustering} to reduce this high-dimensional tensor network into a $\dimthreshold$-dimensional
one. The goal of index clustering is to find a partition $\{\indsi_{k}\}_{k=1}^{d_\theta}$
of the original indices $\indsi$ that reduces the dimensionality without compromising the quality of the final structure search result.

\begin{algorithm}[t]
    \caption{The index clustering heuristics}
    \label{alg:index-cluster}
    \begin{algorithmic}[1]
        \Require A tensor network $\net_{\indsi}$ with free indices $\indsi$, a dimensionality threshold $\dimthreshold$, the number of candidates $\clustercands$, and the exploration probability $\epsilon$
        \Ensure A tensor network with at most $\dimthreshold$ clusters of indices 
        \Function{\mergeindices}{$\net_{\indsi}, \dimthreshold, \clustercands, \epsilon$}
            \IfReturn{$|\indsi| \leq \dimthreshold$}{$\net_{\indsi}$}
            \label{alg:cluster:trivial}
            \State $\clustercost_{i,j}\gets \effrank{\matric{\net_\indsi}{I_i, I_j}}$ for all $I_{i}, I_{j}\in \indsi$
            \Comment{All-pair entropy calculation}
            \label{alg:cluster:cost}
            \For{$c = 1, 2, \ldots, \clustercands$}
            \Comment{Enumerate $T_{\theta}$ candidate partitions}
            \label{alg:cluster:create-start}
                \State $\indsi_{rem}\gets \indsi$ \quad $\indsi_{t,k}\gets \emptyset$ for all $k=1,2,\ldots, \dimthreshold$
                \For{$k = 1, 2, \ldots, \dimthreshold$}\Comment{Create a partition of $\dimthreshold$ clusters}
                    \While{$|\indsi_{c,k}| < |\indsi| / \dimthreshold$}\Comment{Create a cluster of $|\indsi| / \dimthreshold$ indices}
                    \label{alg:cluster:cluster-start}
                        \IfThen{$\Call{UniformRandom}{0, 1}< \epsilon$}{continue} 
                        \State $\indsi_{c,k} \gets \indsi_{c,k}\cup \arg\min_{I_i, I_j \in \indsi_{rem}, I_i \not\in \indsi_{c,k} \lor I_j \not\in \indsi_{c,k}}\clustercost_{i,j}$
                    \EndWhile\label{alg:cluster:cluster-end}
                    \State $\indsi_{rem}\gets \indsi_{rem}\setminus \indsi_{c,k}$ 
                \EndFor
            \EndFor 
            \label{alg:cluster:create-end}
            \State \Return{$\min_{c}\net_{\{\indsi_{c,k}\}_{k=1}^{\dimthreshold}}$}\label{alg:cluster:smallest}
        \EndFunction
    \end{algorithmic}
\end{algorithm}

The proposed process is formalized in \cref{alg:index-cluster}.
The function \textsc{\mergeindices} accepts the network $\net_{\indsi}$, 
a dimensionality threshold $\dimthreshold$,
the number of candidate partitions $\clustercands$, and an exploration probability $\epsilon$.
If the initial dimensionality $|\indsi|$ of $\net_{\indsi}$ is already below
$\dimthreshold$, the network is returned unchanged (Line \ref{alg:cluster:trivial}).
Otherwise, the algorithm decomposes the index partitioning task into two sequential sub-tasks:
\begin{enumerate}[(1)]
    \item \textbf{Amortized all-pair entropy computation} (Line \ref{alg:cluster:cost}). To quantify index
        correlations, the algorithm computes the cost $\clustercost_{i,j}$ as
        the singular value entropy of each unfolding $\matric{\net_{\indsi}}{I_{i}, I_{j}}$.
        Because calculating singular values in a tensor network requires the network to be in a canonical (orthonormal) form relative to the indices of interest~\cite{evenbly2022practical},
        a na\"ive approach would require numerous index swapping and orthonormalization for
        all index pairs. We instead implement a sweep-based method that
        moves the orthognality center across the network to \emph{amortize} the
        cost of swapping and orthonormalization.

    \item \textbf{$\epsilon$-greedy cluster selection} (Lines \ref{alg:cluster:create-start}--\ref{alg:cluster:create-end}). Using the precomputed
        costs $\clustercost$, the algorithm executes $\clustercands$ trials to
        construct candidate partitions. The $\epsilon$-greedy mechanism is used to
        balance the greedy clustering of highly correlated indices (to increase
        compression potential) with stochastic exploration (to avoid suboptimality).
\end{enumerate}

\subsubsection{Amortized All-Pair Cost Computation}
The efficacy of clustering relies on the singular value entropy $\mathcal{H}$, also known as the effective rank \cite{roy2007effective}, of
an unfolding matrix $M$:
\begin{equation}
    \effrank{M}= \exp\left\{-\sum_{i}p_{i}\log p_{i}\right\},\quad p_{i}= \frac{\sigma_{i}}{\sum_{i}\sigma_{i}}
    ,
\end{equation}
where $\sigma_{i}$ is the $i^{th}$ largest singular value of $M$. Lower entropy
indicates a stronger correlation, signaling that these indices should be grouped
to maximize compression.

\begin{figure}[t]
    \centering
    \includegraphics[width=0.9\linewidth]{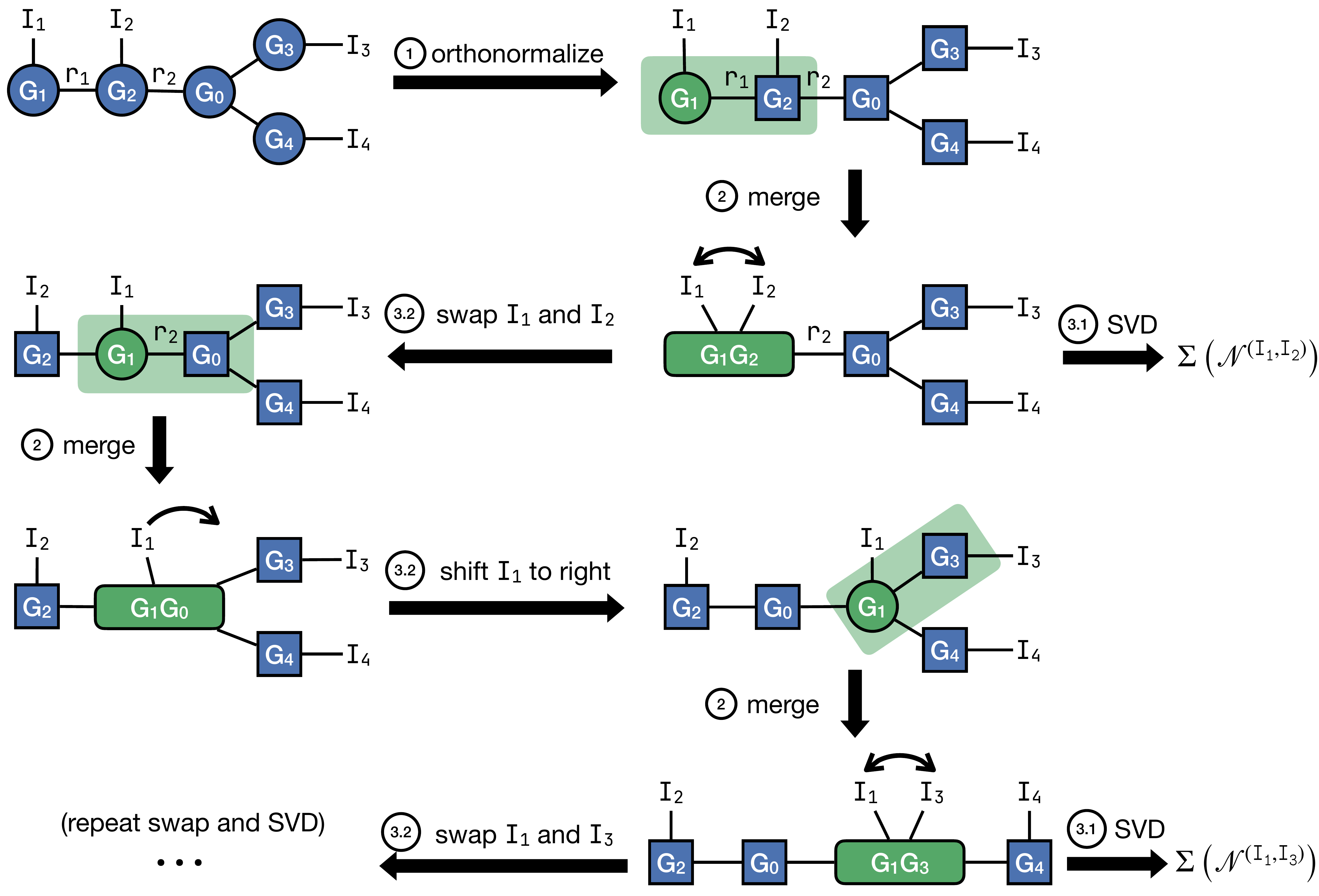}
    \caption{Cost computation for index pairs. Given a tree network, the process orthonormalizes the nodes relative to $\ten{G}_1$.
    Then, it merges two nodes $\ten{G}_1$ and $\ten{G}_2$, computes singular values for index pair $(I_1, I_2)$, swaps the index $I_1$ to the right of $I_2$, and repeats the previous steps until the costs of all index pairs involving $I_1$ have been computed.
    Orthonormalized nodes are marked as squares, and operating nodes are highlighted in green.}
    \label{fig:merge-cost}
\end{figure}

Given the tensor network $\net_{\indsi}$, computing the singular values of an unfolding $\matric
{\net_\indsi}{I_i,I_j}$ requires that indices $I_{i}$ and $I_{j}$ to be on the same node
while the remaining cores are orthonormalized~\cite{evenbly2022practical}.
A na\"ive approach contains redundant computations because the paths between different index pairs share the same internal nodes; for instance, the sequence of orthonormalizations required to bring $I_{1}$ and $I_{3}$ together in \cref{fig:merge-cost} strictly subsumes
the steps needed to bring $I_{1}$ and $I_{2}$ to be neighbors.
To eliminate such redundant computations, our algorithm maintains an \emph{orthogonal context}
during a systematic sweep, analogous to those used in TT-rounding~\cite{dolgov2012tt}, density matrix renormalization group~\cite{white2005density,white1993density}, or gauge transformations~\cite{evenbly2022practical,
evenbly2018gauge}.

As illustrated in \cref{fig:merge-cost}, the algorithm begins by orthonormalizing the surrounding nodes relative to a fixed index $I_1$ on core $\ten{G}_1$ (step
\textcircled{\scriptsize 1}).
It then contracts $\ten{G}_{1}$ (carrying index $I_{1}$) with its neighbor
$\ten{G}_{2}$ (carrying index $I_{2}$) into an intermediate tensor
$\ten{G}_{1}\ten{G}_{2}\in \real^{n_1 \times n_2 \times r_2}$ (step \textcircled{\scriptsize 2}).
This merged tensor is reshaped into a matrix of size $n_{1}n_{2}\times r_{2}$ to
compute the singular values in step \textcircled{\tiny 3.1}, providing
the clustering cost for index pair $I_{1}$ and $I_{2}$.
To advance to the next pair
without re-calculating the entire orthogonal context, a QR-based swap is
performed in step \textcircled{\tiny 3.2}, which exchanges the positions of
$I_{1}$ and $I_{2}$ while ensuring the core now containing $I_{2}$ remains
orthonormalized.
This process effectively shifts $I_{1}$ toward the next node
$G_{0}$, where the sequence of merging, SVD evaluation, and swapping is repeated
to obtain costs of $(I_{1}, I_{3})$ and all subsequent pairs.
By sliding $I_{1}$
through the network in this manner, the algorithm preserves the orthonormal
condition at each step and reduces the total complexity of the
all-pair cost computation.

\subsubsection{Index Clustering via Entropy}
With the cost $\clustercost$ available for all index pairs, the algorithm constructs a total of $\clustercands$ different partitions over the input indices $\indsi$ (Line \ref{alg:cluster:create-start} of \cref{alg:index-cluster}).
For each candidate, it iteratively groups index pairs with
the minimal clustering costs from the remaining indices $\indsi_{rem}$ and adds
them to the current cluster $\indsi_{s,k}$ (Lines \ref{alg:cluster:cluster-start}--\ref{alg:cluster:cluster-end}).
We enforce a cluster size limit $|\indsi_{s,k}| \approx |\indsi| / \dimthreshold$ to maintain balanced dimensions and prevent the formation of excessively large indices that would bottleneck the
subsequent structure search.
\revtwo{Among the $\clustercands$ partition candidates, the algorithm selects the clustering strategy that produces the network structure with the smallest size (Line \ref{alg:cluster:smallest}).}
Notably, this clustering is a virtual operation, where indices are logically
grouped without physical tensor contraction, avoiding unnecessary storage costs.
Details on how to use the resulting network of such virtual operations are provided
in the next section.
\begin{algorithm}[t]
\caption{The structure search algorithm in prior work~\cite{guo2025tensor}; boxed lines indicate steps to be modified for \tnround.}\label{alg:prior}
\begin{algorithmic}[1]
    \Require A data tensor $\ten{A}$ with indices $\indsi$, and the error tolerance $\error$
    \Function{\structsearch}{$\ten{A}$, $\error$}
        \State $\boxed{\Omega_{\indsi_s} \gets$ $\sv{}{\matric{\ten{A}}{\indsi_s}}$, $\indsi_s \subset \indsi}$
        \Comment{Pre-compute singular values}
        \label{alg:prior:precompute}
        \For{\revone{$\net_i \in \Call{CanonicalDimTrees}{\indsi}$}}\label{alg:prior:enum-partitions}
            \State Solve the constraint equation \cref{eqn:constraints} to get the cost of $\net_i$ using $\Omega$
            \label{alg:prior:solve-constraints}
            \State $\searchres{\net} \gets \min(\searchres{\net}, \net_i)$
        \EndFor
        \State $\boxed{\text{Perform tensor decomposition on }\ten{A}\text{ to get }\searchres{\net}}$
        \label{alg:prior:decompose}
    \EndFunction
\end{algorithmic}
\end{algorithm}

\subsection{\textsc{\structsearch}: Structure Search with Clustered Indices}\label{sec:method:search}
After index clustering, the high-dimensional input $\net_{\indsi}$ is reduced into a lower-dimensional representation $\net_{\{\indsi_k\}_{k=1}^{\dimthreshold}}$.
However, this is a virtual representation: indices are not physically merged onto a single node but logically grouped into clusters.
Though we can treat each index cluster $\indsi_k$ as a single dimension to apply an existing structure search framework (\eg \cref{alg:prior} from Guo \etal~\cite{guo2025tensor}), specific adaptations are necessary to optimize networks with clustered indices.
The remainder of this section reviews the search algorithm from \cite{guo2025tensor} and then details the adaptations.

\begin{figure}[t]
    \centering
    \includegraphics[width=.8\linewidth]{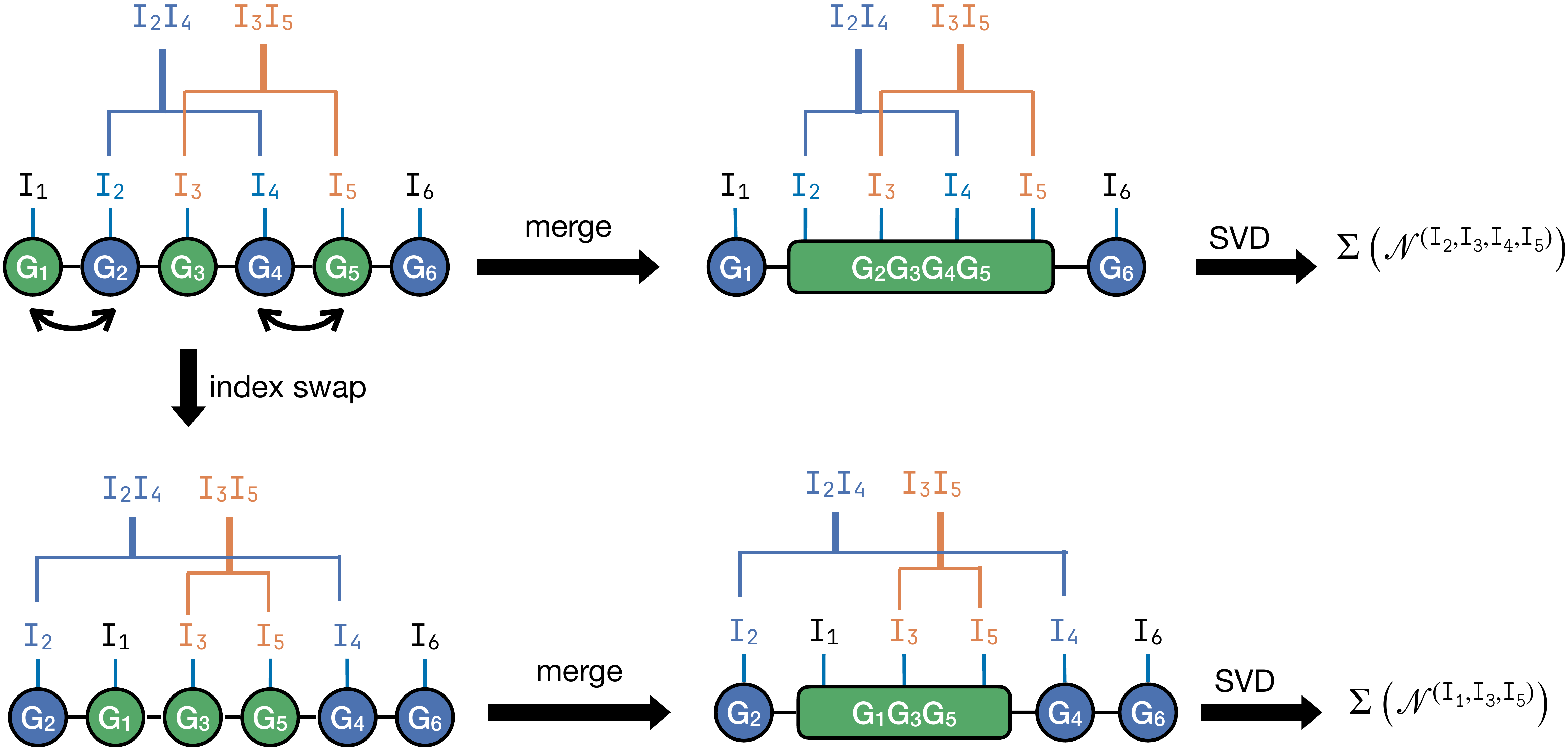}
    \caption{Illustration of singular value computation in tensor networks with clustered indices. The initial tensor network has $6$ free indices that forms $4$ clusters $\{I_1\}, \{I_2, I_4\}, \{I_3, I_5\}, \{I_6\}$. The two rows describe the singular value computation for $\mathcal{N}^{(I_2, I_3, I_4, I_5)}$ and $\mathcal{N}^{(I_1, I_3, I_5)}$ respectively. Swapping nodes are highlighted in green.}
    \label{fig:svals-with-clusters}
\end{figure}

\begin{figure}[t]
    \centering
    \includegraphics[width=0.95\linewidth]{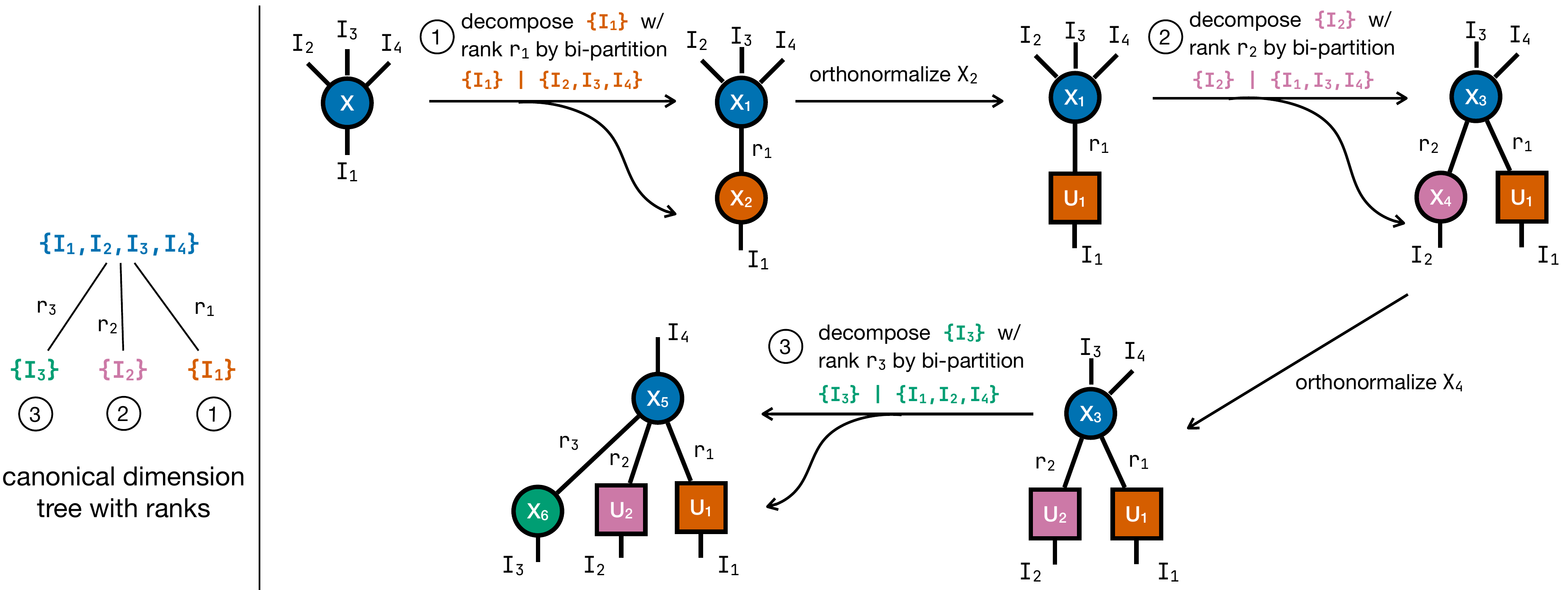}
    \caption{(Left) An example canonical dimension tree. (Right) Step-by-step decomposition of a data tensor into a network described by the canonical dimension tree on the left.}
    \label{fig:cdt}
\end{figure}

\subsubsection{Structure Search with Canonical Dimension Tree Enumeration}
\revone{The framework \cadet, proposed by Guo \etal~\citep{guo2025tensor}, solves TN-SS via a two-phase algorithm: it first enumerates tensor network topologies and then uses constraint solving as a lightweight mechanism to identity the most promising topology together with the near-optimal rank assignment.
In this framework, each tensor network is uniquely represented as a canonical dimension tree (CDT), an example of which is displayed in \cref{fig:cdt} (left). A CDT can be mapped to a set of index bi-partitions, where each partition corresponds to a node split operation during tensor decomposition. Multiple splits of the same node allows us to obtain non-binary tree tensor networks. For instance, \cref{fig:cdt} (right) shows a three-step decomposition process to construct the tensor network described by \cref{fig:cdt} (left). Each step corresponds to a node in the CDT; for example, step \textcircled{\scriptsize 1} isolates index $I_1$, which is represented by the index bi-partition $\{I_1\}\mid\{I_2, I_3, I_4\}$. Subsequent decomposition steps are similarly converted into index bi-partitions, and sequentially satisfying these partitions yields the final tensor network.
Consequently, the search for an optimal topology within the search space is transformed into a combinatorial search for CDTs.
}

As outlined in \cref{alg:prior}, the algorithm takes a data tensor $\ten{A}$ and error tolerance $\error$ as inputs.
It identifies the most efficient topology by enumerating candidate structures $\net_i$ (Line \ref{alg:prior:enum-partitions}) and estimates their costs (Line \ref{alg:prior:solve-constraints}).
To maintain efficiency, the algorithm computes costs without full tensor decomposition.
Instead, it solves a constrained optimization problem to find the minimum bond ranks $r_1, r_2, \ldots, r_k$ that satisfy the total error tolerance $\error$:
\begin{equation}
    \arg\min_{r_1, r_2, \ldots, r_k} \size{\net_i} \quad s.t. \quad \norm{\data{\indsi}{\net_i} - \ten{A}} / \norm{\ten{A}} \leq \error,
    \label{eqn:constraints}
\end{equation}
where the resulting network size---computed from the ranks $\{r_i\}_{i=1}^{k}$---is used as the cost to select candidate structures.

Because solving \cref{eqn:constraints} involves calculating the reconstruction error for every bond in every candidate structure, a straightforward approach would require numerous expensive SVDs.
To make this search feasible, the algorithm performs a one-time pre-computation of singular values for all possible index subsets $\indsi_s \subseteq \indsi$ (Line \ref{alg:prior:precompute}).
These precomputed values, denoted as $\Omega_{\indsi_s}$, serve as a lookup table.
By referencing $\Omega_{\indsi_s}$, the algorithm can instantly estimate the rank and error associated with any bi-partition, shifting the computational burden to the initialization phase and allowing the search to focus entirely on topology enumeration.

While \cref{alg:prior} provides a robust foundation, we adapt it to our specific refinement task at two steps (marked with boxes).
First, since our input consists of clustered indices, the pre-computation (Line \ref{alg:prior:precompute}) must be modified to calculate singular values over index groups rather than individual indices.
Second, as our input is already a tensor network rather than a raw data tensor, we replace the tensor decomposition (Line \ref{alg:prior:decompose}) with a targeted structure transformation that efficiently evolves one tree structure into another.

\subsubsection{Singular Value Pre-computation for Index Clusters}
To compute singular values for combinations of clustered indices, we treat each index cluster as an aggregated dimension (Line \ref{alg:prior:precompute} of \cref{alg:prior}).
Because the number of clusters $d_\theta$ is small, the number of required SVD evaluations is drastically reduced.
Specifically, given a network $\net_{\{\indsi_k\}_{k=1}^{\dimthreshold}}$ and a combination of clusters $(\indsi_i, \indsi_j)$, the singular value computation for this cluster combination requires relocating the indices in $\indsi_i \cup \indsi_j$ to a contiguous region of $\net_{\{\indsi_k\}_{k=1}^{\dimthreshold}}$.
As displayed in \cref{fig:svals-with-clusters}, when the selected clusters lie at adjacent nodes (\eg $\{I_2, I_4\}$ and $\{I_3, I_5\}$), singular values are computed directly through local merging and SVD. For non-adjacent clusters such as $\{I_1\}$ and $\{I_3, I_5\}$, the algorithm determines an optimal plan of adjacent swaps to bring the target indices into proximity, which enables the local contraction and subsequent SVD of the desired unfolding $\matric{\net}{I_1, I_3, I_5}$.

\begin{figure}[t]
    \centering
    \includegraphics[width=.9\linewidth]{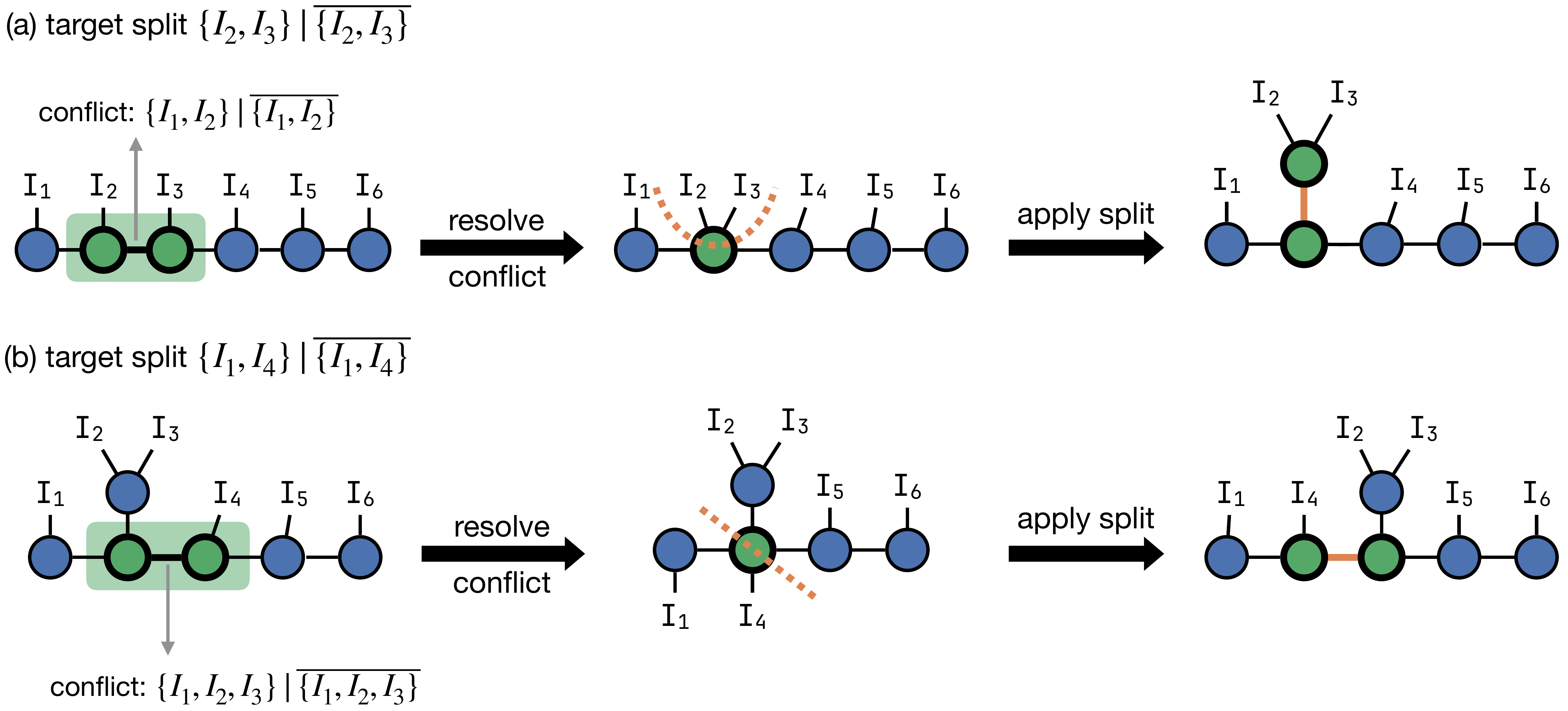}
    \caption{Resolution of structural conflicts via lazy merge during structure transformations. The modifying sub-networks are bolded and highlighted in green. Node splits are marked with orange lines, and the resulting edges are also highlighted in orange. (a) The algorithm resolves the conflict by merging the overlapping core tensors, and the intermediate node is split to satisfy the target bi-partition. (b) The algorithm only resolves the conflicting bi-partition $\bipar{\{I_1,I_2,I_3\}}$ without merging $\bipar{\{I_1\}}$.}
    \label{fig:transform}
\end{figure}

\subsubsection{Structure Transformation via Lazy Merge}
Once a result structure $\searchres{\net}$ is found (Line 4 of \cref{alg:prior}) by the enumerative search, prior work decomposes the data tensor $\ten{A}$ into $\searchres{\net}$ via a series of SVDs (Line \ref{alg:prior:decompose}).
When the input is already a tensor network, we should avoid contracting the input tensor network to a single core to save memory and reduce time.
Therefore, we utilize a lazy algorithm to transform the input structure to the target one, which maximizes reuse of the existing network topology and reduces computational cost during the transformation.
Suppose the input tensor network has free indices $\indsi$, we follow the correspondence between tensor network structures and index bi-partition sets and characterize both the input and target networks by their respective sets of index partitions, $\{\bipar{\indsi^{in}_{i}}\}_{i=1}^{k}$ and $\{\bipar{\indsi^{tar}_{j}}\}_{j=1}^{l}$, where $\indsi^{in}_i \subset \indsi$ for $i \in [1,k]$ and $\indsi^{tar}_j \subset \indsi$ for $j \in [1,l]$.
The structure transformation continues by sequentially applying index bi-partitions (each associated with a node split) specified in the target set. \revone{A target partition $\bipar{\indsi^{tar}_j}$ is executed only if it is nested or disjoint with all current partitions $\{\bipar{\indsi^{in}_{i}}\}_{i=1}^{k}$. In other words, the target partition should satisfy that $\forall i \in [1, k].\ \indsi^{in}_{i} \subset \indsi^{tar}_{j}$ $\lor\ \indsi^{tar}_{j} \subset \indsi^{in}_{i}$ $\lor\ \indsi^{in}_{i} \cap \indsi^{tar}_{j} = \emptyset$.} If a partition $\bipar{\indsi^{in}_{i}}$ does not satisfy this condition, meaning the sets partially overlap without one being a subset of the other, the algorithm resolves the conflict by contracting the affected nodes before executing the required split.

\cref{fig:transform} presents two examples of such resolutions.
In \cref{fig:transform}(a), applying the bi-partition $\bipar{\{I_2,I_3\}}$ conflicts with an existing $\bipar{\{I_1, I_2\}}$.
The algorithm reverts the conflict bi-partition through node merging before applying the target split.
Similarly, in \cref{fig:transform}(b), a conflict between the target split $\bipar{\{I_1, I_4\}}$ and the existing $\bipar{\{I_1, I_2, I_3\}}$ is removed by merging the node with index $I_4$ to its left neighbor.
However, the target split is performed without merging indices $I_1$ and $I_4$ onto the same node.
While the algorithm may fall back to contracting the entire network to a single node in rare, complex conflict scenarios, these incremental resolutions typically suffice in practice.
\subsection{\textsc{\topreshape}: Enabling Index Reshaping in Structure Search}\label{sec:method:reshape}

\begin{algorithm}[t]
\caption{The index reshaping heuristic}\label{alg:index-reshape}
\begin{algorithmic}[1]
    \Require A tensor $\ten{G}$ with indices $I_1, I_2, \ldots, I_k$ where $\texttt{size}(I_i) = n_i$ for $i \in [1,k]$, and a factorization threshold $\factthreshold$
    \Ensure A set of tensors that indices of each are factorized into at most $\factthreshold$ parts
    \Function{\topreshape}{$\ten{G}, \factthreshold$}
        \For{$i \in [1, k]$}
            \State $\mathscr{P}_i \gets \kargmin_{\mathbf{p} \in \mathcal{F}(I_i, \factthreshold)}\mathscr{C}(I_i, \mathbf{p})$\label{alg:reshape:kmin}
        \EndFor
        \State \Return{$\{\reshape{\ten{G}}{p_1, p_2, \ldots, p_d} \mid p_i \in \mathscr{P}_i\}$}
    \EndFunction
\end{algorithmic}
\end{algorithm}

As outlined in Line \ref{alg:hiss:line-reshape} of \cref{alg:recsearch}, the \textsc{\topreshape} procedure is invoked on each node with orthonormalized surrounding context to allow exploring topologies that were previously inaccessible in the native indices.
Recognizing that reshaping introduces another decision layer, we incorporate it into the structure refinement process \textsc{\recsearch} using the heuristic described in \cref{alg:index-reshape} to ensure efficient space traversal.

\subsubsection{Heuristic Filtering of Reshaping Candidates}
Exhaustive search over all possible index reshaping is impractical due to the combinatorial explosion of factorization and permutation options.
To address this, we introduce a cost-based heuristic in \cref{alg:index-reshape} to filter out unpromising reshaping candidates.
For each free index $I_i$ of size $n_i$ in a tensor $\ten{G}$, the algorithm generates a set of candidate factorizations
\begin{equation}
    \mathscr{F}(I_i, \factthreshold) = \left\{ \left(p_{1}, p_{2}, \ldots, p_{m} \right)~\Big|~n_i = \prod_{k=1}^{m}p_{k},  m \leq \factthreshold,  p_{k} \in \nat_{> 1} \right\}
\end{equation}
where $m$ is the variable factorization depth bounded by $\factthreshold$, and $\nat_{> 1}$ denotes natural numbers greater than $1$.

To evaluate these candidates, we define a reshaping cost $\mathscr{C}(I_i, \mathbf{p})$ that estimates the complexity of a specific factorization $\mathbf{p} = (p_{1}, p_{2}, \ldots, p_{\factthreshold})$ of index $I_i$.
The cost is modeled as the size of a local TT sub-network where each core node corresponds to a factor $p_j$. The internal bond ranks are determined by effective ranks at each cumulative splitting point of the original index:
\begin{subequations}
\begin{align}
\mathscr{C}(I_i, \mathbf{p}) &= \sum_{p_{j} \in \mathbf{p}} \effrank{M_{i,j-1}} \times p_{j} \times \effrank{M_{i,j}} \\
M_{i,j} &= \reshape{\matric{\ten{G}}{I_i}}{\prod_{k \leq j} p_{k}, -1}
\end{align}
\end{subequations}
In this formulation, $M_{i,j}$ represents the unfolding of the tensor $\ten{G}$ at the $j^{th}$ split of index $I_i$, with $\effrank{M_{i,0}} = \effrank{M_{i,m}} = 1$.
By calculating the effective rank at these junctions, the heuristic identifies \revone{factorizations} that effectively slice the index at points of low information density.

As shown in \cref{alg:index-reshape} (Line \ref{alg:reshape:kmin}), we retain only the top $K$ reshaping candidates per index to be passed into the subsequent structure search.
This filtering ensures that the search space remains manageable.
Note that this process is reserved for free indices; internal indices remain at their original sizes to ensure the integrity of tree structures.

\subsubsection{Efficiency via Computational Caching}
To further optimize the structure search process, we implement a caching mechanism for the computed effective ranks at all splitting points.
This is particularly effective because different ordered factorizations often share identical prefix products, leading to overlapping split points.
For instance, an index with size $12$ admits factorizations such as $(2,2,3)$, $(2,3,2)$.
Though these reshapings result in different data layouts, the initial split at size 2 is identical for both.
Similarly, two factorizations $(2,3,2)$ and $(3,2,2)$ both require the effective rank for a split at total size $6$.
By caching these results, we eliminate redundant SVD-based computations.

\subsubsection{Recursive Refinement with Index Reshaping}
The individual components of clustering, refining, and reshaping are unified through the recursive refinement procedure \textsc{\recsearch}.
Once the structure search identifies an optimal structure on the clustered indices, the algorithm zooms into each resulting network fragments (\cref{alg:recsearch} Lines \ref{alg:hiss:line-rec-start}--\ref{alg:hiss:line-rec-end}).
If a fragment contains indices that can be further decomposed, the process reshapes those indices and triggers a new iteration of clustering and structure refinement.
The recursion terminates when further splitting yields no reduction in any of the fragments.
This hierarchical strategy effectively bypasses both memory and time bottlenecks of flat search methods, allowing for the discovery of complex, deep-tree topologies that were previously computationally inaccessible.

\subsection{Complexity Analysis}\label{sec:complexity}
In this section, we analyze the computational complexity for a single iteration of structure refinement for a sampled sub-network $\subnet$, and we assume $\subnet$ is a tensor train with $\subnetsize$ free indices of size $n$ and internal ranks of size $r$.

\mypara{Index Clustering}
To compute the clustering cost for an index pair, the algorithm performs $\subnetsize$ rounds of orthonormalization over $\subnet$, each round of complexity $\bigo{\subnetsize n r^3}$. 
Then, the cost computation requires one SVD with complexity $\bigo{n^2 r^4}$ and one QR with complexity $\bigo{n^3 r^3}$.
Hence, the total time complexity for index clustering is $\mathcal{O}\big(\subnetsize^2 n r^3 + \subnetsize^2 (n^2 r^4 + n^3 r^3)\big) = \bigo{\subnetsize^2n^2r^3 (r+n)}$.

\mypara{Structure Search with Clustered Indices}
After index clustering, the structure search is performed over a reduced net of $\dimthreshold$ clustered indices.
The complexity of this phase is divided into two steps.
First, the pre-computation of singular values takes
$\bigo{\subnetsize n^3 r^3 + 2^{\dimthreshold} n^{4+\subnetsize/\dimthreshold} r^4}$
as the worst case needs to move an index from one end of the tensor train to the other end, and each swap requires a QR decomposition.
Second, the enumeration of candidate structures and the following constraint solving takes
$\bigo{2^{\dimthreshold} \dimthreshold^{2.5}}$ time~\cite{morrison2016branch}.
Note that, while the $n^{4+\subnetsize / \dimthreshold}$ term in the first step indicates exponential growth relative to the sub-network dimensionality $\subnetsize$, it does not imply exponential scaling for the global network.
At the sub-network level, we can easily pick small $\subnetsize$ and $\dimthreshold$ values to keep the structure optimization computationally manageable.

\mypara{Index Reshaping}
The time complexity of index reshaping depends on the index sizes.
For an index of size $\prod_{i=1}^{m} p_i^{k_i}$ where $p_i$ are prime factors, $k_i \in \nat$, and $m$ is the number of prime factors, the total number of reshape cost computation is $\prod_{i=1}^{m} (k_i + 1)$, and each individual reshape cost is computed in $\bigo{\subnetsize n r^2}$ time, which adds up to a total of $\bigo{\subnetsize n r^2 \prod_{i=1}^{m} (k_i + 1)}$.

\mypara{Overall Complexity}
To summarize, the overall complexity for optimizing a sub-network at one recursion level is
$
\mathcal{O}\big(\subnetsize^2( n^2 r ^4 + n^3 r^3 ) + n^{4 + \subnetsize/\dimthreshold} r^4  2^{\dimthreshold} + 2^{\dimthreshold} \dimthreshold^{2.5} + \subnetsize n r^2 \prod_{i=1}^{m} (k_i + 1)\big).
$

\section{Numerical Experiments}\label{sec:eval}
This section presents empirical evaluations of the proposed \algoname algorithm.
We utilize analytical functions and three real-world applications --- thermal radiation transport, neutron diffusion, and computational fluid dynamics --- to verify \algoname's ability of discovering new structures with enhanced compression performance, and to establish its scalability regarding higher-dimensional data tensors.
The experiments are performed on a MacBook Pro with an Apple M3 Max chip and 128 GB memory. In all experiments, we choose the number of nodes in sampled sub-networks $\subnetsize=4$, dimensionality threshold $\dimthreshold=4$, number of clustering candidates $\clustercands=5$, exploration probability $\epsilon=0.1$, factorization threshold $\factthreshold=4$, and we select the top $10$ reshape options for each index.
The sampling iterations $\maxiters$ is default to be $5$ except where specifically noted.

Since no existing literature supports structure discovery for tensor network rounding, we evaluate \algoname against fixed-structure TT-round and HT-round.
Our comparisons focus on the compression ratio within error tolerance rather than runtime, so we utilize standard implementations for these baselines without specialized acceleration techniques.

\mypara{Evaluation Metrics}
For all experiments, given the input structure $\net$ with indices $\indsi$ and 
the discovered structure $\searchres{\net}$,
we collect four metrics:
\begin{enumerate}[(1)]
\item CR over Input: the compression ratio relative to the input structure $\net$ computed by $\size{\net} / \size{\searchres{\net}}$;
\item CR over data: the compression ratio relative to the represented data by $\size{\data{\indsi}{\net}} / \size{\searchres{\net}}$;
\item Search time: the end-to-end structure search time in seconds;
\item Reconstruction error: the relative error computed on a random sample of $3,000$ data points $V$, 
$\norm{\data{\indsi}{\net}(V) - \data{\indsi}{\searchres{\net}}(V)} / \norm{\data{\indsi}{\net}(V)}$.
\end{enumerate}

\subsection{Analytical Functions}\label{sec:analytical}

\begin{table}[t]
    \renewcommand{\arraystretch}{1.25}
    \centering
    \caption{List of analytical functions}
    \label{tab:analytical}
    \begin{tabular}{ll}
        \toprule
        Name         &  Analytical formula \\
        \midrule
        \dixon        & $(x_1 - 1) ^ 2 + \sum_{i=2}^{d} i(2x_i^2 - x_{i-1})^2$ \\
        \pathological & $\sum_{i=1}^{d-1} \left(0.5 + \frac{\sin^2 \sqrt{100x_i^2 + x_{i-1}^2} - 0.5}{1 + 0.001(x_i^2 - 2x_ix_{i+1}+x_{i+1}^2)^2}\right)$ \\
        \pinter       & $\sum_{i=1}^d \left(ix_i^2 + 20i \sin^2 A_i + i\log_{10}(1 + iB_i^2)\right)$ \\
        \qing         & $\sum_{i=1}^{d} (x_i^2 - i)^2$ \\
        \schaffer     & $\sum_{i=1}^{d-1} \left(0.5 + \frac{\sin^2 \sqrt{x_i^2 + x_{i-1}^2} - 0.5}{1 + 0.001(x_i^2+x_{i+1}^2)^2}\right)$ \\
        \trigonometric & $\sum_{i=1}^d \left(d - \sum_{j=1}^{d}\cos x_j + i (1 - \cos x_i - \sin x_i)\right)^2$ \\
        \sqsum        & $\left(\sum_{i=1}^{d} x_i^2\right)^{-\frac{1}{2}}$\\
        \hilbert      & $\frac{1}{\sum_{i=1}^{d} x_i}$ \\
        \bottomrule
    \end{tabular}
\end{table}

In this section, we evaluate the compression efficiency and scalability of \algoname using a suite of analytical functions characterized by complex landscapes.
We select $8$ benchmark functions from literature~\cite{oseledets2010tt, ballani2013black, ryzhakov2024black}, specifically excluding those that can be trivially represented as TTs.
These functions, described in \cref{tab:analytical}, are discretized on a uniform grid with $8$ nodes per dimension.
To provide a baseline, we first employ TT-Cross~\cite{oseledets2010tt} to construct an initial TT representation, which \algoname then refines into a more optimized structure.

\begin{figure}[!t]
    \centering
    \includegraphics[width=\linewidth]{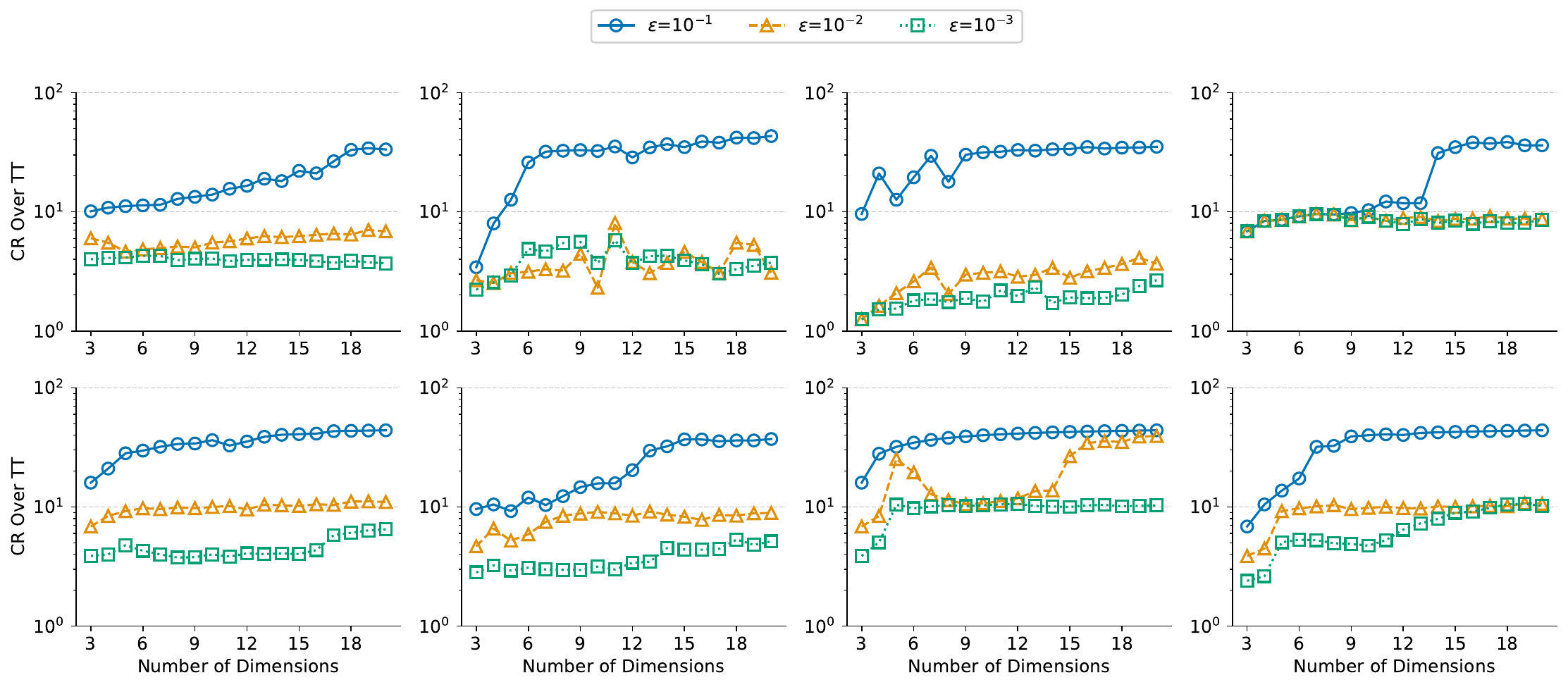}
    \caption{Compression ratio over TT versus dimensions. We choose dimensions from $3$ to $20$, and error tolerance $\error = 10^{-1}$, $10^{-2}$, and $10^{-3}$. Top row: \dixon, \pathological, \pinter, \qing; bottom row: \schaffer, \trigonometric, \sqsum, \hilbert.
    }
    \label{fig:analytical:cr}
\end{figure}

\begin{figure}[!t]
    \includegraphics[width=\linewidth]{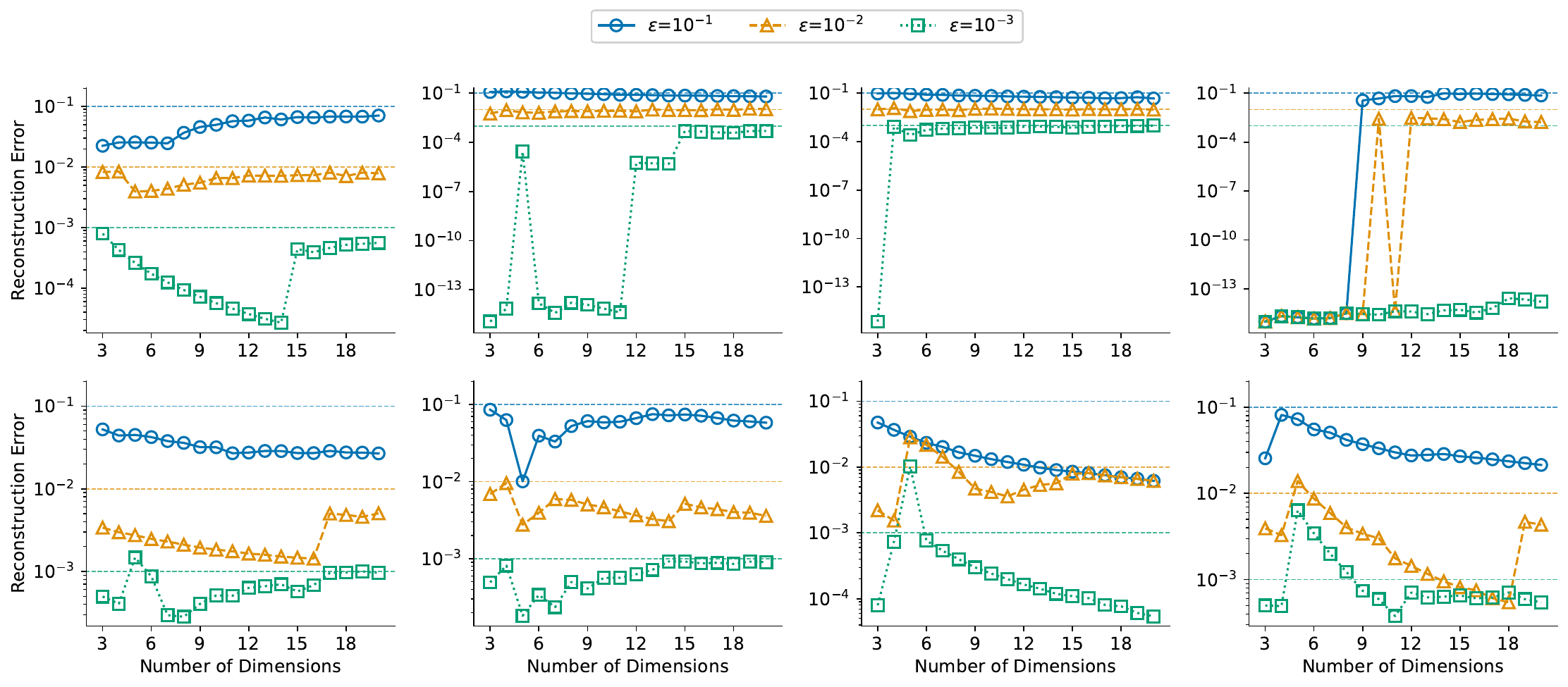}
    \caption{Reconstruction error versus dimensions. We choose dimensions from $3$ to $20$, and error tolerance $\error = 10^{-1}$, $10^{-2}$, and $10^{-3}$. Top row: \dixon, \pathological, \pinter, \qing; bottom row: \schaffer, \trigonometric, \sqsum, \hilbert.
    }
    \label{fig:analytical:error}
\end{figure}

\begin{figure}[t]
    \includegraphics[width=\linewidth]{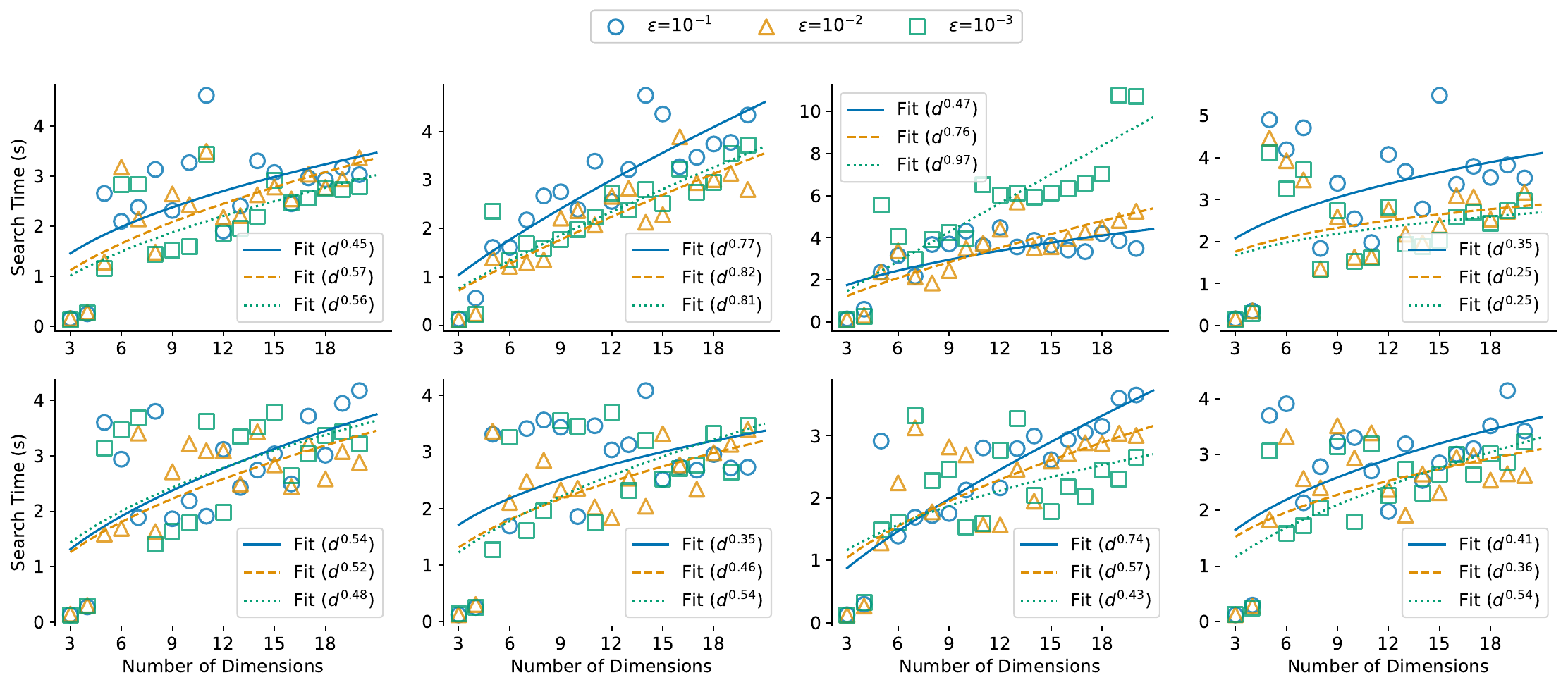}
    \caption{Search time versus dimensions. We choose dimensions from $3$ to $20$, and error tolerance $\error = 10^{-1}$, $10^{-2}$, and $10^{-3}$. Top row: \dixon, \pathological, \pinter, \qing; bottom row: \schaffer, \trigonometric, \sqsum, \hilbert.
    }
    \label{fig:analytical:time}
\end{figure}

We compare compression ratios (\cref{fig:analytical:cr}), reconstruction errors (\cref{fig:analytical:error}), and execution time (\cref{fig:analytical:time}) for the benchmark functions across dimensions $d \in [3, 20]$ and varying error tolerances ($\error = 10^{-1}, 10^{-2}, 10^{-3}$).
Across all benchmarks, the proposed algorithm consistently discovers topologies that achieve compression ratios of $2\times$ to $30\times$ over input tensor trains.
Reconstruction error is verified on a validation set of 3,000 points against the original function instead of the TT-cross results.
While the measured errors occasionally exceed the prescribed tolerance $\error$ slightly, they consistently remain within the same order of magnitude.
This minor variance is due to the inherent approximation bounds of the initial TT-Cross construction and the stochastic nature of sampling over a finite validation set.

\algoname demonstrates favorable runtime performance, with average execution time remaining under $6$ seconds even for high-dimensional cases.
As shown in the trend lines of \cref{fig:analytical:time}, the computational cost scales sub-linearly with dimensionality.
We observe an abrupt transition in runtime between $d=4$ and $d=5$, which corresponds to the algorithm's shift from an exhaustive heuristic search to incorporating stochastic sampling once the dimensionality exceeds the predefined threshold ($\dimthreshold=4$).
This scalability is a direct result of the hierarchical search architecture.
By systematically pruning the search space and focusing on the most promising index clusters, the algorithm maintains tractability in high-dimensional spaces.
Notably, this reduction in search complexity does not lead to a huge degradation in model performance, as compression ratios remain high across the board.
This suggests that our heuristic pruning effectively discards unpromising structural regions \revone{without compromising the optimality} of the final results.
\subsection{Thermal Radiation Transport}\label{sec:eval:radiation}

In this section, we evaluate the performance of \algoname on thermal radiation transport solutions.
This physical process describes the propagation of energy through a medium via electromagnetic radiation, a phenomenon central to astrophysics and nuclear engineering~\cite{rybicki2024radiative, jiang2021implicit, kurzer2024radiation}.
The primary quantity of interest is the frequency-integrated specific intensity $I(\Vec{\mathbf{x}}, \Vec{\mathbf{\Omega}}; t)$,
which represents the radiative energy flux at a specific location $\Vec{\mathbf{x}}$, moving in direction $\Vec{\mathbf{\Omega}}$ at time $t$.
Its evolution is governed by the radiative transport equation, which balances the local change in intensity with transport, scattering, and absorption/emission processes:
\begin{equation}
    \partial_{t} I + c \mathbf{n} \cdot \nabla I = c \left[ \rho \kappa_{s} (J - I) + \rho \kappa_{a} \left( \frac{c}{4\pi} a_{r} T^{4} - I \right) \right].
\end{equation}
In this formulation, $J$ is the mean intensity, $c$ is the speed of light, and $\Vec{\mathbf{n}}$ is the unit direction vector.
The radiation constant is denoted by $a_r$, while $\rho$, $\kappa_s$, and $\kappa_a$ correspond to the density, specific scattering opacity, and specific absorption opacity of the medium, respectively.
The material temperature $T$ is governed by the energy balance equation:
\begin{equation}
    \partial_{t} T = -\frac{c \kappa_{a}}{c_{v}} (a_{r} T^{4} - E),
\end{equation}
where $c_v$ represents the specific heat capacity and $E$ is the radiation energy density.
The resulting solution space is 5-dimensional, comprising two spatial coordinates $\Vec{\mathbf{x}} \in \real^2$, two angular coordinates $\Vec{\mathbf{\Omega}} = (\theta, \phi)$, and time $t$.
Prior work~\cite{gorodetsky2025thermal} addressed the curse of dimensionality in this high-dimensional domain by representing the solution in a fixed TT format and advancing it in time using a step-and-truncate strategy~\cite{einkemmer2025review}.

In this paper we investigate whether alternative network designs can better mitigate rank growth across different spatial geometries.
We propose to use \algoname as a generalized truncation that adaptively modifies both ranks and the underlying tensor network structure according to a specified error tolerance.
This structural flexibility offers a promising foundation for the next generation of PDE solvers.
In time-dependent systems, the solution's correlation structure often shifts as physical features evolve.
While static formats like the TT are restrictive, \algoname can dynamically reconfigure its architecture to match the current state of the physics.
By allowing the network to evolve alongside the solution, we can maintain high fidelity with significantly lower storage overhead, paving the way for more efficient large-scale simulations of complex dynamical systems.

\begin{figure}[t]
    \centering
    \includegraphics[width=\linewidth]{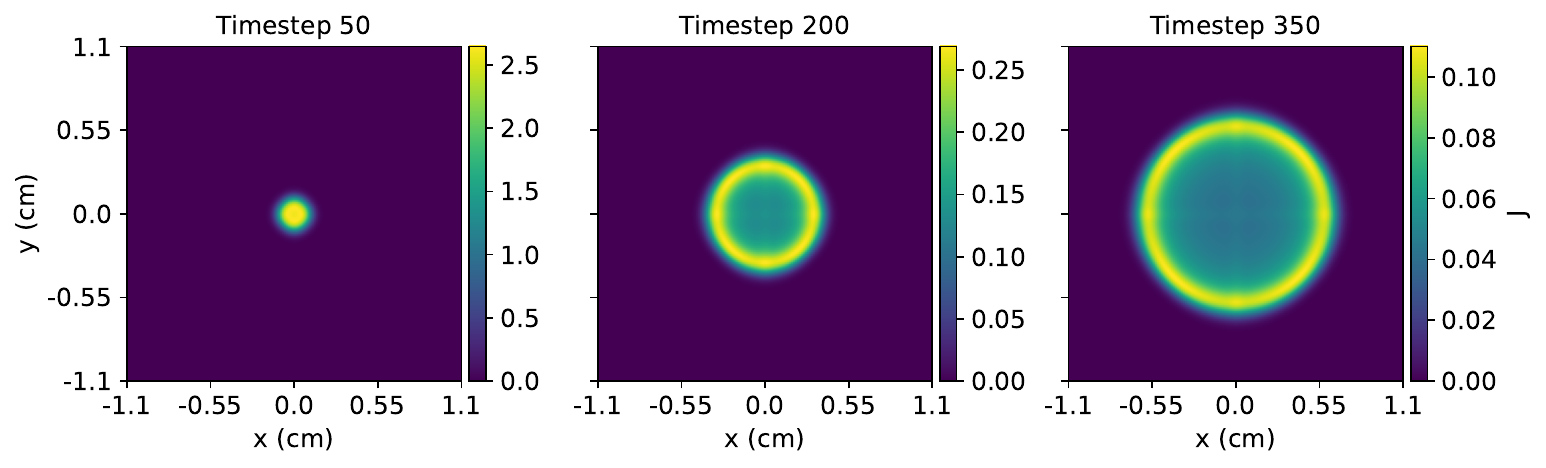}
    
    \includegraphics[width=0.3\linewidth]{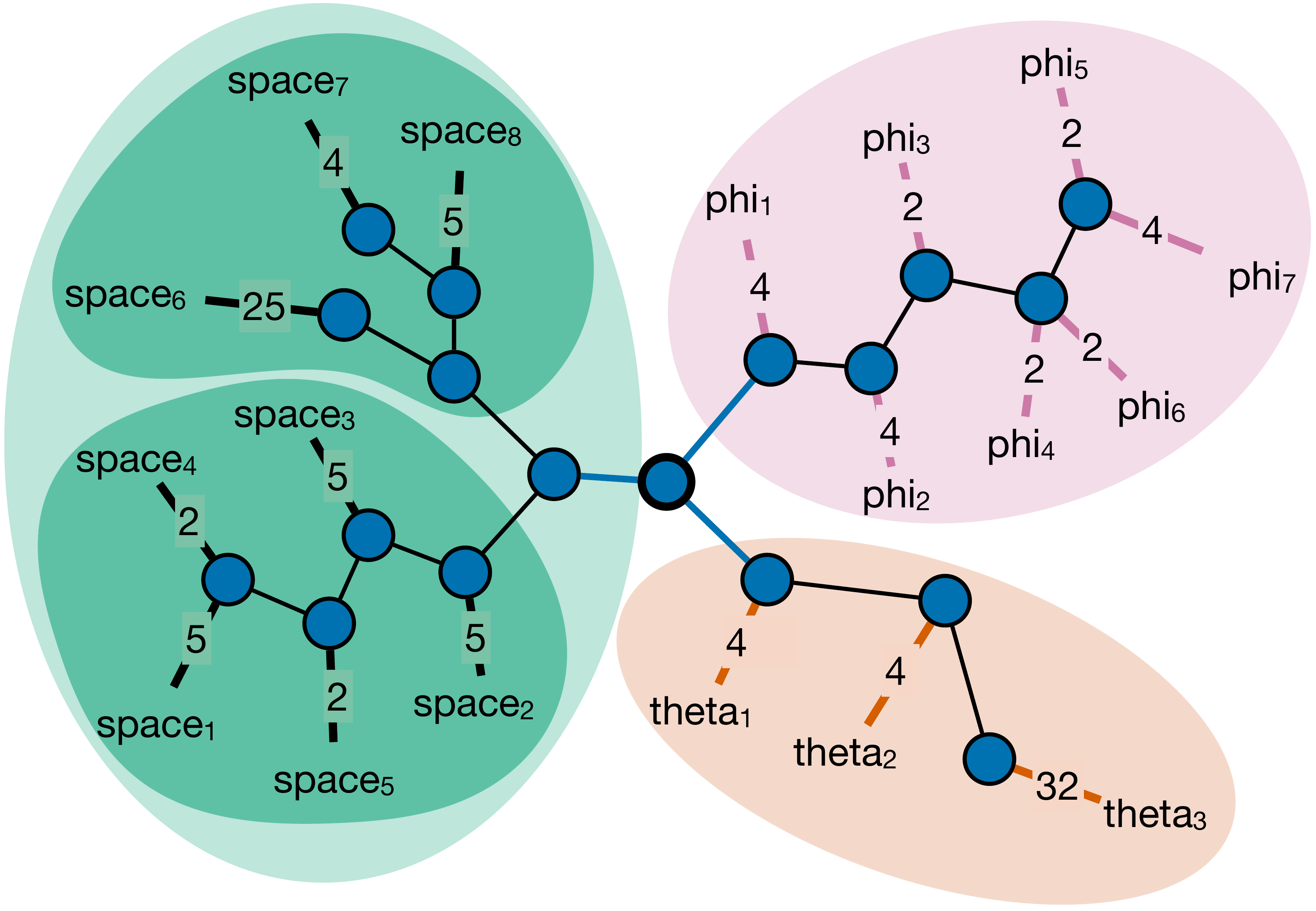}
    ~
    \includegraphics[width=0.3\linewidth]{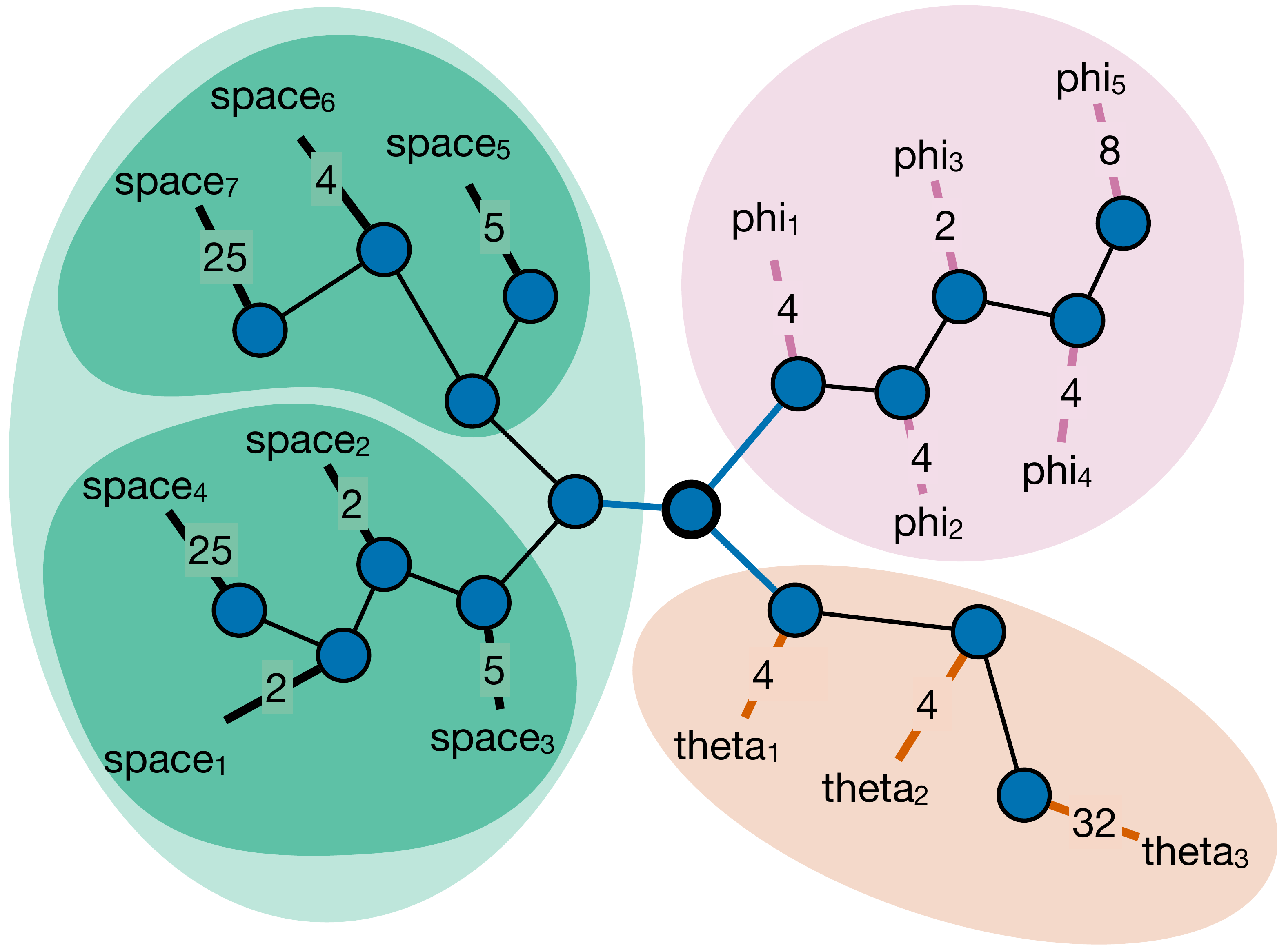}
    ~
    \includegraphics[width=0.3\linewidth]{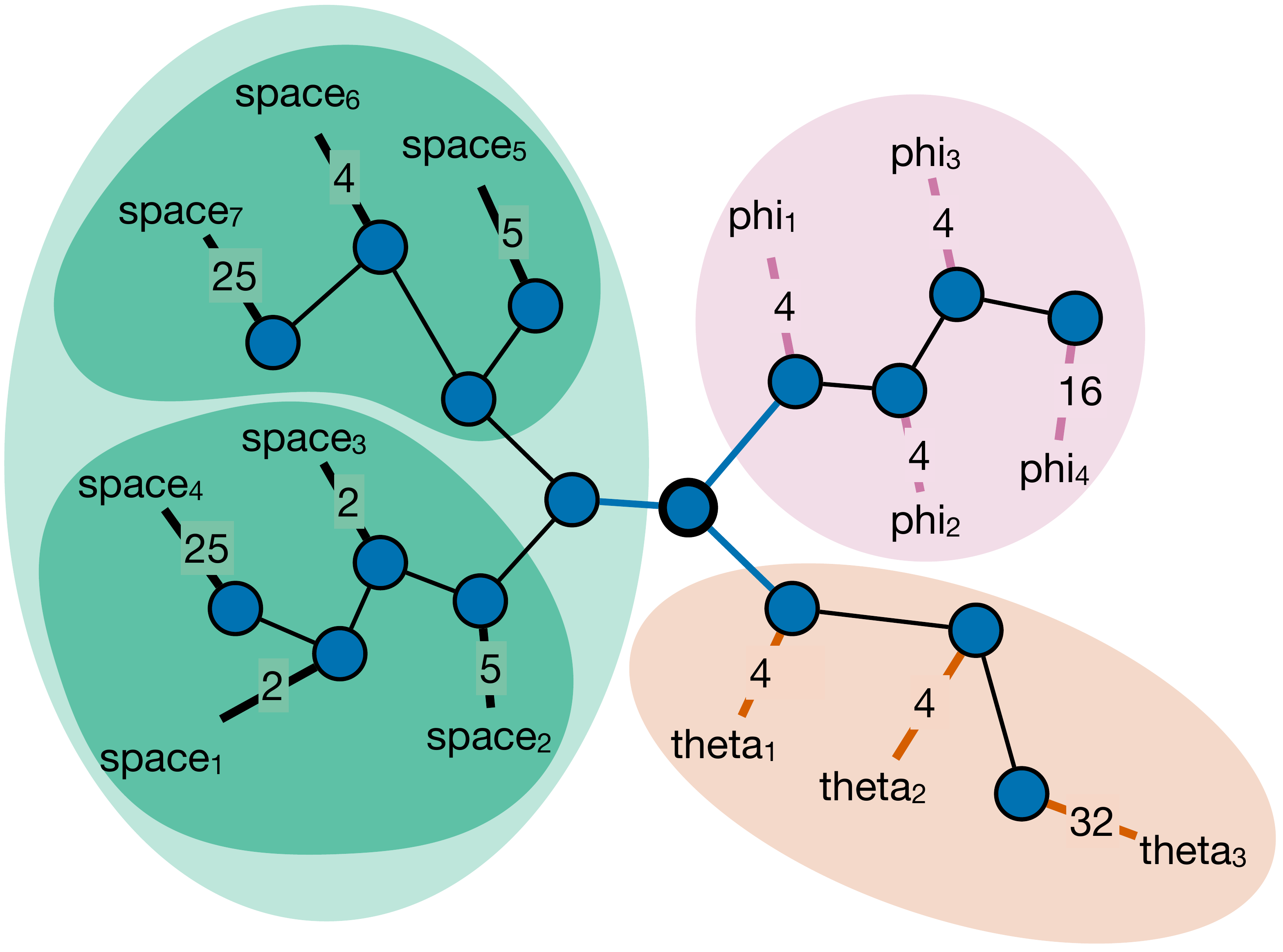}
    \caption{
    Examples of solutions (top) and discovered structures (bottom) for the line source problem at three timesteps $50$, $200$, and $350$ with error tolerance $\error=10^{-4}$.
    Three colored regions represent the native indices: green for the spatial index $X$, pink and orange for the angular indices $\Phi$ and $\Theta$ respectively.
    The three regions are linked through an internal node (with bold edges).
    In the green-shaded spatial domain, \algoname splits the indices into two groups (dark green) representing the $x$-axis (dimension 500) and the $y$-axis (dimension 500). 
    While TT-round yields modest compression ratios of 5.8, 7.0, and 7.9 across these timesteps, \algoname with reshaping achieves significantly higher ratios of 4590.0, 1117.8, and 645.1, respectively.
    }
    \label{fig:thermal:line-source-example}
\end{figure}

\begin{figure}[t]
    \centering
    \subfloat[Compression ratio over TT]{
    \includegraphics[width=0.45\linewidth]{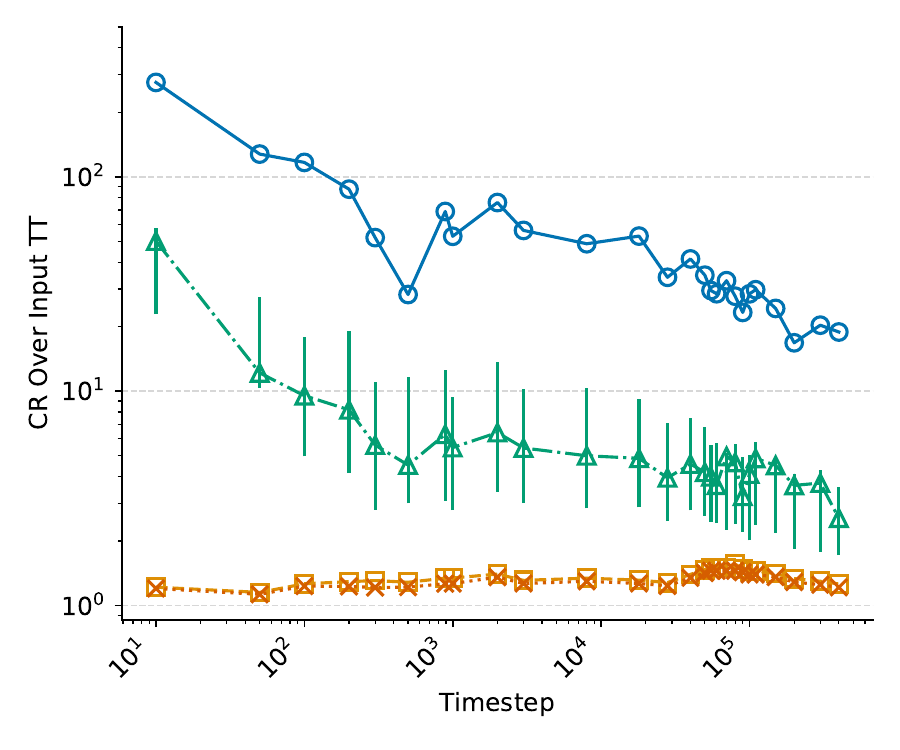}
    \label{fig:thermal:crooked-pipe-rel}
    }
    \subfloat[Compression ratio over data]{
    \includegraphics[width=0.45\linewidth]{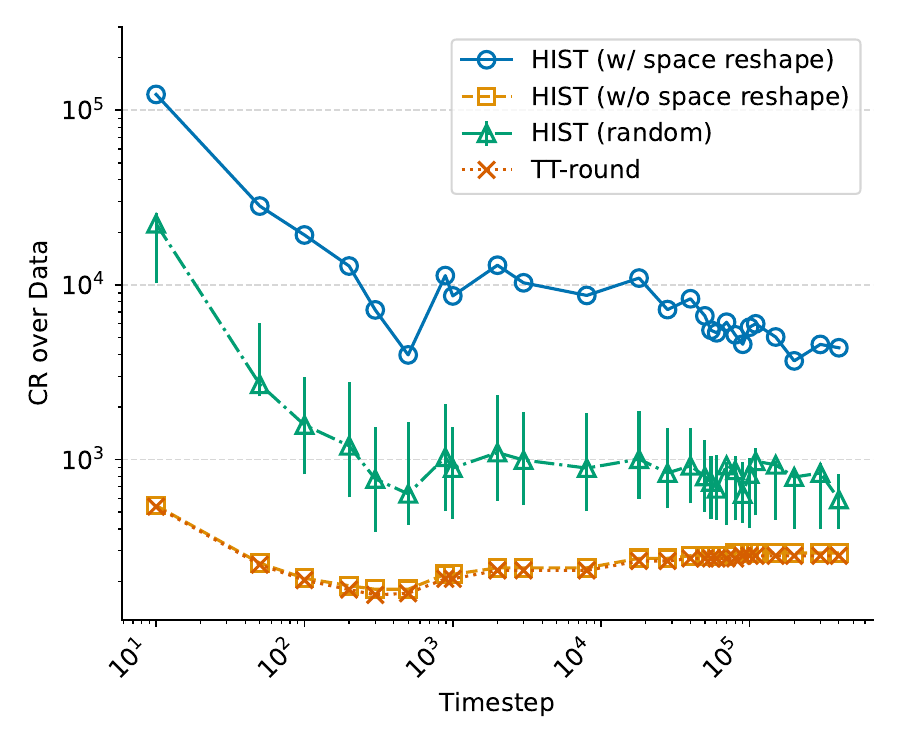}
    \label{fig:thermal:crooked-pipe-abs}
    }
    \caption{Comparison of compression ratios over TT (left) and compression ratios over data (right) 
    across timesteps for \algoname and its two variants: 
    one disables index reshaping for the space dimension (\algoname w/o space reshape),
    and the other replaces heuristics by random choices (\algoname random).
    The central markers and error bars for the random structure baselines represent the $50\%$, $25\%$, and $75\%$ quantiles across $5$ random seeds. 
    \algoname with space reshaping achieves compression ratios an order of magnitude higher than the average of random structures, and disabling reshaping leads to almost no compression. 
    }
    \label{fig:thermal:crooked-pipe}
\end{figure}

We evaluate this idea using 2-D, time-dependent, gray thermal transport simulations.
Following the discretization framework established in~\cite{gorodetsky2025thermal}, the solution at timestep $t$ is represented in a 3-D tensor train format:
\begin{equation}
    \hat{I}^{t}(i,l,p) = X^{t}(:,i) \Theta^{t}(:,l,:) \Phi^{t}(p,:)
\end{equation}
where $X^{t}$ represents the flattened spatial domain on an $N_x \times N_y$ grid, and $\Theta^{t}$ and $\Phi^{t}$ represent the polar and azimuthal angles discretized on grids of size $N_\theta$ and $N_\phi$, respectively.
The underlying TT-based solver is structured as an iterative time-stepping procedure where a rounding algorithm is applied at the end of each step to prevent uncontrollable rank growth.
To assess the impact of replacing the standard TT-rounding with \algoname, we utilize the \emph{unrounded} solutions produced by the solver as our input into \algoname.
\algoname then identifies an optimized network structure that preserves the solution within a target error tolerance $\error$, effectively acting as a structurally adaptive truncation operator.
We select two test benchmarks: the line source~\cite{bhattacharyya2023finite} and the crooked pipe problem~\cite{southworth2024one}.
They expose different behaviors under the standard TT representation: in the line source problem, the compression ratio fails to stabilize when using traditional TT-rounding (\cref{fig:thermal:line-source-abs} (TT-round));
in contrast, the crooked pipe benchmark shows a sharp initial decrease in compression performance which then converges to a steady state as the solution evolves (\cref{fig:thermal:crooked-pipe-abs} (TT-round)).

We evaluate \algoname on simulation solutions across various timesteps to assess compression efficiency, structural optimality, and generalization.
To isolate the impact of our proposed heuristics, we compare \algoname against three baselines:
(1) a variant that restricts spatial index reshaping to benchmark performance in scenarios requiring fixed spatial partitions, such as domain decomposition~\cite{wang2015parallel,brunner2026domain} or Adaptive Mesh Refinement~\cite{jessee1998adaptive,velarde2005radiation};
(2) a variant that replaces \mergeindicescall and \topreshapecall in \cref{alg:recsearch} with random selection to validate the effectiveness of our heuristic-guided search;
and
(3) the conventional TT-rounding used in the solver.
We report absolute compression ratios of discovered structures across all timesteps, as well as their values normalized against the unrounded TT.

We also evaluate the optimality and generalizability of the learned structures.
We compress the TT solution at each timestep using structures discovered at other timesteps and compute the \emph{relative compression ratio} $CR_{i,j}/CR_{i,i}$  where $CR_{i,j}$ is the compression ratio obtained by \emph{structural rounding the unoptimized TT solution} at timestep $t_i$ using the structure discovered at timestep $t_j$.
We analyze the distribution of these ratios from different perspectives to understand how closely the discovered structures approach optimality and whether they can generalize across the simulation's evolution.
We conclude by relating these trends to the underlying physics of each problem.

\subsubsection{Line Source}
\begin{figure}[t]
    \centering
    \subfloat[Compression ratio over TT]{
    \includegraphics[width=0.45\linewidth]{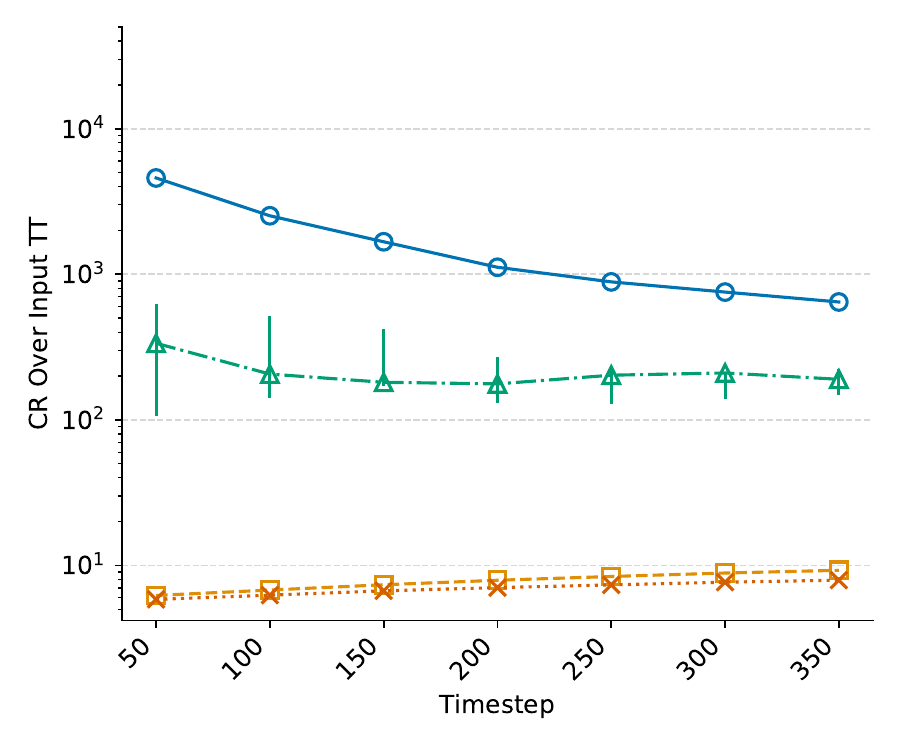}
    \label{fig:thermal:line-source-rel}
    }
    \subfloat[Compression ratio over data]{
    \includegraphics[width=0.45\linewidth]{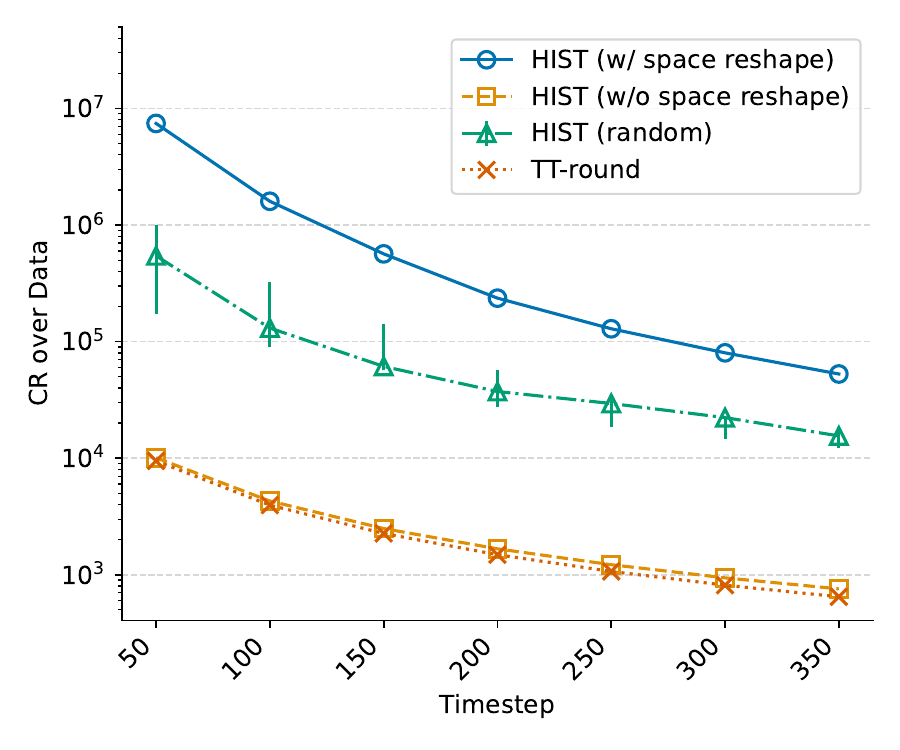}
    \label{fig:thermal:line-source-abs}
    }
    \caption{CR over input TT across timesteps for line source data at $\error=10^{-4}$. The central markers and error bars for the random structure baselines represent the $50\%$, $25\%$, and $75\%$ quantiles across $5$ random seeds. Structures with space reshaping achieves CRs one and two orders of magnitude higher than \algoname (random) and TT-round respectively. Disabling reshaping (\algoname (w/o space reshape)) leads to almost no compression.}
    \label{fig:thermal:line-source}
\end{figure}

The 2-D line source problem in \cite{gorodetsky2025thermal} is initialized with a Gaussian pulse at the center, which produces an axisymmetric ring of radiation propagating outward at speed $c$.
As a vacuum transport problem without a source term, the solution does not reach a steady state, and the TT ranks consequently do not stabilize (as shown in \cref{fig:thermal:line-source-abs}).
This scenario represents a ``worst-case" setting for a TT-based solver, as the ranks continue to grow over time.

The problem requires a large amount of angles to mitigate ray effects inducing a very fine discretization of $(N_x, N_y, N_\theta, N_\phi)=(500, 500, 512, 1024)$ and $\error=10^{-4}$.
The line source solutions obtained from \cite{gorodetsky2025thermal} are illustrated in \Cref{fig:thermal:line-source-example} for 3 different timesteps.

\begin{figure}[t]
    \centering
    \subfloat[Relative CR distribution for each data across all discovered structures.]{
    \includegraphics[width=.75\linewidth]{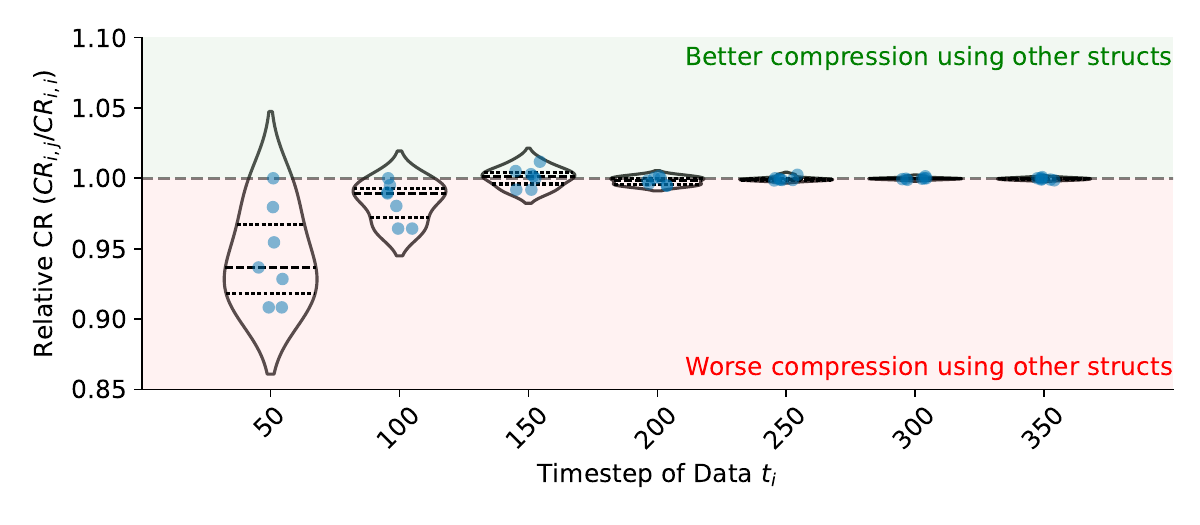}
    \label{fig:line-source:gen-data}
    }

    \subfloat[Relative CR distribution for each structure when applied to all data.]{
    \includegraphics[width=.75\linewidth]{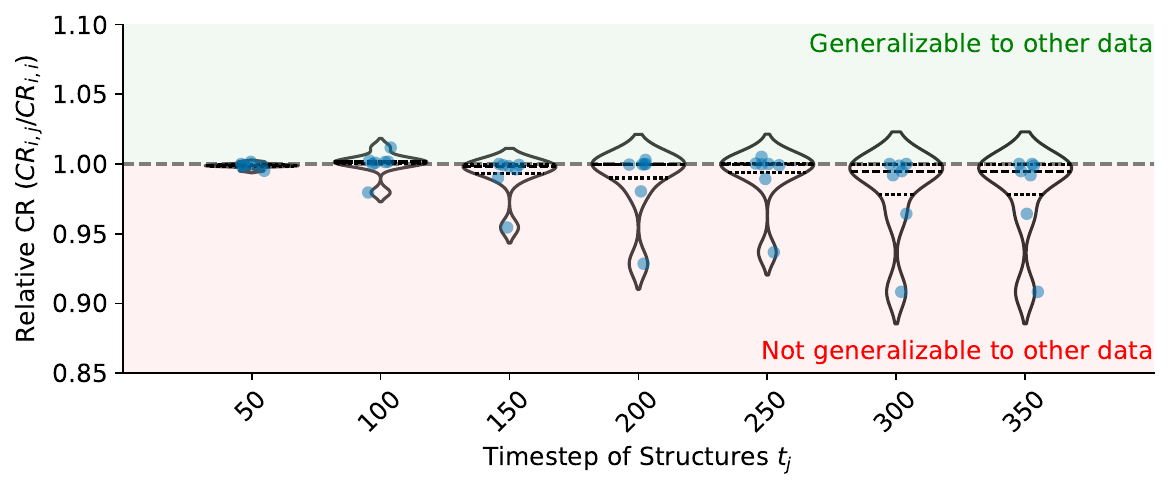}
    \label{fig:line-source:gen-struct}
    }
    \caption{Distribution of relative compression ratios ($CR_{i, j} / CR_{i, i}$) for the line source problem. Quartiles are marked as dashed lines.
    (a) Fixed Data ($t_i$): Applying structures discovered at all other timesteps $t_j$ to a fixed data tensor at $t_i$. Most ratios lie below 1.0, confirming that local structural searches achieve near-peak optimality for their specific timesteps.
    (b) Fixed Structure ($t_j$): Applying a single discovered structure from $t_j$ to data across all other timesteps $t_i$. The clustering of ratios near 1.0 demonstrates strong structural generalization at early timesteps, while points below 1.0 reveal poor generalization for structures discovered at later timesteps.
    }
    \label{fig:line-source:gen}
\end{figure}

The compression performance of \algoname for the line source problem is shown in \Cref{fig:thermal:line-source}.
The structures identified by \algoname consistently outperform randomly sampled structures due to the effective guidance of entropy-based heuristics.
They also achieve significantly higher compression ratios compared to standard TT-rounding, and the improvement of compression ranges from $80\times$ (timestep $350$) to $790\times$ (timestep $50$).
When the solution is represented in a Cartesian spatial coordinate system, the tensor ranks grow monotonically over time, which is reflected in the progressively decreasing compression ratios in \Cref{fig:thermal:line-source-abs}.

\Cref{fig:line-source:gen-data} evaluates structural optimality by comparing the compression achieved by structures discovered at different timesteps against the structure specifically optimized for a fixed TT solution.
Consider the violin at timestep $50$ in \cref{fig:line-source:gen-data}, each point inside the violin represents the compression ratio when structures found at other timesteps are applied to unrounded TT data at timestep $50$ relative to the compression of the structure discovered from timestep $50$.
More points around or below $1.0$ indicates that structures discovered at other timesteps exhibit poor performance at timestep $50$.
The fact that many points fall below the $1.0$ threshold is a strong indicator of the search algorithm's near-optimality; it demonstrates that the discovered structure specialized to a timestep is substantially superior to structures derived from other timesteps.
Notably, the variance in relative compression ratios peaks at $t=50$ but shrinks as time advances; this suggests that as the physical data increases in complexity, the solution becomes less sensitive to specific network topologies.

\cref{fig:line-source:gen-struct} assesses how a fixed structure from $t_j$ performs on TT solutions across the entire simulation.
For example, each point inside the violin at timestep $50$ represents the compression ratio of structure found at timestep $50$ is applied to data at other timesteps relative to the compression of the structure discovered from timestep $50$.
More points around or above $1.0$ indicates that structures discovered at timestep $50$ exhibit good performance on data at other timesteps.
Many points staying around the $1.0$ threshold suggests high generalizability of our search results; the discovered structure specialized to a timestep performs well on data at other timesteps.
We also observe a temporal asymmetry: structures discovered at early timesteps maintain high efficiency on later data, exhibiting strong generalization capabilities;
whereas structures derived from late-stage data perform poorly when applied to early-timestep data, as early-timestep data leverages specialized spatial and angular decompositions, as we can see in \cref{fig:thermal:line-source-example} (bottom).

\begin{figure}[t]
    \centering
    \includegraphics[width=\linewidth]{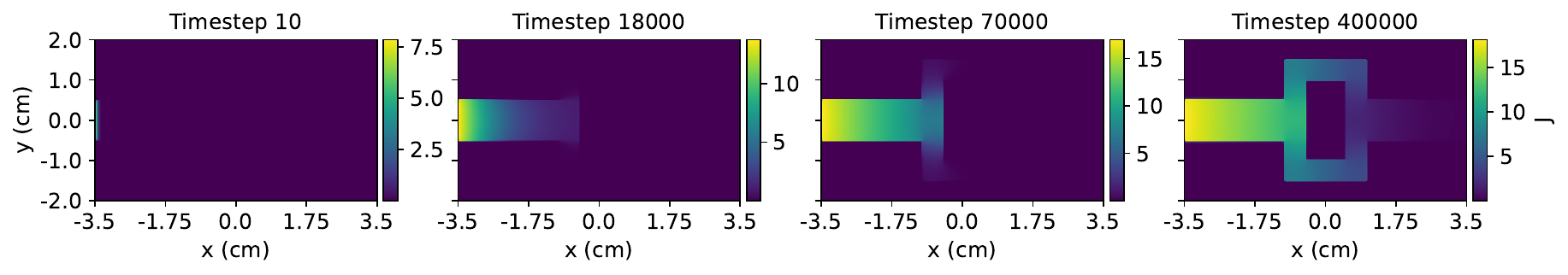}
    
    \includegraphics[width=.24\linewidth]{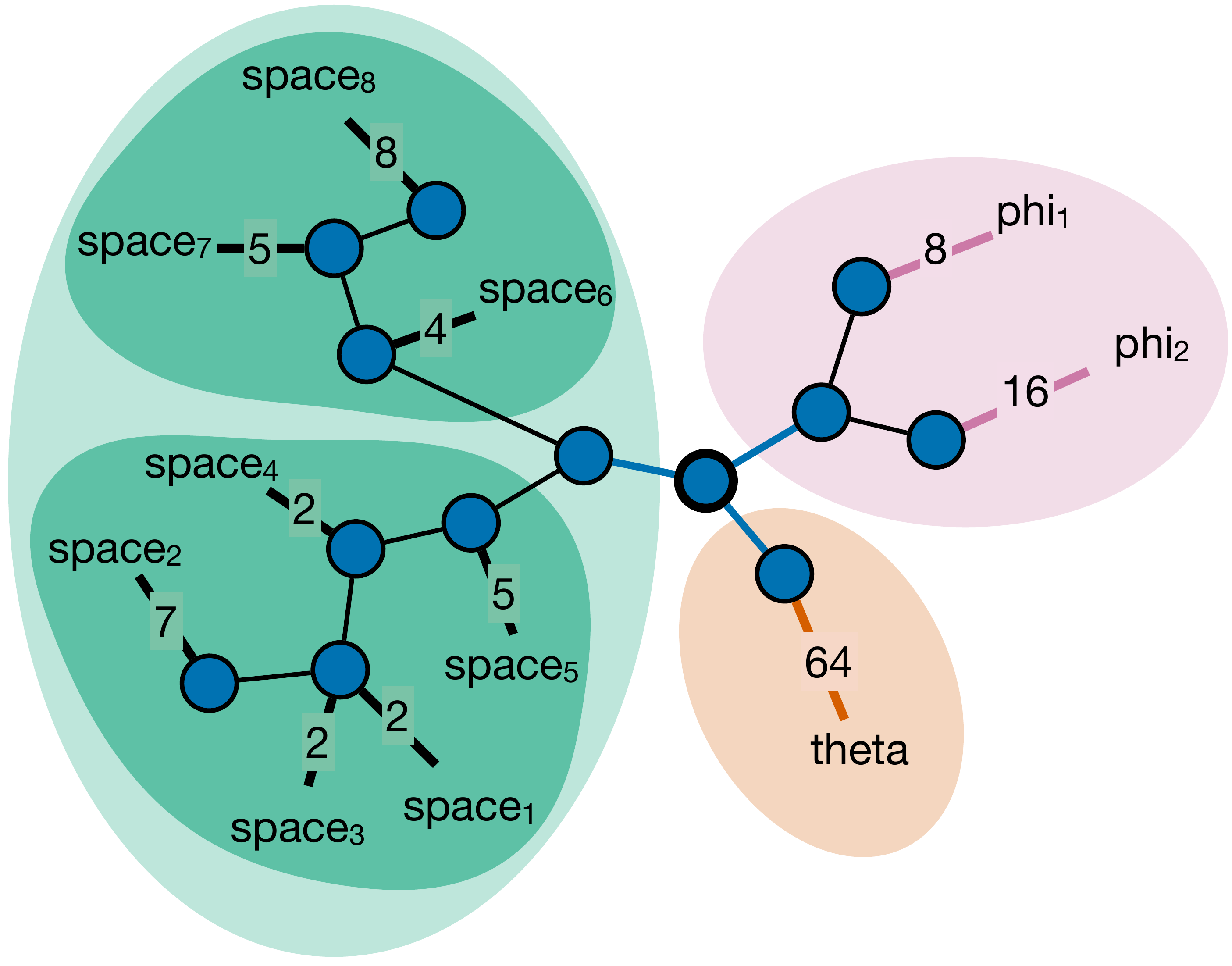}
    \includegraphics[width=.24\linewidth]{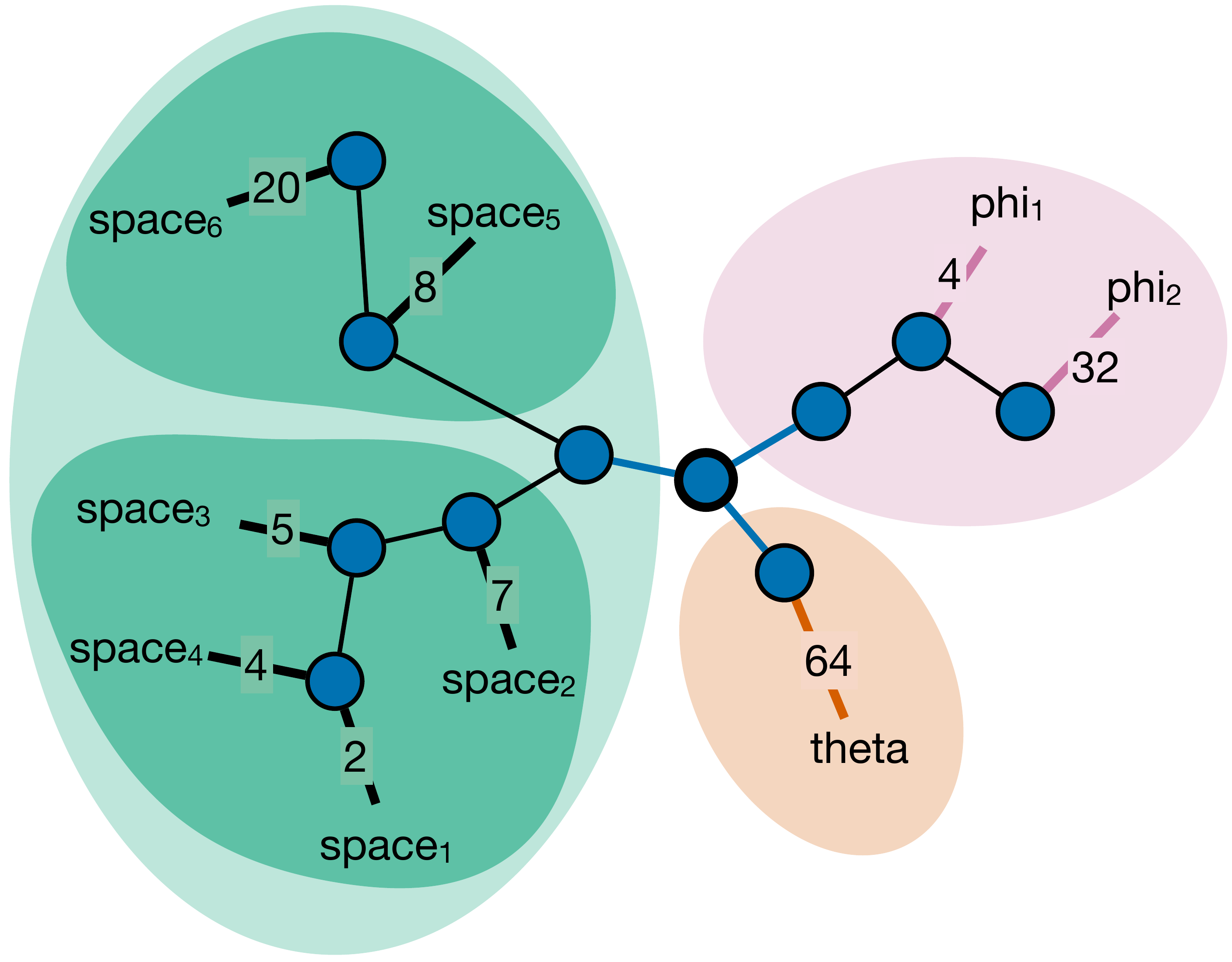}
    \includegraphics[width=.24\linewidth]{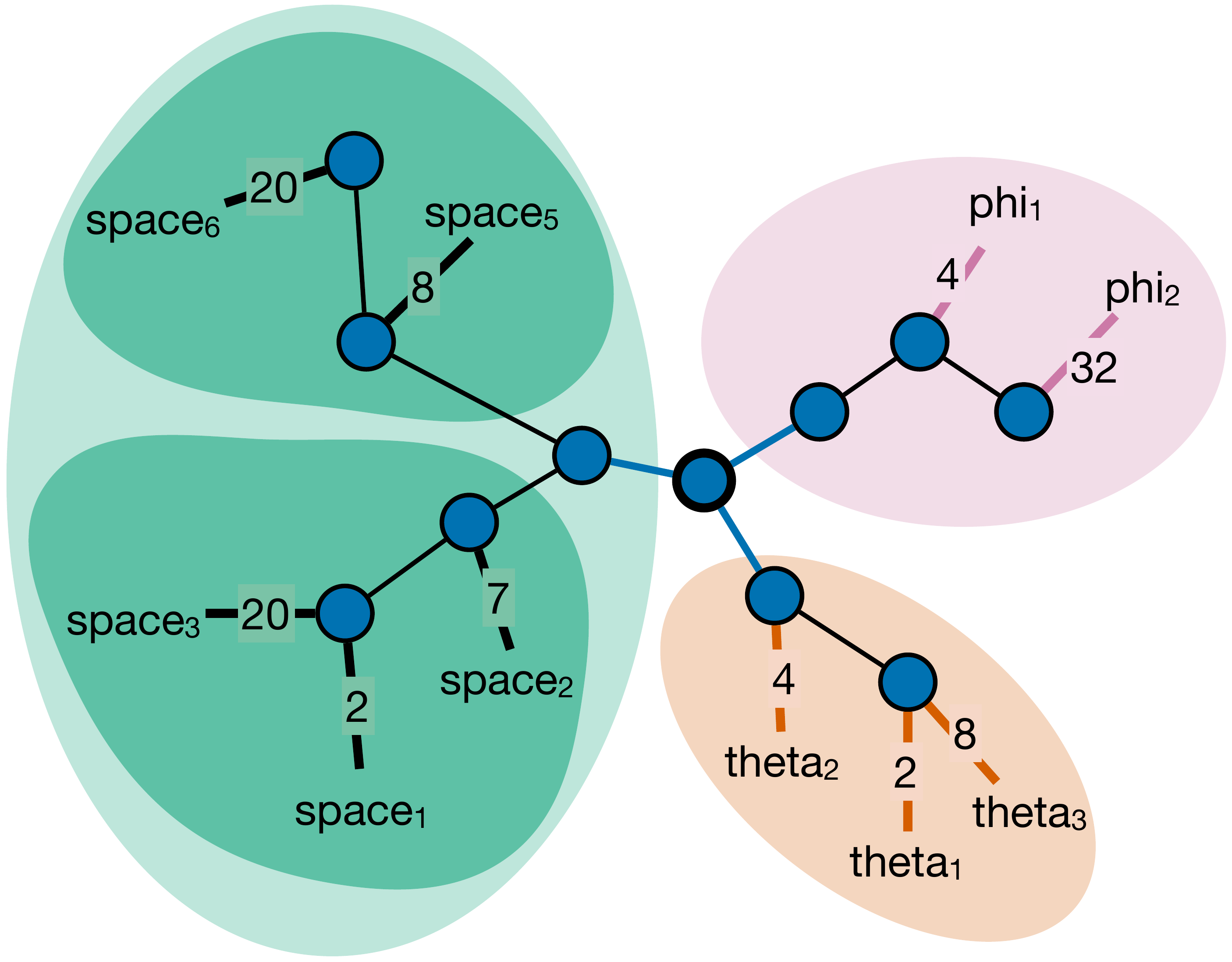}
    \includegraphics[width=.24\linewidth]{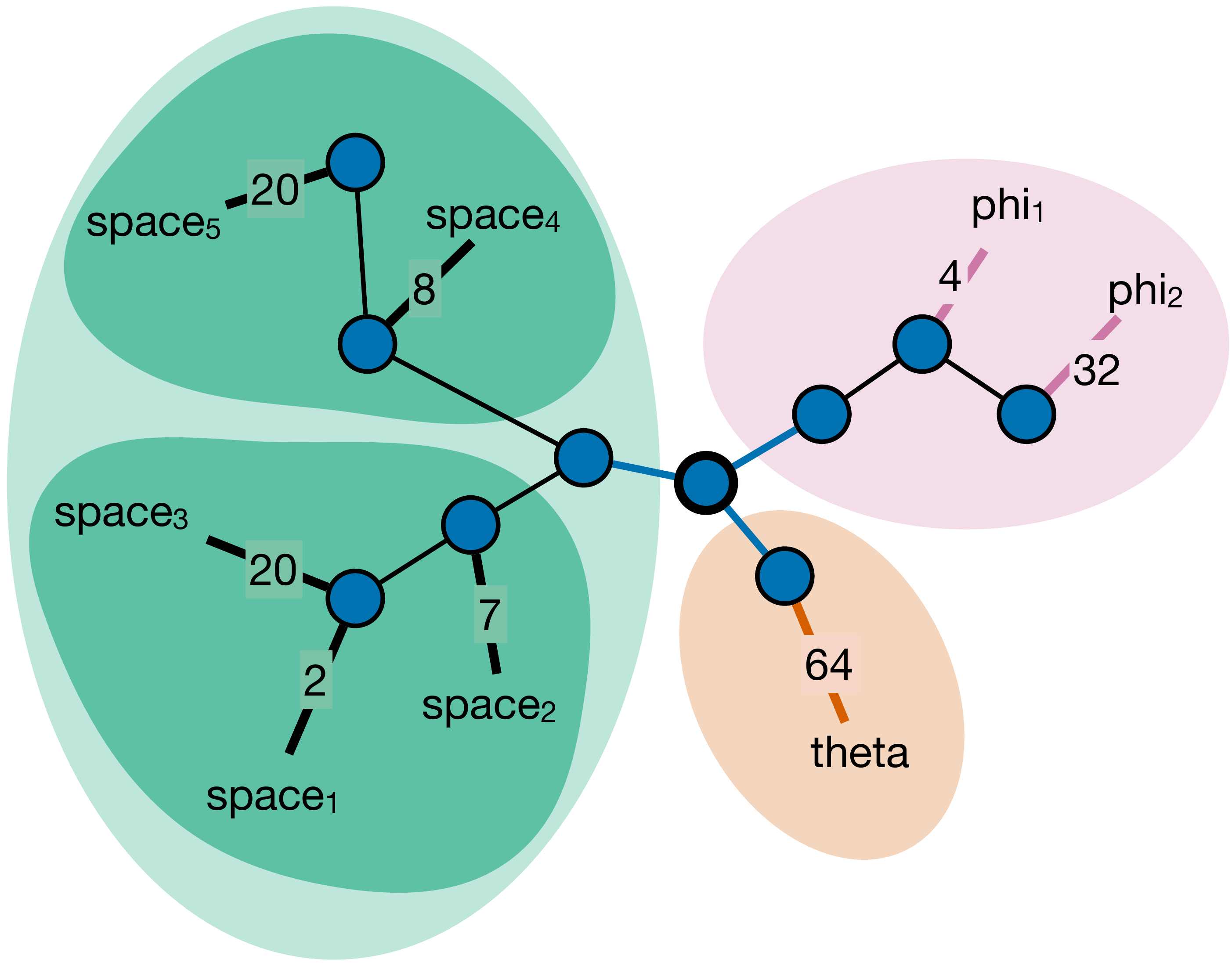}
    \caption{
    Thermal radiation transport equation solutions (top) and discovered structures (bottom) for the crooked pipe problem~\cite{gorodetsky2025thermal} at time steps $10$, $18000$, $70000$, and $400000$ with error tolerance $\error=10^{-3}$. 
    The TT-round yields compression ratios of $1.2$, $1.3$, $1.5$, and $1.2$, while \algoname with index reshaping achieves $271.1$, $52.0$, $32.8$, and $18.9$ respectively.
    Color definitions are the same as that in \cref{fig:thermal:line-source}.
    }
    \label{fig:thermal:crooked-pipe-example}
\end{figure}

\subsubsection{Crooked Pipe}
The crooked pipe problem is different from the line source test in that it examines the solution in both the diffusive and free-streaming regimes.
An isotropic light source is applied at the left boundary of the pipe, allowing radiation to propagate through the optically thin interior while heating the optically thick walls.
The heated walls subsequently re-emit radiation, creating a mechanism that enables radiation to bend around the corners of the pipe.
The crooked pipe data is obtained from \cite{gorodetsky2025thermal} at a discretization of $(N_x, N_y, N_\theta, N_\phi)=(280, 160, 64, 128)$ and the corresponding $J$ plots are depicted in \Cref{fig:thermal:crooked-pipe-example} for 4 different timesteps.

\Cref{fig:thermal:crooked-pipe} illustrates the evolution of the compression ratios identified by \algoname, its variants, and conventional TT-round across simulation timesteps.
Similar to the settings in the line source problem, we compare \algoname to two variants by restricting index reshaping to only angular indices, and replacing heuristic guidance with random choices.
At initial stages, compression ratios are remarkably high---approximately $200\times$---due to extreme spatial homogeneity.
As shown in \cref{fig:thermal:crooked-pipe-example} ($t=10$), meaningful data patterns are confined to a small area near $x=-3.5, y=0$, while the remainder of the domain is empty.
The discovered structure for this phase leverages this sparsity through complex spatial decomposition.
As the simulation evolves, the data becomes increasingly heterogeneous in space, and the compression ratio stabilize around $15\times$ by $t=400,000$.
This trend correlates with the observed increase of non-zero values in the $J$ plots of \cref{fig:thermal:crooked-pipe-example} (top) and a corresponding transition to simpler spatial decompositions in resulting structures (\cref{fig:thermal:crooked-pipe-example} (bottom)).

The ability to decompose the spatial index via reshaping is the primary driver of these efficiency gains.
Without this component (\algoname (w/o space reshape)), compression performance degrades to levels nearly identical to the input TT.
Furthermore, while random index clustering follows a similar temporal trend, \algoname consistently outperforms these baselines at every timestep, validating the effectiveness of the entropy-based heuristics.

\begin{figure}[t]
    \centering
    \subfloat[Distribution of relative CRs for each data tensor across all discovered structures.]{
    \includegraphics[width=.99\linewidth]{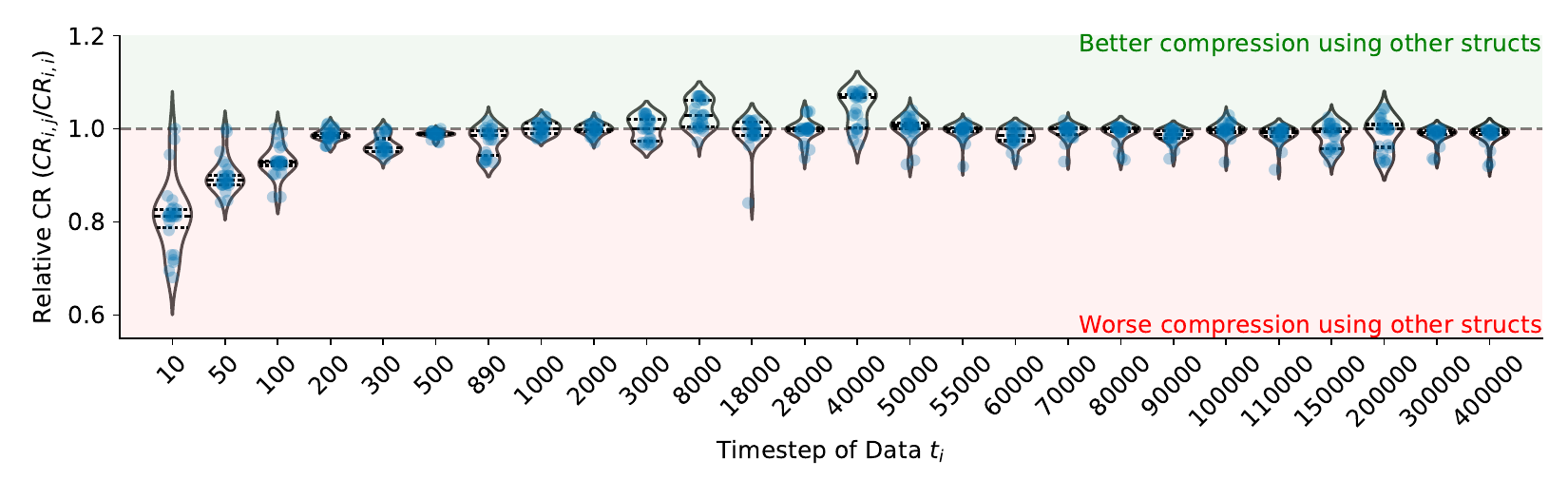}
    \label{fig:crooked-pipe:gen-data}
    }

    \subfloat[Distribution of relative CRs for fixed network structures across multiple solution data.]{
    \includegraphics[width=.99\linewidth]{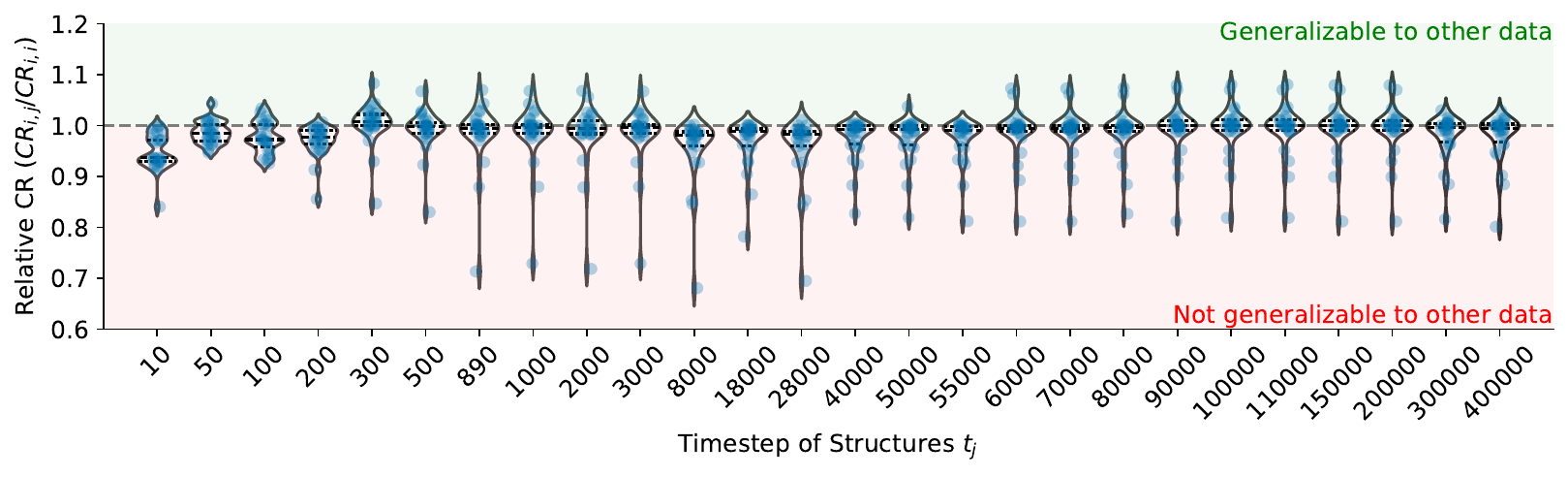}
    \label{fig:crooked-pipe:gen-struct}
    }
    \caption{Distribution of relative compression ratios ($CR_{i, j} / CR_{i, i}$) for the crooked pipe test. Quartiles are marked as dashed lines.
    (a) Fixed Data ($t_i$): Applying structures discovered at all other timesteps $t_j$ to a fixed data tensor at $t_i$. Most ratios lie below 1.0, confirming that local structural searches achieve near-peak optimality for their specific timesteps.
    (b) Fixed Structure ($t_j$): Applying a single discovered structure from $t_j$ to data across all other timesteps $t_i$. The clustering of ratios near 1.0 demonstrates strong structural generalization at early timesteps, while points below 1.0 reveal poor generalization for structures discovered at later timesteps.
        }
    \label{fig:crooked-pipe:gen}
\end{figure}

We further evaluate the robustness of these topologies through relative compression ratios.
\Cref{fig:crooked-pipe:gen-data} confirms that the discovered structures are near-optimal, with most relative CRs remaining below 1.0.
Minor exceptions at $t=8,000$ and $t=40,000$ indicate rare instances where the heuristics converged to sub-optimal configurations.
Similar to the line source problem, variance is highest at initial timesteps where data is most sensitive to structural patterns, stabilizing as the solution complexity increases.
In terms of generalization, \cref{fig:crooked-pipe:gen-struct} reveals that early-stage structures generalize poorly to later timesteps due to the rapid evolution of initial data patterns.
However, as the solution stabilizes, structures discovered at later timesteps exhibit strong generalizability, with quartiles clustered around 1.0.
The observed long tails in these distributions are primarily due to the unique structural sensitivity of the early-stage data.
This shift from high sensitivity to stable generalization aligns with the underlying physics: the initial transient phase requires highly specialized topologies, while the later, more stable phases share a common structural information that persists over time.

We conclude this section by highlighting several properties of the solutions identified by our algorithm in both test problems (bottom rows of \cref{fig:thermal:line-source-example} and \cref{fig:thermal:crooked-pipe-example}). 
First, the discovered ``optimal'' structures consistently exhibit a Tucker decomposition~\cite{kolda2009tensor} for the native indices, with an internal node connecting the spatial and the two angular dimensions. 
The discrepancy between machine-discovered structures and expert-defined structures suggests the advantages of automatic structure search, as expert intuition might fail to precisely capture complex data characteristics. 
Second, the algorithm always recovers hidden dimension separability by reshaping the flattened spatial dimension into its $x$-axis and $y$-axis components, reflecting the inherent separability of the coordinate system in the transport equations.
This is automatically achieved by our proposed heuristics without prior structural knowledge. 
\subsection{Neutron Diffusion}\label{sec:neutron}
\begin{figure}[t]
    \centering
    \subfloat[Example neutron core layout]{
    \includegraphics[width=0.38\linewidth]{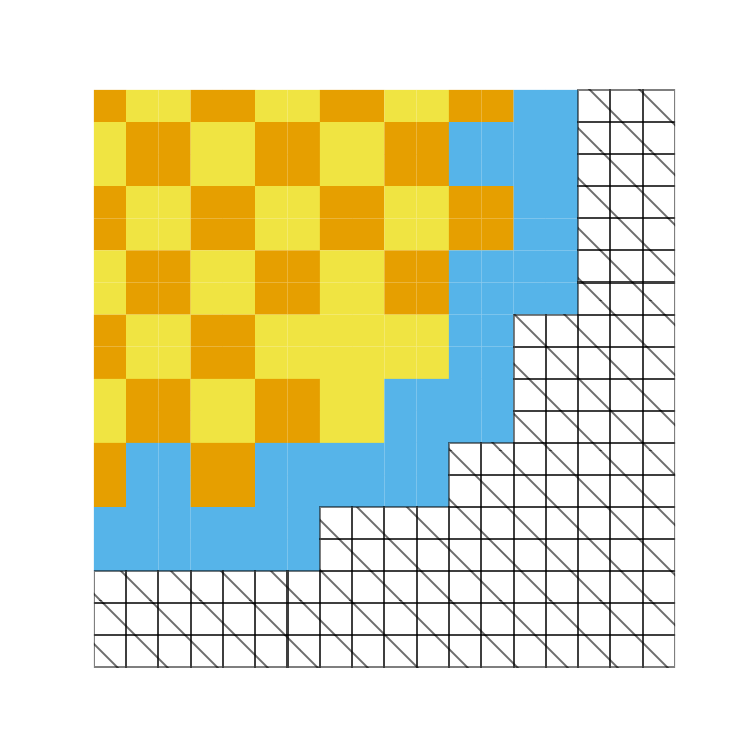}
    \label{fig:neutron:layout}
    }
    ~
    \subfloat[$k_{\text{eff}}$ values for three materials]{
    \includegraphics[width=0.52\linewidth]{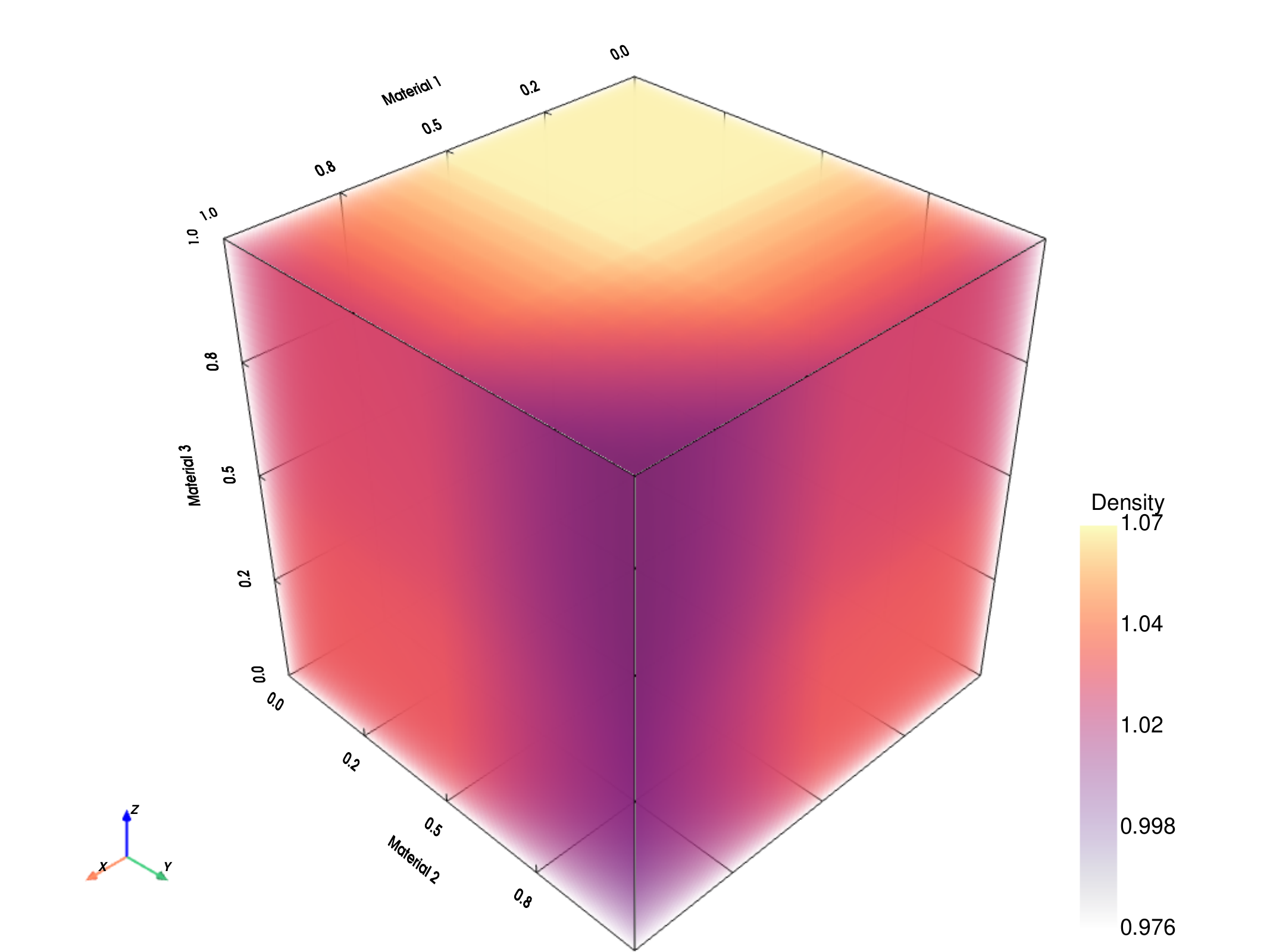}
    \label{fig:neutron:keff}
    }
    \caption{Example core layout (left) and the $k_{\text{eff}}$ values (right) for three materials. Each color stands for a different material.}
    \label{fig:neutron:example}
\end{figure}

In this section, we apply \algoname to a parametric multi-group neutron diffusion problem, which is a mathematical simplification of neutron transport~\cite{allaire2000homogenization,xiao2024operator,wilson2024symmetric}.
This problem is of central importance in nuclear reactor physics~\cite{stamm1983methods, duderstadt1975nuclear}, where predicting the steady-state distribution of neutrons is essential for safety and fuel cycle optimization.
The physics is characterized by localized material couplings and high-dimensional parameter spaces, providing a robust testbed for discovering non-trivial low-rank representations.

In this setup, we simulate a heterogeneous 2-D reactor core partitioned into $d$ discrete material zones ($d \in [3, 8]$).
Each zone $i$ is associated with a parameter $m_i \in [0, 2]$ uniformly discretized into a grid of size $21$, representing localized perturbations in the macroscopic absorption and fission cross-sections.
The governing physics is defined by the multi-group steady-state neutron diffusion eigenvalue problem:
\begin{equation}
-\nabla \cdot (D_g \nabla\phi_g) + \Sigma_{r,g}\phi_g = \sum_{g'=1, g'\not= g}^{G} \Sigma_{s,g' \rightarrow g} \phi_{g'} + \frac{\chi_g}{k_{\text{eff}}} \sum_{g'=1}^{G} \nu \Sigma_{f,g'}\phi_{g'},
\end{equation}
where $\phi_g$ is the neutron flux, $D_g$ is the diffusion coefficient, $\Sigma_{r,g}$, $\Sigma_{s,g' \to g}$, and $\Sigma_{f,g}$ are removal cross-section, scattering cross-section, and fission probability respectively.
Our primary objective is to approximate the effective multiplication factor $k_{\text{eff}}$ across the parametric manifold.
As illustrated in \cref{fig:neutron:example}, different material perturbations significantly alter the system's criticality.
Similar to the processing of analytical functions, we first represent this black-box parametric function as a TT using MaxVol-based cross approximation~\cite{oseledets2010tt}.
We set a baseline relative tolerance of $\error=10^{-4}$.
\algoname then operates on this initial TT to discover more efficient structures without the need for further expensive queries to the physics solver.

Following the experiment design in the previous section, we evaluate the compression performance of the \algoname framework against its two ablation variants, \algoname (w/o Reshape) and \algoname (random), and the standard TT and HT.
For the TT baseline, standard rounding is applied directly to the initial cross approximation result at the target error.
For the HT baseline, the initial TT is transformed into a balanced Hierarchical Tucker format~\cite{handschuh2015numerical}, which is then rounded at the target error.
In this experiment, we set $\maxiters=3$ for best performance, and
we collect search time, reconstruction errors, and compression ratios for all variants.

\begin{figure}[t]
    \centering
    \subfloat[Search Time]{
    \includegraphics[height=0.3\linewidth]{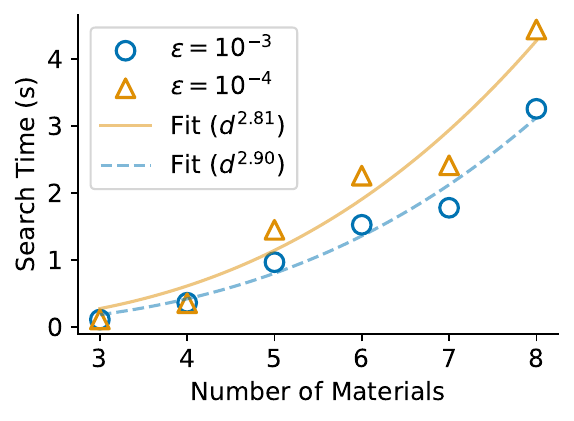}
    \label{fig:neutron:time}
    }
    ~
    \subfloat[Reconstruction Error]{
    \includegraphics[height=0.3\linewidth]{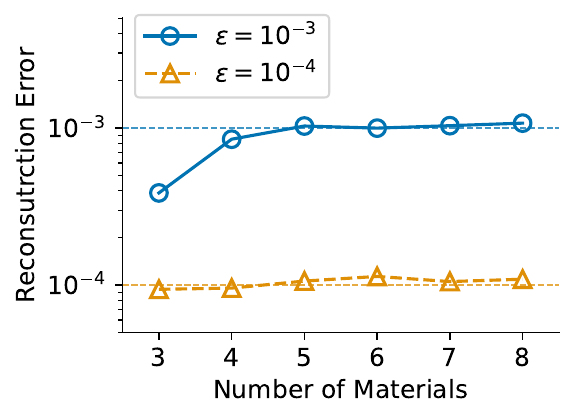}
    \label{fig:neutron:error}
    }

    \subfloat[Compression over TT ($\error = 10^{-3}$)]{
    \includegraphics[width=0.47\linewidth]{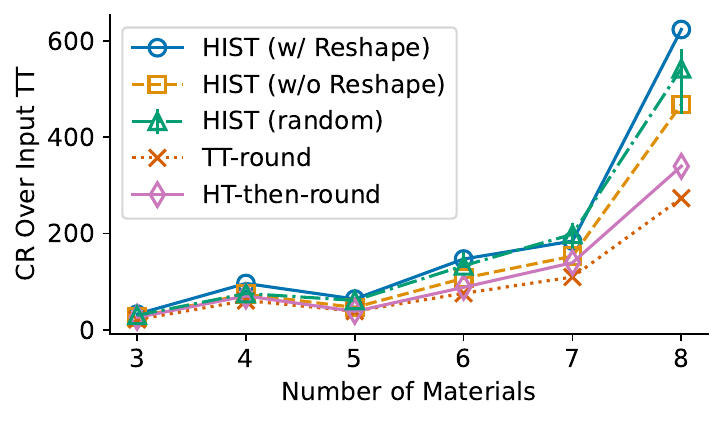}
    \label{fig:neutron:eps-0.001}
    }
    ~
    \subfloat[Compression over TT ($\error = 10^{-4}$)]{
    \includegraphics[width=0.47\linewidth]{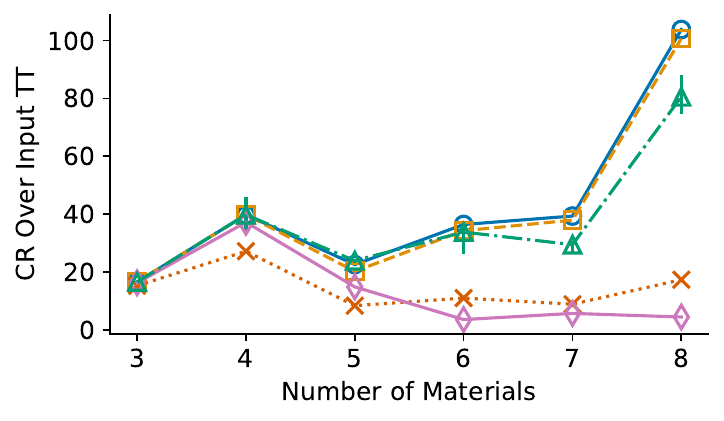}
    \label{fig:neutron:eps-0.0001}
    }
    \caption{(a) Search time scales polynomially with the number of materials.
    (b) Reconstruction error around pre-defined error tolerance. 
    (c, d) Compression ratio comparison between \algoname, \algoname without index reshaping, \algoname with random decisions, and TT and HT, at error bounds $10^{-3}$ and $10^{-4}$, 
    showing that \algoname with index reshaping provides best compression performance, 
    particularly as the number of materials grows.
    }
    \label{fig:neutron:data}
\end{figure}

\begin{figure}[t]
    \centering
    \subfloat[Compression over data ($\error=10^{-3}$)]{
    \includegraphics[width=\linewidth]{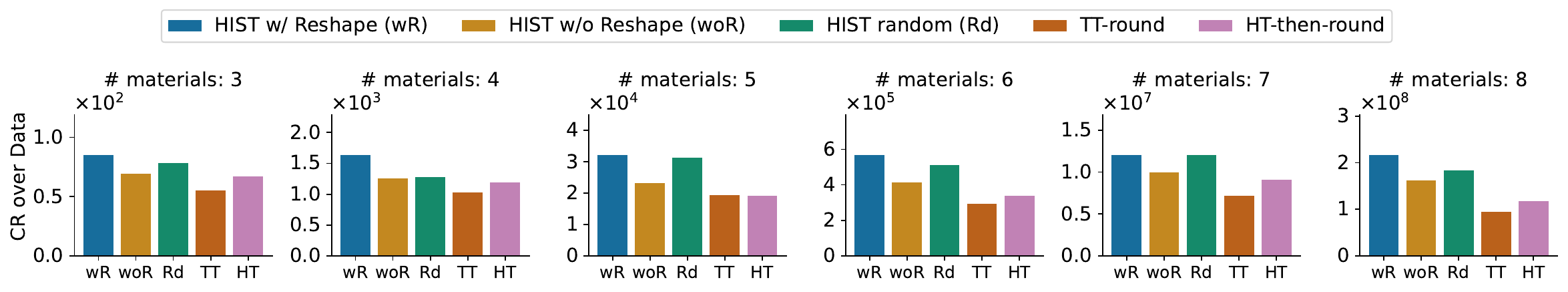}
    \label{fig:neutron:abs-eps-0.001}
    }

    \subfloat[Compression over data ($\error=10^{-4}$)]{
    \includegraphics[width=\linewidth]{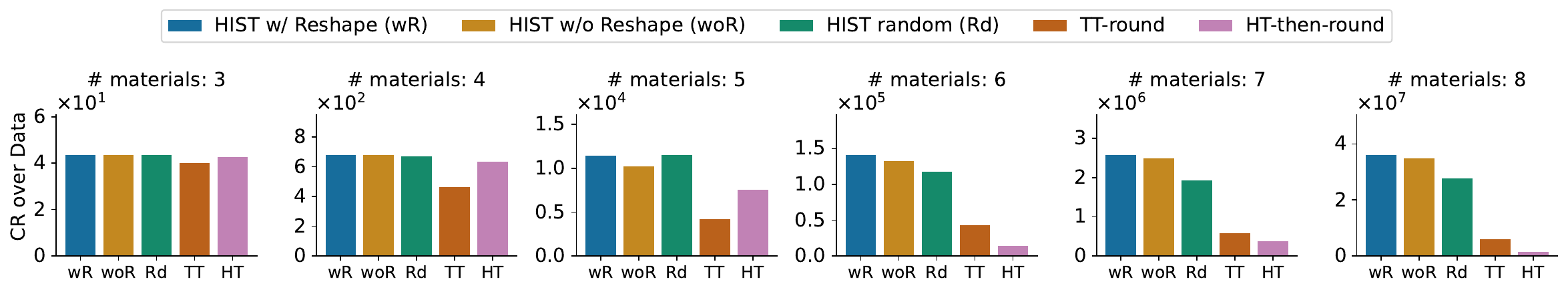}
    \label{fig:neutron:abs-eps-0.0001}
    }
    \caption{CR over data for error bounds $\error=10^{-3}$ (top) and $\error=10^{-4}$ (bottom).
    Compression ratios scale exponentially with the number of materials.
    \algoname with index reshaping consistently delivers the highest compression efficiency at both error tolerances, 
    while the HT and TT methods provide lower ratios, particularly at higher material counts.}
    \label{fig:neutron:abs}
\end{figure}

As summarized in \cref{fig:neutron:data}, the search time for identifying optimal structures (\cref{fig:neutron:time}) demonstrates a polynomial growth of degree around $3$ relative to the number of materials.
While the overall search space for tensor network structures is grows exponentially, \algoname maintains tractable scaling by restricting systematic searches to small, sampled sub-networks over a fixed number of sampling iterations.
This empirical efficiency validates the effectiveness of our stochastic sampling and hierarchical refinement strategies.

\algoname also maintains excellent control over reconstruction accuracy, as shown in \cref{fig:neutron:error}.
Regardless of the material count, the reconstruction error remains consistently close to the prescribed bounds.
Minor instances where the final error slightly exceeds the target tolerance are due to the cumulative nature of the approximation process:
\algoname performs compression at $\varepsilon = 10^{-4}$ using a TT-cross result that itself carries an initial error of $10^{-4}$ relative to the ground-truth black-box function, and therefore the total error reflects the nested tolerances of both stages.
Overall, \algoname consistently operates within the intended bounds, confirming that it respects the error budget without over-compressing at the expense of accuracy.

As illustrated in \cref{fig:neutron:eps-0.001} and \cref{fig:neutron:eps-0.0001}, the structures discovered by \algoname consistently outperform standard fixed formats.
This performance gap widens as the problem complexity increases, emphasizing the advantage of customized topologies over static representations.
The magnitude of this improvement scales with both the number of materials and the stringency of the error tolerance.
At a large tolerance of $\error = 10^{-3}$ (\cref{fig:neutron:eps-0.001}), \algoname achieves a massive compression increase, reaching more than $500\times$ compression over the input TT at $8$ materials, while TT-round remains below $300\times$.
The structural advantage is more evident under a strict error tolerance $\error=10^{-4}$ (\cref{fig:neutron:eps-0.0001}).
While standard TT and HT formats plateau or even degrade, and fail to exceed $20\times$ and $10\times$ compression respectively, \algoname identifies structures that achieve up to $100\times$ compression.
This represents a $5\times$ improvement over the TT-round baseline and a $20\times$ improvement over the HT format at high dimensionality ($d=8$).
These results demonstrate that \algoname effectively breaks the efficiency ceilings inherent to fixed-topology networks.

We also observe the critical role of index reshaping in revealing hidden correlations in the data.
By enabling index reshaping, the algorithm can explore factorized representations of the material parameters, leading to higher compression than the ``w/o Reshape'' variant in most cases of $\error=10^{-3}$ (\cref{fig:neutron:eps-0.001}).
This effect is particularly evident at $d=8$, where reshaping improves the compression ratio over input TT from $540\times$ to $600\times$.
Interestingly, \algoname and \algoname (w/o Reshape) show similar performance at $\error=10^{-4}$ (\cref{fig:neutron:eps-0.0001}), suggesting that index reshaping offers diminishing returns at this high-precision regime.

Furthermore, the comparison with a random baseline (\algoname random) validates the effectiveness of the entropy-based clustering and reshaping heuristics.
While the random search explores non-standard topologies without heuristic guidance, it outperforms fixed TT and HT formats but falls short of the full \algoname framework.
This performance gap is most visible at $d=8$ with $\error=10^{-4}$, where the guided heuristics achieves a $100\times$ compression ratio but the random baseline achieves $80\times$.
This result confirms that the proposed heuristics provide necessary guidance during the search, steering the algorithm toward near-optimal topologies in high-dimensional search space. 

Finally, \cref{fig:neutron:abs} demonstrates the compression ratios relative to the raw data.
As the number of materials grows, the compression ratios scale exponentially, reaching approximately $2 \times 10^8$ and $3.5 \times 10^7$ for $d=8$ at $\error=10^{-3}$ and $\error=10^{-4}$ respectively.
This result suggests that while the complexity of the reactor core increases, the underlying physical correlations remain highly compressible if the correct topology is discovered.
\algoname with index reshaping consistently delivers the highest efficiency, particularly in higher dimensions, demonstrating that a flexible, learned topology is far superior to rigid, predefined structures for complex physical manifolds.
\begin{figure}[t]
    \centering
    \includegraphics[width=0.65\linewidth]{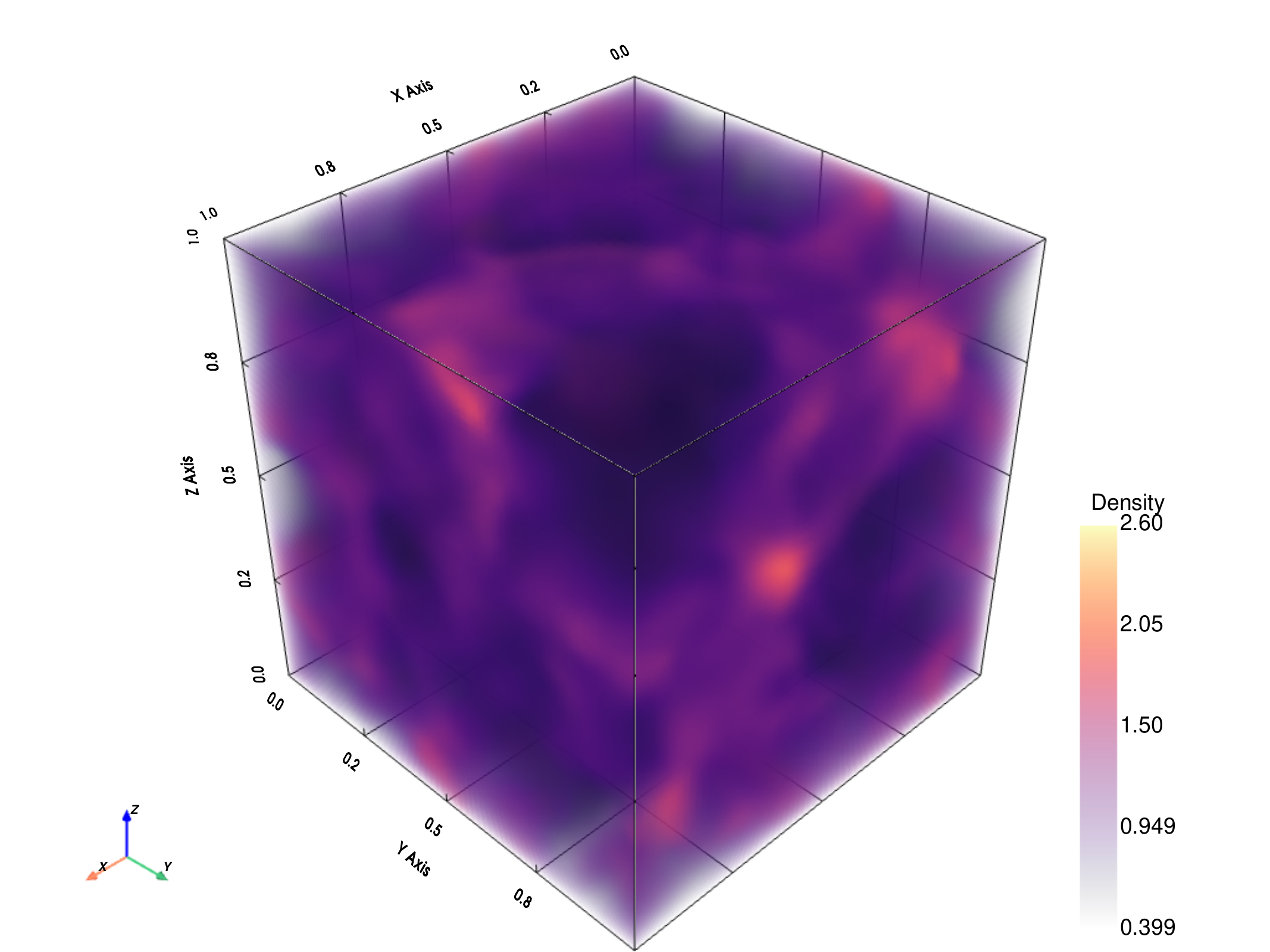}
    \caption{An example solution of 3-D Navier-Stokes problems~\cite{takamoto2022pdebench}}
    \label{fig:cfd:example}
\end{figure}

\subsection{Compressible Navier-Stokes}\label{sec:eval:cfd}
In this section, we assess the ability of \algoname in handling transient, multi-scale physical phenomena by evaluating its performance on high-fidelity simulation data from computational fluid dynamics (CFD)~\cite{anderson1995computational}.
Unlike steady-state problems, CFD involves complex, time-evolving interactions across multiple physical fields.
We consider compressible Navier-Stokes (NS) equations~\cite{rowley2004model,balajewicz2016minimal,zucatti2021data}, which govern the conservation of mass, momentum, and energy:
\begin{subequations}
    \begin{align}
    \partial_t \rho + \nabla \cdot (\rho \boldsymbol{v}) &= 0, \\
    \rho (\partial_t \boldsymbol{v} + \boldsymbol{v}\cdot \nabla\boldsymbol{v}) &= -\nabla p + \eta \Delta v + (\zeta + \eta/3) \nabla (\nabla \cdot \boldsymbol{v}),\\
    \partial_t\left[\epsilon + \frac{\rho v^2}{2}\right] &+ \nabla \cdot \left[ \left( \epsilon + p + \frac{\rho v^2}{2} \right) \mathbf{v} - \mathbf{v}\cdot \sigma' \right] = 0,
\end{align}
\end{subequations}
where $\rho$, $\mathbf{v}$, and $p$ denote the mass density, velocity, and gas pressure, $\epsilon = p / (\Gamma - 1)$ is the internal energy, $\sigma'$ is the viscous stress tensor, and $\eta, \zeta$ are the shear and bulk viscosity. Simulation data come from PDEBench~\cite{takamoto2022pdebench}, with a sample in \cref{fig:cfd:example}.

\begin{figure}[t]
    \centering
    \subfloat[Compression over Data ($\error=0.1$)]{\includegraphics[width=0.45\linewidth]{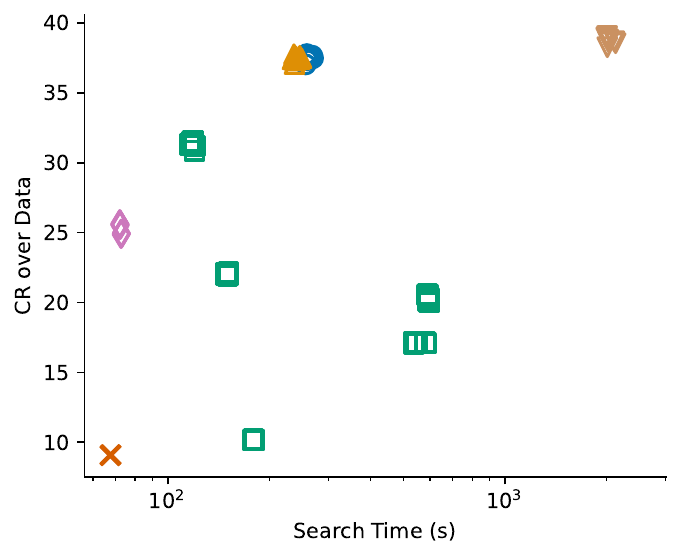}
    \label{fig:cfd:scatter:0.1}
    }
    ~
    \subfloat[Compression over Data ($\error=0.01$)]{\includegraphics[width=0.45\linewidth]{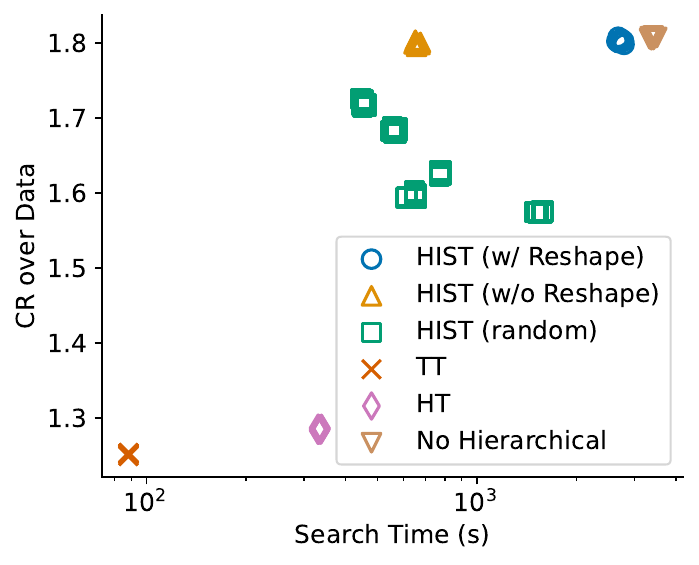}
    \label{fig:cfd:scatter:0.01}
    }
    \caption{Ablation analysis of algorithmic performance, illustrating the trade-off between search time and compression ratio across \algoname, \algoname without index reshaping, and \algoname with random configurations, alongside fixed TT and HT structures at error bounds $\error=0.1$ (left) and $\error=0.01$ (right). 
    \algoname achieves $1.5\times$ to $3\times$ better compression than fixed structures with one order of magnitude higher search time.
    Compared to the non-hierarchical search, \algoname yields similar compression while operating approximately $2\times$ faster at $\error=0.1$.
    }
    \label{fig:cfd:scatter}
\end{figure}

We follow \cite{guo2025tensor}, and randomly sample and stack $10$ data points from the dataset to get a 6th-order tensor, with dimensions $10 \times 5 \times 21 \times 64 \times 64 \times 64$ representing the sample size, physical fields, timesteps, and 3D spatial discretizations, respectively.
A total of $10$ such tensors are collected to form a comprehensive benchmark set.
We compare \algoname against two fixed structures, Tensor Train (TT) and Hierarchical Tucker (HT), alongside three variants: \algoname (w/o reshape), \algoname (random), and No hierarchical.
The first two variants are defined the same as previous experiments where \algoname (w/o reshape) disables index reshaping, and \algoname (random) replaces \mergeindicescall and \topreshapecall with random selection.
\revtwo{The no-hierarchical variant serves as the baseline for non-hierarchical tensor decomposition performance from prior work~\cite{guo2025tensor}, where both hierarchical search and index reshaping are disabled.}

\begin{figure}[t]
    \centering
    \subfloat[Distribution of relative CRs for each data tensor across all discovered structures.]{
    \includegraphics[width=.5\linewidth]{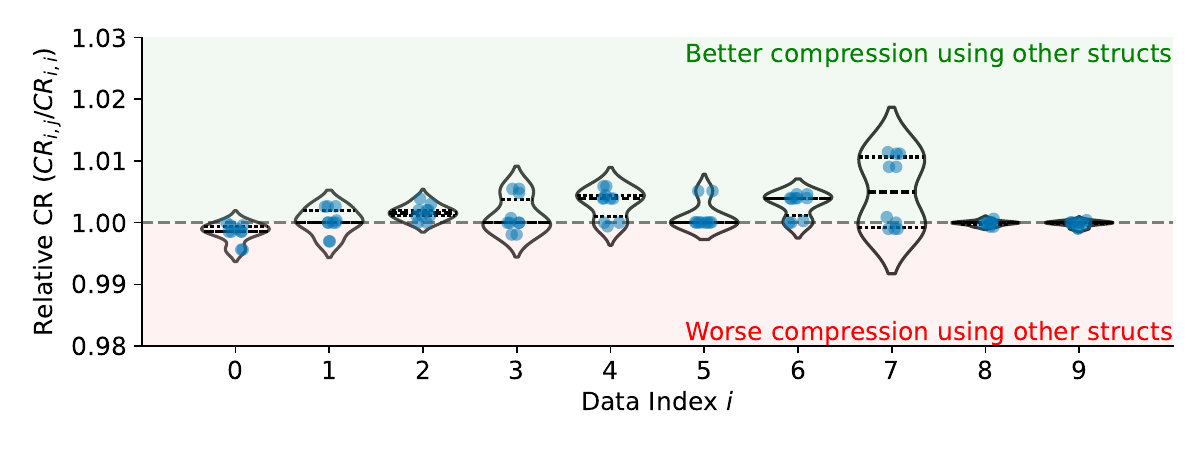}
    \includegraphics[width=.5\linewidth]{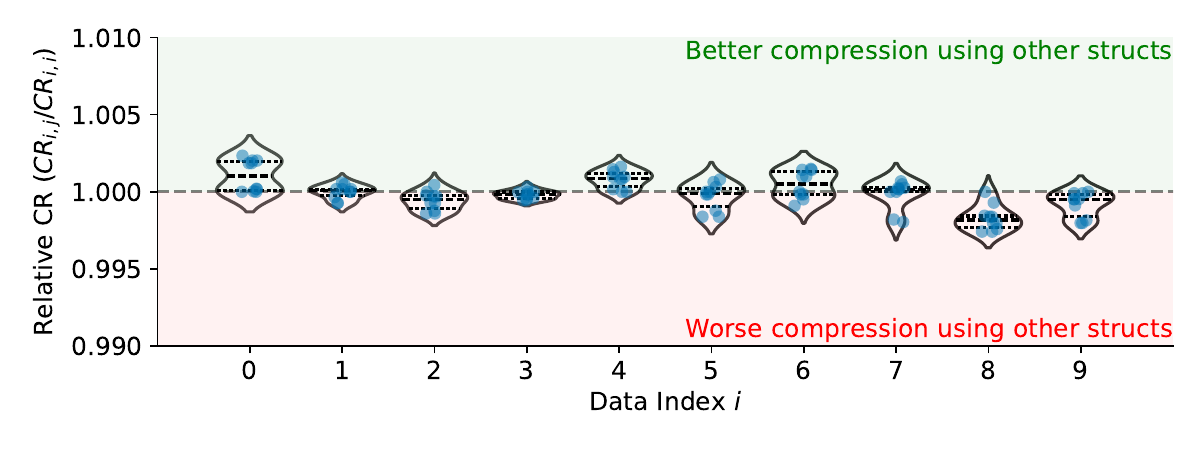}
    \label{fig:cfd:gen-data}
    }

    \subfloat[Performance distribution of each discovered structure when applied to all data.]{
    \includegraphics[width=.5\linewidth]{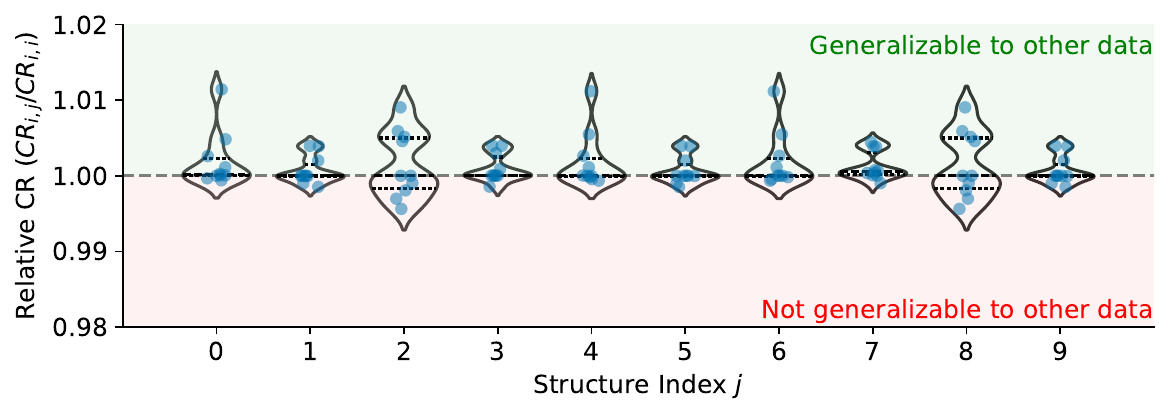}
    \includegraphics[width=.5\linewidth]{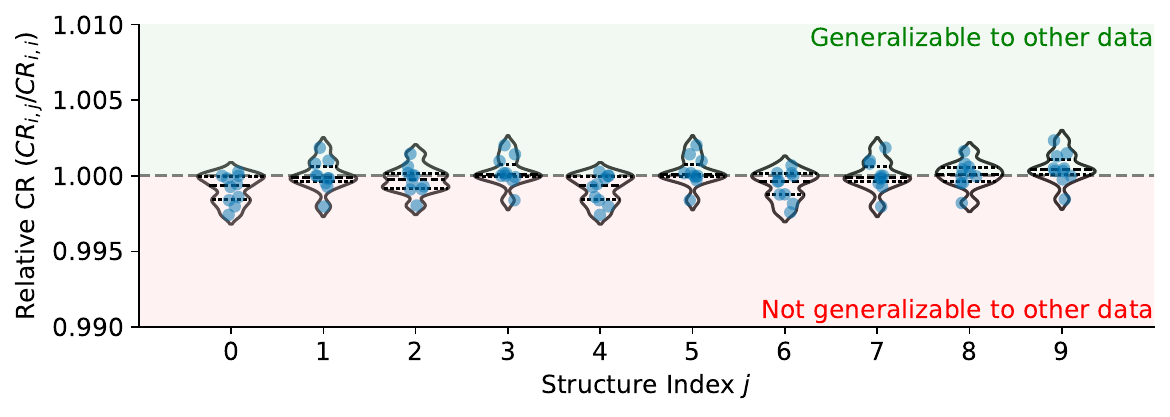}
    \label{fig:cfd:gen-struct}
    }
    \caption{Distribution of relative compression ratios (left: $\error=0.1$, right: $\error=0.01$). The top plot shows that data-specific structures $i$ achieve near-peak performance for their respective tensors (structural optimality). The bottom plot displays the performance of structure $j$ across all data tensors $i$, indicating high cross-data generalizability as ratios remain largely $\geq 1.0$.}
    \label{fig:cfd:gen}
\end{figure}

As visualized in \cref{fig:cfd:scatter}, \algoname identifies superior topologies across both evaluated error tolerances.
Compared to prior work~\cite{guo2025tensor} that does not use hierarchical search, \algoname demonstrates similar compression performance but runs $7.5\times$ faster at $\error = 0.1$ and $1.25\times$ faster at $\error=0.01$.
For a loose error bound ($\error=0.1$), the algorithm identifies structures with compression ratios up to $1.5\times$ higher than HT and $3.5\times$ higher than TT.
Compared to random sampling, our hierarchical search achieves compression improvement ranging from $1.25\times$ to $3\times$.
This confirms that the algorithm effectively navigates the topology space toward near-optimal configurations rather than relying on stochastic luck.
Index reshaping provides modest gains for this specific dataset, but the search overhead is low, accounting for only 10\% of the total search time.
When the error tolerance is tightened to $\error=0.01$, the identified structures for all variants tend to converge, though \algoname consistently maintains higher compression than fixed formats and random structures.

To evaluate the optimality and cross-data generalizability of the discovered structures, we perform the same analysis where the structure optimized for one data tensor is applied to all other tensors within the benchmark set.
The performance metric, relative compression ratio ($CR_{i,j}/CR_{i,i}$), is represented by the ratio between the compression achieved on tensor $i$ using a structure optimized for tensor $j$ ($CR_{i,j}$) and the baseline compression achieved using its own optimized structure ($CR_{i,i}$).
As illustrated in \cref{fig:cfd:gen}, the majority of these ratios cluster tightly around $1.0$, with a narrow variance typically ranging between $0.99$ and $1.02$ across both error tolerances.
These results confirm that the structures discovered for the NS data are near-optimal (\cref{fig:cfd:gen-data}).
Furthermore, the high performance of transferred structures (relative CR $\geq 1.0$) demonstrates significant generalizability across multiple benchmark data (\cref{fig:cfd:gen-struct}).
This suggests that the structural search can be treated as a one-time offline cost; once a structure is optimized for a representative tensor, it can be directly applied to similar datasets with negligible loss in compression performance.

\section{Conclusion}\label{sec:conclusion}

We introduced \tnround, a structural optimization problem for \revone{tree tensor networks} within a prescribed error tolerance.
\revone{A hierarchical search algorithm for tree networks, \algoname, is designed to automatically discover optimal tree structures and index reshapes for high-dimensional tensor networks.}
By combining global stochastic sampling with local hierarchical refinement guided by entropy-based heuristics, the algorithm effectively navigates the combinatorial complexity of the structural search space.
Unlike conventional tensor network rounding, which is restricted to a fixed skeleton, \algoname dynamically reconfigures the network structure to uncover latent data correlations and enhance compression.
Our numerical evaluations across diverse benchmarks---including thermal radiation transport, parametric neutron diffusion, and computational fluid dynamics (CFD)---demonstrate the broad applicability of \algoname.
\algoname consistently achieves compression ratios $\mincr\times$ to $\maxcr\times$ higher than standard fixed formats like TT and HT. 
In the line source problem, \algoname-discovered structures reach a maximum compression ratio of over $\peakcr\times$ against the input TT, which improves standard TT-rounding by around $500\times$.
These gains are driven by two primary factors. First, the integration of index reshaping proves critical in structure search, as it uncovers hidden correlations in original indices.
Second, the flexibility of the structural search allows the network to adapt to the specific features of each test case.
Beyond static compression, we find that these discovered structures exhibit strong generalization across multiple simulation timesteps or data batches.

Despite the advantages, we note several limitations of the current algorithm, and leave them for future work.
One primary constraint involves the current decoupling of index clustering and reshaping, which prevents the search space with reshaping from strictly subsuming the space without it.
Future work will consider better orchestration of these search decisions to form a composable and nested search space.
Furthermore, while our entropy-based heuristics provide effective guidance, the occasional identification of sub-optimal structures suggests that these metrics could be further tuned to improve performance stability across diverse datasets.

The framework's interaction with black-box tensors also presents an opportunity for tighter integration.
Currently, the reliance on an intermediate structure generated via cross approximation introduces a dependency on the quality of that initial format.
We aim to move toward a structure-aware sampling approach, where the topological search directly drives the selection of query entries in the black-box tensor, thereby eliminating any sub-optimality inherited from pre-defined formats.

Finally, the scope of \algoname is limited to acyclic structures (\ie tree tensor networks).
The fundamental mechanism of our sub-network refinement relies on the correspondence between tensor network edges and index bi-partitions, but this relationship does not trivially generalize to cyclic structures, as edges in loopy graphs are not necessarily associated with a unique cut~\cite{guo2025tensor}.
Extending these structure search techniques to support cyclic structures remains a significant and compelling challenge for the next phase of this work.

\section*{Acknowledgments}
This work was supported by Schmidt Sciences, LLC., the AFOSR Computational Mathematics Program under the Award \#FA9550-24-1-0246,
and the National Science Foundation under Grant Numbers
CCF-2236233,
CCF-2210832,
CCF-2318937.

\bibliographystyle{elsarticle-num} 
\bibliography{main.bib}

@article{gorodetsky2025thermal,
  title={Thermal Radiation Transport with Tensor Trains},
  author={Gorodetsky, Alex A and Mullen, Patrick D and Deshpande, Aditya and Dolence, Joshua C and Meyer, Chad D and Miller, Jonah M and Roberts, Luke F},
  journal={arXiv preprint arXiv:2503.18056},
  year={2025}
}

@article{takamoto2022pdebench,
  title={Pdebench: An extensive benchmark for scientific machine learning},
  author={Takamoto, Makoto and Praditia, Timothy and Leiteritz, Raphael and MacKinlay, Daniel and Alesiani, Francesco and Pfl{\"u}ger, Dirk and Niepert, Mathias},
  journal={Advances in Neural Information Processing Systems},
  volume={35},
  pages={1596--1611},
  year={2022}
}

@article{ryzhakov2024black,
  title={Black-box approximation and optimization with hierarchical tucker decomposition},
  author={Ryzhakov, Gleb and Chertkov, Andrei and Basharin, Artem and Oseledets, Ivan},
  journal={arXiv preprint arXiv:2402.02890},
  year={2024}
}

@article{ballani2013black,
  title={Black box approximation of tensors in hierarchical Tucker format},
  author={Ballani, Jonas and Grasedyck, Lars and Kluge, Melanie},
  journal={Linear algebra and its applications},
  volume={438},
  number={2},
  pages={639--657},
  year={2013},
  publisher={Elsevier}
}

@article{oseledets2010tt,
  title={TT-cross approximation for multidimensional arrays},
  author={Oseledets, Ivan and Tyrtyshnikov, Eugene},
  journal={Linear Algebra and its Applications},
  volume={432},
  number={1},
  pages={70--88},
  year={2010},
  publisher={Elsevier}
}

@article{oseledets2011tensor,
  title={Tensor-train decomposition},
  author={Oseledets, Ivan V},
  journal={SIAM Journal on Scientific Computing},
  volume={33},
  number={5},
  pages={2295--2317},
  year={2011},
  publisher={SIAM}
}

@techreport{bassett2019comparison,
  title={Comparison of Meshless and High-Order Polynomial Functions for Neutron Transport with Streamline-Upwind Petrov-Galerkin Stabilization},
  author={Bassett, BR and Kiedrowski, BC},
  year={2019},
  institution={Lawrence Livermore National Lab.(LLNL), Livermore, CA (United States)}
}

@article{oseledets2009breaking,
  title={Breaking the curse of dimensionality, or how to use SVD in many dimensions},
  author={Oseledets, Ivan V and Tyrtyshnikov, Eugene E},
  journal={SIAM Journal on Scientific Computing},
  volume={31},
  number={5},
  pages={3744--3759},
  year={2009},
  publisher={SIAM}
}

@article{altman2018curse,
  title={The curse (s) of dimensionality},
  author={Altman, Naomi and Krzywinski, Martin},
  journal={Nat Methods},
  volume={15},
  number={6},
  pages={399--400},
  year={2018}
}

@article{bachmayr2023low,
  title={Low-rank tensor methods for partial differential equations},
  author={Bachmayr, Markus},
  journal={Acta Numerica},
  volume={32},
  pages={1--121},
  year={2023},
  publisher={Cambridge University Press}
}

@article{dolgov2012tt,
  title={TT-GMRES: on solution to a linear system in the structured tensor format},
  author={Dolgov, Sergey V},
  journal={arXiv preprint arXiv:1206.5512},
  year={2012}
}

@article{rodgers2024tensor,
  title={Tensor approximation of functional differential equations},
  author={Rodgers, Abram and Venturi, Daniele},
  journal={Physical Review E},
  volume={110},
  number={1},
  pages={015310},
  year={2024},
  publisher={APS}
}

@article{truong2024tensor,
  title={Tensor networks for solving the time-independent Boltzmann neutron transport equation},
  author={Truong, Duc P and Ortega, Mario I and Boureima, Ismael and Manzini, Gianmarco and Rasmussen, Kim {\O} and Alexandrov, Boian S},
  journal={Journal of Computational Physics},
  volume={507},
  pages={112943},
  year={2024},
  publisher={Elsevier}
}

@article{Tucker_1966, 
title={Some Mathematical Notes on Three-Mode Factor Analysis}, 
volume={31}, 
DOI={10.1007/BF02289464}, 
number={3}, 
journal={Psychometrika}, 
author={Tucker, Ledyard R}, 
year={1966}, 
pages={279–311}
}

@article{ht,
author = {Grasedyck, Lars},
title = {Hierarchical Singular Value Decomposition of Tensors},
journal = {SIAM Journal on Matrix Analysis and Applications},
volume = {31},
number = {4},
pages = {2029-2054},
year = {2010},
doi = {10.1137/090764189}
}

@article{guo2025tensor,
  title={Tensor Network Structure Search with Program Synthesis},
  author={Guo, Zheng and Deshpande, Aditya and Kiedrowski, Brian and Wang, Xinyu and Gorodetsky, Alex},
  journal={arXiv preprint arXiv:2502.02711},
  year={2025}
}

@article{hashemizadeh2020adaptive,
  title={Adaptive learning of tensor network structures},
  author={Hashemizadeh, Meraj and Liu, Michelle and Miller, Jacob and Rabusseau, Guillaume},
  journal={arXiv preprint arXiv:2008.05437},
  year={2020}
}

@inproceedings{li2020evolutionary,
  title={Evolutionary topology search for tensor network decomposition},
  author={Li, Chao and Sun, Zhun},
  booktitle={International conference on machine learning},
  pages={5947--5957},
  year={2020},
  organization={PMLR}
}

@inproceedings{zheng2024svdinstn,
  title={SVDinsTN: A Tensor Network Paradigm for Efficient Structure Search from Regularized Modeling Perspective},
  author={Zheng, Yu-Bang and Zhao, Xi-Le and Zeng, Junhua and Li, Chao and Zhao, Qibin and Li, Heng-Chao and Huang, Ting-Zhu},
  booktitle={Proceedings of the IEEE/CVF Conference on Computer Vision and Pattern Recognition},
  pages={26254--26263},
  year={2024}
}

@inproceedings{li2023alternating,
  title={Alternating local enumeration (tnale): Solving tensor network structure search with fewer evaluations},
  author={Li, Chao and Zeng, Junhua and Li, Chunmei and Caiafa, Cesar F and Zhao, Qibin},
  booktitle={International conference on machine learning},
  pages={20384--20411},
  year={2023},
  organization={PMLR}
}

@inproceedings{li2022permutation,
  title={Permutation search of tensor network structures via local sampling},
  author={Li, Chao and Zeng, Junhua and Tao, Zerui and Zhao, Qibin},
  booktitle={International conference on machine learning},
  pages={13106--13124},
  year={2022},
  organization={PMLR}
}

@inproceedings{ghadiri2023approximately,
  title={Approximately optimal core shapes for tensor decompositions},
  author={Ghadiri, Mehrdad and Fahrbach, Matthew and Fu, Gang and Mirrokni, Vahab},
  booktitle={International Conference on Machine Learning},
  pages={11237--11254},
  year={2023},
  organization={PMLR}
}

@phdthesis{handschuh2015numerical,
  title={Numerical methods in tensor networks},
  author={Handschuh, Stefan},
  year={2015},
  school={Dissertation, Leipzig, Universit{\"a}t Leipzig, 2015}
}

@article{wang2025renormalization,
  title={Renormalization Group Guided Tensor Network Structure Search},
  author={Wang, Maolin and Yu, Bowen and Zhang, Sheng and Mi, Linjie and Wang, Wanyu and Wang, Yiqi and Jia, Pengyue and Wei, Xuetao and Xu, Zenglin and Guo, Ruocheng and others},
  journal={arXiv preprint arXiv:2512.24663},
  year={2025}
}

@article{ke2023tree,
  title={Tree tensor network state approach for solving hierarchical equations of motion},
  author={Ke, Yaling},
  journal={The Journal of Chemical Physics},
  volume={158},
  number={21},
  year={2023},
  publisher={AIP Publishing}
}

@article{sands2025high,
  title={High-order adaptive rank integrators for multiscale linear kinetic transport equations in the hierarchical tucker format},
  author={Sands, William A and Guo, Wei and Qiu, Jing-Mei and Xiong, Tao},
  journal={SIAM Journal on Scientific Computing},
  volume={47},
  number={6},
  pages={A3383--A3412},
  year={2025},
  publisher={SIAM}
}

@article{ghahremani2024deim,
  title={A DEIM Tucker tensor cross algorithm and its application to dynamical low-rank approximation},
  author={Ghahremani, Behzad and Babaee, Hessam},
  journal={Computer Methods in Applied Mechanics and Engineering},
  volume={423},
  pages={116879},
  year={2024},
  publisher={Elsevier}
}

@article{ghahremani2024cross,
  title={Cross interpolation for solving high-dimensional dynamical systems on low-rank Tucker and tensor train manifolds},
  author={Ghahremani, Behzad and Babaee, Hessam},
  journal={Computer Methods in Applied Mechanics and Engineering},
  volume={432},
  pages={117385},
  year={2024},
  publisher={Elsevier}
}

@article{hikihara2023automatic,
  title={Automatic structural optimization of tree tensor networks},
  author={Hikihara, Toshiya and Ueda, Hiroshi and Okunishi, Kouichi and Harada, Kenji and Nishino, Tomotoshi},
  journal={Physical Review Research},
  volume={5},
  number={1},
  pages={013031},
  year={2023},
  publisher={APS}
}

@article{hillar2013most,
  title={Most tensor problems are NP-hard},
  author={Hillar, Christopher J and Lim, Lek-Heng},
  journal={Journal of the ACM (JACM)},
  volume={60},
  number={6},
  pages={1--39},
  year={2013},
  publisher={ACM New York, NY, USA}
}

@inproceedings{linderman2014discovering,
  title={Discovering latent network structure in point process data},
  author={Linderman, Scott and Adams, Ryan},
  booktitle={International conference on machine learning},
  pages={1413--1421},
  year={2014},
  organization={PMLR}
}

@article{borm2003introduction,
  title={Introduction to hierarchical matrices with applications},
  author={B{\"o}rm, Steffen and Grasedyck, Lars and Hackbusch, Wolfgang},
  journal={Engineering analysis with boundary elements},
  volume={27},
  number={5},
  pages={405--422},
  year={2003},
  publisher={Elsevier}
}

@article{chen2024learning,
  title={Learning discrete latent variable structures with tensor rank conditions},
  author={Chen, Zhengming and Cai, Ruichu and Xie, Feng and Qiao, Jie and Wu, Anpeng and Li, Zijian and Hao, Zhifeng and Zhang, Kun},
  journal={Advances in Neural Information Processing Systems},
  volume={37},
  pages={17398--17427},
  year={2024}
}

@article{anandkumar2014tensor,
  title={Tensor decompositions for learning latent variable models.},
  author={Anandkumar, Animashree and Ge, Rong and Hsu, Daniel J and Kakade, Sham M and Telgarsky, Matus and others},
  journal={J. Mach. Learn. Res.},
  volume={15},
  number={1},
  pages={2773--2832},
  year={2014}
}

@inproceedings{roy2007effective,
  title={The effective rank: A measure of effective dimensionality},
  author={Roy, Olivier and Vetterli, Martin},
  booktitle={2007 15th European signal processing conference},
  pages={606--610},
  year={2007},
  organization={IEEE}
}

@article{falco2021tree,
  title={Tree-based tensor formats},
  author={Falc{\'o}, Antonio and Hackbusch, Wolfgang and Nouy, Anthony},
  journal={SeMA Journal},
  volume={78},
  number={2},
  pages={159--173},
  year={2021},
  publisher={Springer}
}

@article{kolda2009tensor,
  title={Tensor decompositions and applications},
  author={Kolda, Tamara G and Bader, Brett W},
  journal={SIAM review},
  volume={51},
  number={3},
  pages={455--500},
  year={2009},
  publisher={SIAM}
}

@inproceedings{zawawi2018review,
  title={A review: Fundamentals of computational fluid dynamics (CFD)},
  author={Zawawi, Mohd Hafiz and Saleha, A and Salwa, A and Hassan, NH and Zahari, Nazirul Mubin and Ramli, Mohd Zakwan and Muda, Zakaria Che},
  booktitle={AIP conference proceedings},
  volume={2030},
  number={1},
  pages={020252},
  year={2018},
  organization={AIP Publishing LLC}
}

@article{cox2022monte,
  title={Monte Carlo methods for the neutron transport equation},
  author={Cox, Alexander MG and Harris, Simon C and Kyprianou, Andreas E and Wang, Minmin},
  journal={SIAM/ASA Journal on Uncertainty Quantification},
  volume={10},
  number={2},
  pages={775--825},
  year={2022},
  publisher={SIAM}
}

@article{white2005density,
  title={Density matrix renormalization group algorithms with a single center site},
  author={White, Steven R},
  journal={Physical Review B—Condensed Matter and Materials Physics},
  volume={72},
  number={18},
  pages={180403},
  year={2005},
  publisher={APS}
}

@article{white1993density,
  title={Density-matrix algorithms for quantum renormalization groups},
  author={White, Steven R},
  journal={Physical review b},
  volume={48},
  number={14},
  pages={10345},
  year={1993},
  publisher={APS}
}

@article{evenbly2022practical,
  title={A practical guide to the numerical implementation of tensor networks i: Contractions, decompositions, and gauge freedom},
  author={Evenbly, Glen},
  journal={Frontiers in Applied Mathematics and Statistics},
  volume={8},
  pages={806549},
  year={2022},
  publisher={Frontiers Media SA}
}

@article{evenbly2018gauge,
  title={Gauge fixing, canonical forms, and optimal truncations in tensor networks with closed loops},
  author={Evenbly, Glen},
  journal={Physical Review B},
  volume={98},
  number={8},
  pages={085155},
  year={2018},
  publisher={APS}
}

@article{dektor2023tensor,
  title={Tensor rank reduction via coordinate flows},
  author={Dektor, Alec and Venturi, Daniele},
  journal={Journal of Computational Physics},
  volume={491},
  pages={112378},
  year={2023},
  publisher={Elsevier}
}

@article{rodgers2022adaptive,
  title={Adaptive integration of nonlinear evolution equations on tensor manifolds},
  author={Rodgers, Abram and Dektor, Alec and Venturi, Daniele},
  journal={Journal of Scientific Computing},
  volume={92},
  number={2},
  pages={39},
  year={2022},
  publisher={Springer}
}

@article{dektor2025coordinate,
  title={Coordinate-adaptive integration of PDEs on tensor manifolds},
  author={Dektor, Alec and Venturi, Daniele},
  journal={Communications on Applied Mathematics and Computation},
  volume={7},
  number={4},
  pages={1562--1579},
  year={2025},
  publisher={Springer}
}

@article{watanabe2025ttnopt,
  title={TTNOpt: Tree tensor network package for high-rank tensor compression},
  author={Watanabe, Ryo and Manabe, Hidetaka and Hikihara, Toshiya and Ueda, Hiroshi},
  journal={arXiv preprint arXiv:2505.05908},
  year={2025}
}

@article{khoromskij2011d,
  title={O (d log N)-quantics approximation of N-d tensors in high-dimensional numerical modeling},
  author={Khoromskij, Boris N},
  journal={Constructive Approximation},
  volume={34},
  number={2},
  pages={257--280},
  year={2011},
  publisher={Springer}
}

@article{stamm1983methods,
  title={Methods of steady-state reactor physics in nuclear design},
  author={Stamm'Ler, Rudi JJ and Abbate, M{\'a}ximo Julio},
  journal={(No Title)},
  year={1983}
}

@article{duderstadt1975nuclear,
  title={Nuclear reactor analysis},
  author={Duderstadt, James J and Hamilton, Louis J},
  year={1975},
  publisher={John Wiley and Sons, Inc., New York}
}

@article{allaire2000homogenization,
  title={Homogenization of a spectral problem in neutronic multigroup diffusion},
  author={Allaire, Gr{\'e}goire and Capdeboscq, Yves},
  journal={Computer methods in applied mechanics and engineering},
  volume={187},
  number={1-2},
  pages={91--117},
  year={2000},
  publisher={Elsevier}
}

@article{xiao2024operator,
  title={Operator inference driven data assimilation for high fidelity neutron transport},
  author={Xiao, Wei and Liu, Xiaojing and Zu, Jianhua and Chai, Xiang and He, Hui and Zhang, Tengfei},
  journal={Computer Methods in Applied Mechanics and Engineering},
  volume={430},
  pages={117214},
  year={2024},
  publisher={Elsevier}
}

@article{wilson2024symmetric,
  title={A symmetric interior-penalty discontinuous Galerkin isogeometric analysis spatial discretization of the self-adjoint angular flux form of the neutron transport equation},
  author={Wilson, SG and Eaton, MD and K{\'o}ph{\'a}zi, J},
  journal={Computer Methods in Applied Mechanics and Engineering},
  volume={432},
  pages={117414},
  year={2024},
  publisher={Elsevier}
}

@book{anderson1995computational,
  title={Computational fluid dynamics},
  author={Anderson, John David and Wendt, John and others},
  volume={206},
  year={1995},
  publisher={Springer}
}

@article{rowley2004model,
  title={Model reduction for compressible flows using POD and Galerkin projection},
  author={Rowley, Clarence W and Colonius, Tim and Murray, Richard M},
  journal={Physica D: Nonlinear Phenomena},
  volume={189},
  number={1-2},
  pages={115--129},
  year={2004},
  publisher={Elsevier}
}

@article{balajewicz2016minimal,
  title={Minimal subspace rotation on the Stiefel manifold for stabilization and enhancement of projection-based reduced order models for the compressible Navier--Stokes equations},
  author={Balajewicz, Maciej and Tezaur, Irina and Dowell, Earl},
  journal={Journal of Computational Physics},
  volume={321},
  pages={224--241},
  year={2016},
  publisher={Elsevier}
}

@article{zucatti2021data,
  title={Data-driven closure of projection-based reduced order models for unsteady compressible flows},
  author={Zucatti, Victor and Wolf, William},
  journal={Computer Methods in Applied Mechanics and Engineering},
  volume={386},
  pages={114120},
  year={2021},
  publisher={Elsevier}
}

@article{einkemmer2025review,
  title={A review of low-rank methods for time-dependent kinetic simulations},
  author={Einkemmer, Lukas and Kormann, Katharina and Kusch, Jonas and McClarren, Ryan G and Qiu, Jing-Mei},
  journal={Journal of Computational Physics},
  volume={538},
  pages={114191},
  year={2025},
  publisher={Elsevier}
}

@article{kurzer2024radiation,
  title={Radiation and heat transport in divergent shock--bubble interactions},
  author={Kurzer-Ogul, Kelin and Haines, Brian M and Montgomery, David S and Pandolfi, Silvia and Sauppe, Joshua P and Leong, Andrew FT and Hodge, Daniel and Kozlowski, Pawel M and Marchesini, Stefano and Cunningham, Eric and others},
  journal={Physics of plasmas},
  volume={31},
  number={3},
  year={2024},
  publisher={AIP Publishing}
}

@article{bhattacharyya2023finite,
  title={A finite element method for angular discretization of the radiation transport equation on spherical geodesic grids},
  author={Bhattacharyya, Maitraya K and Radice, David},
  journal={Journal of Computational Physics},
  volume={491},
  pages={112365},
  year={2023},
  publisher={Elsevier}
}

@article{southworth2024one,
  title={One-sweep moment-based semi-implicit-explicit integration for gray thermal radiation transport},
  author={Southworth, Ben S and Olivier, Samuel and Park, HyeongKae and Buvoli, Tommaso},
  journal={Journal of Computational Physics},
  volume={517},
  pages={113349},
  year={2024},
  publisher={Elsevier}
}

@article{wang2015parallel,
  title={Parallel algorithm and its convergence of spatial domain decomposition of discrete ordinates method for solving radiation heat transfer problem},
  author={Wang, Zhenhua and He, Zhihong and Mu, Lei and Dong, Shikui},
  journal={Chinese Journal of Aeronautics},
  volume={28},
  number={1},
  pages={77--85},
  year={2015},
  publisher={Elsevier}
}

@article{brunner2026domain,
  title={Domain decomposition dynamical low-rank for multi-dimensional radiative transfer equations},
  author={Brunner, Stefan and Einkemmer, Lukas and Haut, Terry},
  journal={arXiv preprint arXiv:2602.14854},
  year={2026}
}

@inproceedings{velarde2005radiation,
  title={Radiation Transport in AMR},
  author={Velarde, P and Ogando, F},
  booktitle={Adaptive Mesh Refinement-Theory and Applications: Proceedings of the Chicago Workshop on Adaptive Mesh Refinement Methods, Sept. 3--5, 2003},
  pages={271--280},
  year={2005},
  organization={Springer}
}

@article{jessee1998adaptive,
  title={An adaptive mesh refinement algorithm for the radiative transport equation},
  author={Jessee, J Patrick and Fiveland, Woodrow A and Howell, Louis H and Colella, Phillip and Pember, Richard B},
  journal={Journal of computational Physics},
  volume={139},
  number={2},
  pages={380--398},
  year={1998},
  publisher={Elsevier}
}

@article{nakatani2013efficient,
  title={Efficient tree tensor network states (TTNS) for quantum chemistry: Generalizations of the density matrix renormalization group algorithm},
  author={Nakatani, Naoki and Chan, Garnet Kin},
  journal={The Journal of chemical physics},
  volume={138},
  number={13},
  year={2013},
  publisher={AIP Publishing}
}

@article{murg2015tree,
  title={Tree tensor network state with variable tensor order: An efficient multireference method for strongly correlated systems},
  author={Murg, Valentin and Verstraete, Frank and Schneider, Reinhold and Nagy, Peter R and Legeza, O},
  journal={Journal of Chemical Theory and Computation},
  volume={11},
  number={3},
  pages={1027--1036},
  year={2015},
  publisher={ACS Publications}
}

@article{gunst2019three,
  title={Three-legged tree tensor networks with SU (2) and molecular point group symmetry},
  author={Gunst, Klaas and Verstraete, Frank and Van Neck, Dimitri},
  journal={Journal of chemical theory and computation},
  volume={15},
  number={5},
  pages={2996--3007},
  year={2019},
  publisher={ACS Publications}
}

@article{ferrari2022adaptive,
  title={Adaptive-weighted tree tensor networks for disordered quantum many-body systems},
  author={Ferrari, Giovanni and Magnifico, Giuseppe and Montangero, Simone},
  journal={Physical Review B},
  volume={105},
  number={21},
  pages={214201},
  year={2022},
  publisher={APS}
}

@article{seki2020tensor,
  title={Tensor-network strong-disorder renormalization groups for random quantum spin systems in two dimensions},
  author={Seki, Kouichi and Hikihara, Toshiya and Okunishi, Kouichi},
  journal={Physical Review B},
  volume={102},
  number={14},
  pages={144439},
  year={2020},
  publisher={APS}
}

@article{pan2022simulation,
  title={Simulation of quantum circuits using the big-batch tensor network method},
  author={Pan, Feng and Zhang, Pan},
  journal={Physical Review Letters},
  volume={128},
  number={3},
  pages={030501},
  year={2022},
  publisher={APS}
}

@article{kalachev2021multi,
  title={Multi-tensor contraction for XEB verification of quantum circuits},
  author={Kalachev, Gleb and Panteleev, Pavel and Yung, Man-Hong},
  journal={arXiv preprint arXiv:2108.05665},
  year={2021}
}

@article{huang2020classical,
  title={Classical simulation of quantum supremacy circuits},
  author={Huang, Cupjin and Zhang, Fang and Newman, Michael and Cai, Junjie and Gao, Xun and Tian, Zhengxiong and Wu, Junyin and Xu, Haihong and Yu, Huanjun and Yuan, Bo and others},
  journal={arXiv preprint arXiv:2005.06787},
  year={2020}
}

@inproceedings{yang2024loretta,
  title={Loretta: Low-rank economic tensor-train adaptation for ultra-low-parameter fine-tuning of large language models},
  author={Yang, Yifan and Zhou, Jiajun and Wong, Ngai and Zhang, Zheng},
  booktitle={Proceedings of the 2024 Conference of the North American Chapter of the Association for Computational Linguistics: Human Language Technologies (Volume 1: Long Papers)},
  pages={3161--3176},
  year={2024}
}

@inproceedings{veeramacheneni2025canonical,
  title={Canonical rank adaptation: An efficient fine-tuning strategy for vision transformers},
  author={Veeramacheneni, Lokesh and Wolter, Moritz and Kuehne, Hilde and Gall, Juergen},
  booktitle={Forty-second International Conference on Machine Learning},
  year={2025}
}

@inproceedings{ghiasvand2025decentralized,
  title={Decentralized low-rank fine-tuning of large language models},
  author={Ghiasvand, Sajjad and Alizadeh, Mahnoosh and Pedarsani, Ramtin},
  booktitle={Proceedings of the 1st Workshop for Research on Agent Language Models (REALM 2025)},
  pages={334--345},
  year={2025}
}

@article{zheng2022subttd,
  title={SubTTD: DOA estimation via sub-Nyquist tensor train decomposition},
  author={Zheng, Hang and Zhou, Chengwei and Shi, Zhiguo and de Almeida, Andr{\'e} LF},
  journal={IEEE Signal Processing Letters},
  volume={29},
  pages={1978--1982},
  year={2022},
  publisher={IEEE}
}

@article{xie2025coarray,
  title={Coarray tensor train decomposition for bistatic MIMO radar with uniform planar array},
  author={Xie, Qianpeng and Wang, Zhanling and Wen, Fangqing and He, Jin and Truong, Trieu-Kien},
  journal={IEEE Transactions on Antennas and Propagation},
  year={2025},
  publisher={IEEE}
}

@article{wang2023tensor,
  title={Tensor decompositions for hyperspectral data processing in remote sensing: A comprehensive review},
  author={Wang, Minghua and Hong, Danfeng and Han, Zhu and Li, Jiaxin and Yao, Jing and Gao, Lianru and Zhang, Bing and Chanussot, Jocelyn},
  journal={IEEE Geoscience and Remote Sensing Magazine},
  volume={11},
  number={1},
  pages={26--72},
  year={2023},
  publisher={IEEE}
}

@article{ravishankar2022hierarchical,
  title={A hierarchical approach for lossy light field compression with multiple bit rates based on tucker decomposition via random sketching},
  author={Ravishankar, Joshitha and Sharma, Mansi},
  journal={IEEE Access},
  volume={10},
  pages={56677--56690},
  year={2022},
  publisher={IEEE}
}

@article{yang2017loop,
  title={Loop optimization for tensor network renormalization},
  author={Yang, Shuo and Gu, Zheng-Cheng and Wen, Xiao-Gang},
  journal={Physical review letters},
  volume={118},
  number={11},
  pages={110504},
  year={2017},
  publisher={APS}
}

@article{gray2024hyperoptimized,
  title={Hyperoptimized approximate contraction of tensor networks with arbitrary geometry},
  author={Gray, Johnnie and Chan, Garnet Kin-Lic},
  journal={Physical Review X},
  volume={14},
  number={1},
  pages={011009},
  year={2024},
  publisher={APS}
}

@article{gao2025fermionic,
  title={Fermionic tensor network contraction for arbitrary geometries},
  author={Gao, Yang and Zhai, Huanchen and Gray, Johnnie and Peng, Ruojing and Park, Gunhee and Liu, Wen-Yuan and Kj{\o}nstad, Eirik F and Chan, Garnet Kin-Lic},
  journal={Physical Review Research},
  volume={7},
  number={2},
  pages={023193},
  year={2025},
  publisher={APS}
}

@article{juels1995stochastic,
  title={Stochastic hillclimbing as a baseline method for evaluating genetic algorithms},
  author={Juels, Ari and Wattenberg, Martin},
  journal={Advances in Neural Information Processing Systems},
  volume={8},
  year={1995}
}

@book{rybicki2024radiative,
  title={Radiative processes in astrophysics},
  author={Rybicki, George B and Lightman, Alan P},
  year={2024},
  publisher={John Wiley \& Sons}
}

@article{jiang2021implicit,
  title={An implicit finite volume scheme to solve the time-dependent radiation transport equation based on discrete ordinates},
  author={Jiang, Yan-Fei},
  journal={The Astrophysical Journal Supplement Series},
  volume={253},
  number={2},
  pages={49},
  year={2021},
  publisher={The American Astronomical Society}
}

@misc{morrison2016branch,
  title={Branch-and-bound algorithms: a survey of recent advances in searching, branching, and pruning. Discret Optim 19: 79--102},
  author={Morrison, DR and Jacobson, SH and Sauppe, JJ and Sewell, EC},
  year={2016}
}

@article{kaya2019computing,
  title={Computing dense tensor decompositions with optimal dimension trees},
  author={Kaya, Oguz and Robert, Yves},
  journal={Algorithmica},
  volume={81},
  number={5},
  pages={2092--2121},
  year={2019},
  publisher={Springer}
}

@article{harada2025tensor,
  title={Tensor tree learns hidden relational structures in data to construct generative models},
  author={Harada, Kenji and Okubo, Tsuyoshi and Kawashima, Naoki},
  journal={Machine Learning: Science and Technology},
  volume={6},
  number={2},
  pages={025002},
  year={2025},
  publisher={IOP Publishing}
}

@article{akamatsu2026plastic,
  title={Plastic tensor networks for interpretable generative modeling},
  author={Akamatsu, Katsuya O and Harada, Kenji and Okubo, Tsuyoshi and Kawashima, Naoki},
  journal={Machine Learning: Science and Technology},
  volume={7},
  number={1},
  pages={015014},
  year={2026},
  publisher={IOP Publishing}
}

@article{shi2006classical,
  title={Classical simulation of quantum many-body systems with a tree tensor network},
  author={Shi, Y-Y and Duan, L-M and Vidal, Guifre},
  journal={Physical Review A—Atomic, Molecular, and Optical Physics},
  volume={74},
  number={2},
  pages={022320},
  year={2006},
  publisher={APS}
}



\end{document}